\documentclass[11pt]{article}
\pdfoutput=1
\usepackage[utf8]{inputenc}
\usepackage{caption}
\usepackage{braket,slashed,bm}
\usepackage{jheppub}
\usepackage{array,multirow}
\usepackage{amsmath}
\usepackage[normalem]{ulem}
\usepackage{calligra}
\usepackage{floatrow}
\usepackage[T1]{fontenc}
\usepackage{mathrsfs}
\usepackage{subcaption}
\usepackage{lscape}
\usepackage{hyperref}
\usepackage{graphicx}
\usepackage{booktabs}
\newfloatcommand{capbtabbox}{table}[][\FBwidth]
\def\beq{\begin{equation}}
\def\eeq{\end{equation}}
\def\bea{\begin{eqnarray}}
\def\eea{\end{eqnarray}}
\def\nn{\nonumber \\}
\def\hyp{\mathsf{y}}
\renewcommand{\a}{\alpha}

\renewcommand{\d}{\delta}

\newcommand{\s}{\sigma}
\newcommand{\de}{\partial}

\newcommand{\Q}{Q}
\newcommand{\hv}{\hat{v}}
\renewcommand{\to}{\rightarrow}
\newcommand{\cHlt}{C_{Hl}^{(3)}}

\newcommand{\cHls}{C_{Hl}^{(1)}}
\newcommand{\cHqs}{C_{Hq}^{(1)}}
\newcommand{\aem}{\alpha_{ew}}
\newcommand{\hst}{s_{\hat\theta}}
\newcommand{\hct}{c_{\hat\theta}}
\newcommand{\hsdt}{s_{2\hat\theta}}
\newcommand{\hcdt}{c_{2\hat\theta}}
\newcommand{\dsw}{\d s_\theta}
\def\gcb{{\overline g_{1}}}
\def\gcw{{\overline g_{2}}}
\def\gcg{{\overline g_{3}}}

\def\tc{{\overline \theta}}

\def\ckin{c_{H,\text{kin}}}

\newcommand{\FR}{\textsc{FeynRules}}
\newcommand{\MG}{\textsc{MadGraph5}}
\newcommand{\MGNLO}{\textsc{MadGraph5\_aMC@NLO}}

\title{The SMEFTsim package, theory and tools}
\author[a]{Ilaria Brivio, Yun Jiang and Michael Trott}

\affiliation[a]{Niels Bohr International Academy \& Discovery Center,
Niels Bohr Institute, University of Copenhagen,
Blegdamsvej 17, DK-2100, Copenhagen, Denmark}

\abstract{We report codes for the Standard Model Effective Field Theory (SMEFT)
in \FR~ -- the SMEFTsim package.
The codes enable theoretical predictions for
dimension six operator corrections to the Standard Model using numerical tools, where predictions can be made based on
either the electroweak input parameter set $\{\hat{\alpha}_{ew}, \hat{m}_Z, \hat{G}_F \}$
or $\{\hat{m}_{W}, \hat{m}_Z, \hat{G}_F\}$.
All of the baryon and lepton number conserving operators present in the SMEFT dimension six Lagrangian, defined in the Warsaw basis, are included.
A flavour
symmetric ${\rm U}(3)^5$ version with possible non-SM $\rm CP$ violating phases, a (linear) minimal flavour violating version neglecting such phases, and the fully general flavour
case are each implemented. The SMEFTsim package
allows global constraints to be determined on the full Wilson coefficient space of the SMEFT. As the number of parameters present
is large, it is important to develop global analyses on reduced sets of parameters minimizing any UV assumptions
and relying on IR kinematics of scattering events and symmetries.
We simultaneously develop the theoretical framework of a ``W-Higgs-Z pole parameter'' physics program
that can be pursued at the LHC using this approach and the SMEFTsim package. We illustrate this methodology
with several numerical examples interfacing SMEFTsim with \MG.\\
\\
The SMEFTsim package can be downloaded at \href{https://feynrules.irmp.ucl.ac.be/wiki/SMEFT}{https://feynrules.irmp.ucl.ac.be/wiki/SMEFT}}

\begin{document}
\maketitle

\date{\today}

\section{Introduction}

When physics beyond the Standard Model (SM) resides at scales larger than the electroweak
scale ($\Lambda \gg \bar{v}_T$), one can utilise an expansion in this ratio of scales to construct an
Effective Field Theory (EFT).\footnote{Here $\bar{v}_T$ is the vev defined as the gauge independent
vacuum expectation value of the Higgs field including the effect of dimension six operators,
$\langle H^\dagger H \rangle \equiv \bar{v}_T^2/2$.}
Such an EFT can capture the low energy, or infrared (IR), limit of physics beyond the SM so long as
no light hidden states are in the particle spectrum and $\bar{v}_T/\Lambda < 1 $ is assumed/experimentally indicated.
When these conditions are satisfied, and
a $\rm SU_L(2)$ scalar doublet with hypercharge $\hyp_H = 1/2$ is assumed to be present in the IR limit of the underlying sector,
the theory that results from expanding in (currently) experimentally
accessible scales divided by the heavy scales of new physics
is the Standard Model Effective Field Theory (SMEFT).

The SMEFT is well defined and has been studied with increased theoretical sophistication
in recent years. This theory can capture the IR limit of a wide range of possible extensions of the SM, consistent with the stated assumptions.
Such SM extensions can address  the strong evidence for dark matter and neutrino masses in addition to the theoretical issue
of the hierarchy problem motivating $\sim {\rm TeV}$ scale new physics.
The interest in the SMEFT is due to
the significant growth in available experimental data due to the continued operation of the LHC, and
is also due to the theoretical developments reviewed in Ref.~\cite{Brivio:2017vri}.
In recent years, it has has become more widely understood that to gain the most out of studying the
current LHC data set, and the high luminosity LHC data set,
utilizing the SMEFT is valuable. Many LHC measurements will effectively be made below the scale $\Lambda$,
even if new states are discovered with masses $m \sim \Lambda$. This enables a practically useful,
and simplifying, expansion in $\sim \bar{v}_T/\Lambda$ when considering physics beyond the SM.

Putting in place this theoretical framework allows a general constraint program to be systematically developed over the long term, and
also enables the combination of LHC data with the extensive amount of lower energy data in a consistent field theory setting.
Efforts in this direction have been hampered by
the lack of a general coding of the leading SMEFT corrections to the SM in \FR  \, \cite{Christensen:2008py,Degrande:2011ua,Alloul:2013bka}
to date (using the theoretical approach of Sections \ref{Section:canonicalform}-\ref{Section:normalization}), to enable numerical studies.
A major result of this paper is to address this issue by reporting a
series of novel SMEFT implementations into \FR  \, that have been developed and are now released for public use at \href{http://feynrules.irmp.ucl.ac.be/wiki/SMEFT}{http://feynrules.irmp.ucl.ac.be/wiki/SMEFT}.
The codes supplied define the SMEFTsim package, and they cover two different approaches to how the SM Lagrangian parameters are
extracted from experimental measurements; i.e. the two electroweak input parameter
schemes $\{\hat{\alpha}_{ew}, \hat{m}_Z, \hat{G}_F \}$ and $\{\hat{m}_{W}, \hat{m}_Z, \hat{G}_F\}$.
SMEFTsim also consists of three different symmetry assumptions for the SMEFT operator basis, a fully flavour general SMEFT, a $\rm U(3)^5$ - SMEFT with non SM phases,
and a Minimal Flavour Violating (MFV) \cite{Chivukula:1987py,Buras:2000dm,DAmbrosio:2002vsn,Cirigliano:2005ck,Kagan:2009bn} version. We discuss these theories in Section \ref{Section:symassumptions} and
we present details of the structure of the coding of the SMEFTsim package in Section~\ref{codesections}.

Using SMEFTsim, LHC studies using EFT methods are still challenging as the number of real parameters present is very large in the leading lepton and baryon number conserving
corrections to the SM: there are 76 parameters in the
case the number of generations $n_f =1$ and 2499 parameters in the case $n_f =3$.  Even before this precise counting of parameters was determined \cite{Alonso:2013hga},
the understanding that the number of parameters was large led to pessimism that a general EFT approach could be pursued
with collider data. As is well known, to ensure the key point of
the EFT approach is not lost due to theoretical inconsistency,
a full non-redundant set of operators (allowed by the assumed symmetries) must be retained in a consistent EFT at each order in the power counting of the theory.

Retaining all operators in the SMEFT does not imply that global fits to interesting experimental data necessarily involve the full
set of 2499 parameters. Rather remarkably, a SMEFT physics program,
although challenging, can be practically carried out at the LHC. The second main result of this paper is to develop
the theoretical support for leading order (LO) EFT studies in a collider environment with subsets of parameters.
We lay out the theoretical foundation of this approach in Section~\ref{poleparam}  and define
a `WHZ pole parameter' program using this reasoning.

Reduced parameter sets can be adopted, despite neglecting terms the same order in the power counting of the EFT, for two main reasons.
First, flavour symmetry assumptions, well motivated out of low scale experimental constraints, can be used. This leads to consistent alternate theories
in addition to the general SMEFT: a $\rm U(3)^5$-SMEFT and a MFV-SMEFT. A simple corollary that we also systematically exploit defining the `WHZ pole parameters'
is that terms that violate symmetries preserved, or approximately preserved, in the SM interfere in a numerically suppressed fashion.
Second, the number of relevant parameters is dramatically reduced in a global study of processes
involving near on-shell intermediate narrow states of the SM. Exploiting such kinematics is already generic
in well measured processes that are distinguished from large non-resonant backgrounds in a hadron collider environment and we advocate pushing
this approach to its logical, and experimentally attainable limit.

Taking all of this into account, the feasibility of a `WHZ pole parameter' approach is illustrated in Table \ref{tab:parametercounts} -- which shows
a manageable set of parameters to simultaneously study and constrain considering the global data set.
We advocate dedicated experimental analyses be developed along these lines taking advantage of this dramatic simplification
of the SMEFT approach at LHC. We demonstrate this numerically using SMEFTsim and \MG \, \cite{Alwall:2014hca} in Section~\ref{sec:numerics}.

So long as an appropriate theoretical error is assigned
for this reduction in parameters, this approach can be adopted without introducing undue UV bias or blocking the possibility of
building an inverse map to new physics sectors through the SMEFT. This is because these are IR assumptions and simplifications of the SMEFT
projected into well measured LHC observables.
Although the number of parameters is still $23$ in the case of one generation and $46$ in the case of three generations, we note that the number
of models considered and experimentally constrained in the past decades in dedicated particle physics studies is substantially larger.
As soon as a decoupling limit  $\bar{v}_T/\Lambda < 1$ is present, vast arrays of possible extensions to the SM project into the compact and well defined SMEFT formalism.
LHC results indicate that at this time it is reasonable to systematically consider and use the assumption that $\bar{v}_T/\Lambda < 1$ via the SMEFT. In this case, it is
much more efficient
to project experimental results into the SMEFT using SMEFTsim, rather than into a endless series of models based on yet more assumptions. We encourage the LHC experimental collaborations
to develop and study this approach using the tool provided.

\begin{table}[t]\centering
  \begin{tabular}{c|c|c|c}
  Case & $\rm CP$ even & $\rm CP$ odd & WHZ Pole parameters \\ \hline
  General SMEFT ($n_f = 1$) & 53 \, \cite{Alonso:2013hga} & 23 \, \cite{Alonso:2013hga} &   $\sim 23$ \\
  General SMEFT ($n_f = 3$) & 1350 \, \cite{Alonso:2013hga} & 1149  \, \cite{Alonso:2013hga} &  $\sim 46$ \\
  $\rm U(3)^5$ SMEFT & $\sim 52$  & $\sim 17$ & $\sim 24$ \\
  \rm MFV \, \, SMEFT & $\sim 108$ & - & $\sim 30$ \\
  \hline
  \end{tabular}
  \caption{Parameter counts in the general SMEFT flavour cases for $n_f$ generations, and the approximate number of
  parameters that feed into a W-Higgs-Z pole parameter program in the Warsaw basis, as discussed in Section \ref{poleparam}. Also shown are the
  parameter counts in the $\rm U(3)^5$ limit and the MFV SMEFT case. The symbol $\sim$ indicates that these latter results are approximate counts for a leading
  order analysis, with leading flavour breaking spurion insertions, as discussed in Section~\ref{poleparam}.} \label{tab:parametercounts}
\end{table}

\section{Notation, canonical normalization and gauge}\label{Section:canonicalform}
Our formulation of the SMEFT is based upon Refs.~\cite{Grinstein:1991cd,Alonso:2013hga,Berthier:2015oma,Berthier:2015gja,Berthier:2016tkq,Brivio:2017bnu,Brivio:2017vri,deFlorian:2016spz,Passarino:2016pzb}.
We use the Warsaw basis for $\mathcal{L}^{(6)}$ as defined in Ref.~\cite{Grzadkowski:2010es}.
The SMEFT is constructed out of a series of $\rm SU_C(3) \times SU_L(2) \times U_Y(1)$ invariant (local and analytic) higher dimensional operators built out
of the SM fields. The Lagrangian is given as
\bea
\mathcal{L}_{SMEFT} = \mathcal{L}_{SM} + \mathcal{L}^{(5)} + \mathcal{L}^{(6)} + \mathcal{L}^{(7)} + ...,
\quad \quad \mathcal{L}^{(d)}= \sum_{i = 1}^{n_d} \frac{C_i^{(d)}}{\Lambda^{d-4}} Q_i^{(d)} \hspace{0.25cm} \text{ for $d > 4$,}
\eea
with the SM Lagrangian \cite{Glashow:1961tr,Weinberg:1967tq,Salam:1968rm},
defined as
\bea\label{sm1}
\mathcal{L} _{\rm SM} &=& -\frac14 G_{\mu \nu}^A G^{A\mu \nu}-\frac14 W_{\mu \nu}^I W^{I \mu \nu} -\frac14 B_{\mu \nu} B^{\mu \nu}
+ \! \! \! \! \! \sum_{\psi=q,u,d,\ell,e} \overline \psi\, i \slashed{D} \, \psi \\
&+&(D_\mu H)^\dagger(D^\mu H) -\lambda \left(H^\dagger H -\frac12 v^2\right)^2- \biggl[ H^{\dagger j} \overline d\, Y_d\, q_{j}
+ \widetilde H^{\dagger j} \overline u\, Y_u\, q_{j} + H^{\dagger j} \overline e\, Y_e\,  \ell_{j} + \hbox{h.c.}\biggr], \nonumber
\eea
where $H$ is an $\rm SU_L(2)$ scalar doublet and $\tilde{H}_j = \epsilon_{jk} H_k^\dagger$ with $\epsilon_{kj} = - \epsilon_{jk}$ and $\epsilon_{12} = 1$,
$j,k = \{1,2 \}$ and topological Lagrangian terms are neglected. The gauge covariant derivative is defined with the convention $D_\mu = \partial_\mu + i g_3 T^A A^A_\mu + i g_2  t^I W^I_\mu + i g_1 \hyp B_\mu$, where $T^A$ are the $\rm SU_c(3)$ generators,  $t^I=\tau^I/2$ are the $\rm SU_L(2)$ generators, and $\hyp$ is the $\rm U_Y(1)$ hypercharge
generator.\footnote{This covariant derivative convention is the same as adopted in Ref.~\cite{Grzadkowski:2010es}, and
opposite to the usual convention in \FR~\cite{Christensen:2008py,Degrande:2011ua,Alloul:2013bka}.}
The Yukawa matrices are complex with the relation $Y_i = \sqrt{2} M_i/v$ to the complex mass matrices $M_i$, whose real eigenvalues are the fermion masses.
The chiral projectors are defined as $P_{R/L} = (1 \pm \gamma_5)/2$. The fields $\{q,\ell\}$ are left handed and the fields $\{e,u,d\}$ are right handed. We use the definition $\sigma_{\mu \, \nu} = i \, [\gamma_\mu,\gamma_\nu]/2$
and at times the short hand notation $\psi$ for a general fermion field, and $X^\mu$ for a general gauge field is used.

The number of non-redundant operators in $\mathcal{L}^{(5)}$, $\mathcal{L}^{(6)}$, $\mathcal{L}^{(7)}$ and $\mathcal{L}^{(8)}$ is
known \cite{Buchmuller:1985jz,Grzadkowski:2010es,Weinberg:1979sa,Wilczek:1979hc,Abbott:1980zj,Lehman:2014jma,Lehman:2015coa,Henning:2015alf}.\footnote{The general algorithm to determine the number of operators at higher orders in the SMEFT's defining expansion has been developed
in Refs.~\cite{Lehman:2015via,Lehman:2015coa,Henning:2015daa,Henning:2015alf,Henning:2017fpj}.}
The operators  $Q_i^{(d)}$ are suppressed by $d-4$ powers of the cutoff scale $\Lambda$,
and the $C_i^{(d)}$ are the Wilson coefficients.
The explicit definition of the $\mathcal{L}^{(6)}$ operators used here
are given in Ref.~\cite{Grzadkowski:2010es} and listed in Table \ref{op59}. We absorb factors of $1/\Lambda^2$ into the Wilson coefficients
as a notational choice unless otherwise noted. Utilizing the Warsaw basis is theoretically favoured
as it is the only $\mathcal{L}^{(6)}$ basis that has been completely renormalized to date
in Refs.~\cite{Grojean:2013kd,Jenkins:2013zja,Jenkins:2013wua,Alonso:2013hga,Alonso:2014zka}.

We use notation where the parameters of the canonically normalized Lagrangian (i.e. couplings, masses) carry bar superscripts.\footnote{This notation should not be confused with bar notation used to
denote the Dirac adjoint - $\bar{\psi}$.} The canonically normalized fields are generally indicated with
a script font: $\{\mathcal{G}, \mathcal{W}, \mathcal{B} \}$. The procedure for canonically normalizing is the same in both input parameter schemes,
and we follow the approach laid out in Ref.~\cite{Alonso:2013hga}.
In unitary gauge, the Higgs doublet is expanded as
\begin{align}
H &= \frac{1}{\sqrt 2} \left(\begin{array}{c}
0 \\
 \left[ 1+ \ckin \right]  h + \bar{v}_T
 \end{array}\right),
 \label{Hvev}
\end{align}
where
\begin{align}
\ckin &\equiv \left(C_{H\Box}-\frac14 C_{HD}\right)\bar{v}^2, &
\bar{v}_T &\equiv \left( 1+ \frac{3 C_H \bar{v}^2}{8 \lambda} \right) \bar{v}.
\end{align}
This results in a canonically normalized $h$ field when the Lagrangian is written in mass eigenstate fields.
Note that the distinction between $\bar{v}_T$ and $\bar{v}$ is at dimension eight when $\bar{v}$ multiplies a Wilson coefficient $C_i$.
As such we can trade $C_i \, \bar{v}_T^2 \leftrightarrow C_i \, \bar{v}^2$ to the accuracy we are working.
The gauge fields are redefined as
\begin{align}\label{canonical1}
G_\mu^A &= \mathcal{G}_\mu^A \left(1 + C_{HG} \bar{v}_T^2 \right), &
W^I_\mu  &=  \mathcal{W}^I_\mu \left(1 + C_{HW} \bar{v}_T^2 \right), &
B_\mu  &=  \mathcal{B}_\mu \left(1 + C_{HB} \bar{v}_T^2 \right),
\end{align}
to take them to canonical form. The modified coupling constants are also redefined
\begin{align}\label{canonicalagain}
\gcg &= g_3 \left(1 + C_{HG} \, \bar{v}_T^2 \right), & \gcw &= g_2 \left(1 + C_{HW} \, \bar{v}_T^2 \right), & \gcb &= g_1 \left(1 + C_{HB} \, \bar{v}_T^2 \right),
\end{align}
so that the products $g_3 G_\mu^A=\gcg \mathcal{G}_\mu^A$, etc.\ are unchanged (at $\mathcal{O}(1/\Lambda^2)$) when canonically normalizing the theory.

Furthermore, the mass eigenstate basis for $\{\mathcal{W}^3_\mu,\mathcal{B}_\mu\}$ in the SMEFT is given by~\cite{Grinstein:1991cd,Alonso:2013hga}
\begin{align}
\left[ \begin{array}{cc}  \mathcal{W}_\mu^3 \\ \mathcal{B}_\mu \end{array} \right]
&=
\left[ \begin{array}{cc}  1   &  -  \frac{1}{2} \,  v_T^2 \,  C_{HWB} \\
- \frac{1}{2} \,  v_T^2 \,  C_{HWB} & 1 \end{array} \right] \, \left[ \begin{array}{cc} \cos \tc  &  \sin \tc \\
-\sin \tc & \cos \tc \end{array} \right] \left[ \begin{array}{cc}  \mathcal{Z}_\mu \\ \mathcal{A}_\mu \end{array} \right],
\end{align}
while the mass eigenstate fields of the SM $\{Z_\mu,A_\mu \}$ are defined via
\begin{align}
\left[ \begin{array}{cc}  W_\mu^3 \\ B_\mu \end{array} \right]
&=   \left[ \begin{array}{cc} \cos \theta  &  \sin \theta \\
-\sin \theta & \cos \theta \end{array} \right] \left[ \begin{array}{cc}  Z_\mu \\ A_\mu \end{array} \right],
\end{align}
with $\cos \theta = g_2/\sqrt{g_1^2 + g_2^2}$, $\sin \theta = g_1/\sqrt{g_1^2 + g_2^2}$. The relation between the
mass eigenstate fields in the two theories are
\begin{align}
  Z_\mu &=\mathcal Z_\mu\left( 1+ \frac{\hst^2 C_{HB}}{\sqrt{2}\hat{G}_F} + \frac{\hct^2  C_{HW}}{\sqrt{2}\hat{G}_F} + \frac{\hst \hct C_{HWB}}{\sqrt{2}\hat{G}_F}
  \right)\nonumber \\
  &+ \mathcal A_\mu\left( \frac{\hst \hct (C_{HW}-C_{HB})}{\sqrt{2}\hat{G}_F} - \frac{(1- 2 \hst^2) C_{HWB}}{2 \sqrt{2}\hat{G}_F}+\frac{\dsw^2}{2\hst\hct}
  \right),\\
  A_\mu &=\mathcal A_\mu\left( 1+ \frac{\hct^2 C_{HB}}{\sqrt{2}\hat{G}_F} + \frac{\hst^2  C_{HW}}{\sqrt{2}\hat{G}_F} - \frac{\hst \hct C_{HWB}}{\sqrt{2}\hat{G}_F}
  \right)\nonumber \\
  &+ \mathcal Z_\mu\left( \frac{\hst \hct (C_{HW}-C_{HB})}{\sqrt{2}\hat{G}_F} - \frac{(1- 2 \hst^2) C_{HWB}}{2 \sqrt{2}\hat{G}_F}-\frac{\dsw^2}{2\hst\hct}
  \right).
\end{align}
These expressions hold in both input parameter schemes. The notation $\hat\theta$ denotes the weak angle defined in
terms of input parameters (see next section). The three angles $\theta,\,\bar\theta,\,\hat\theta$ differ by quantities
proportional to the $\mathcal{L}^{(6)}$ Wilson coefficients. When such factors multiply a Wilson coefficient $C_i$, the three notations
are equivalent up to neglected dimension eight corrections.

The SMEFTsim codes use unitary gauge and several simplifications that
have taken place are dependent on this gauge choice. Nevertheless, the approach laid out here does not have any intrinsic theoretical assumption
that blocks it being extended to other gauges to enable next to leading order (NLO) SMEFT results.\footnote{See Ref.~\cite{Brivio:2017vri} for more discussion on this point.}
This approach to canonically normalizing the mass eigenstate Lagrangian in the SMEFT has been shown to be extendable to
$R_\xi$ gauge fixing in Ref.~\cite{Dedes:2017zog}.
Gauging the SMEFT is subtle and will not be discussed at length here but we note that utilizing unitary gauge for one loop calculations is well known to
be best avoided. Even
$R_\xi$ gauge requires a careful treatment of novel ghost interactions introduced in the gauge fixing terms rotating between the gauge
and mass eigenstates, as discussed in Ref.~\cite{Hartmann:2015oia,Ghezzi:2015vva,Dedes:2017zog}. It has been shown that the related issues involving ghosts in one loop calculations
can be overcome when the leading order (LO - i.e. only retaining $\mathcal{L}^{(6)}$ corrections) approach of this work is adopted.

\section{Treatment of inputs}\label{Section:inputschemes}
\subsection{\texorpdfstring{$\{\hat{\alpha}_{ew}, \hat{m}_Z, \hat{G}_F \}$}{\{alpha\_ew, m\_Z, G\_f\}} input parameter scheme}
We use notation where the input parameters used to define the numerical values of Lagrangian parameters, and quantities derived from input parameters carry hat superscripts.
To determine the numerical value of the SM Lagrangian parameters from the $\{\hat{\alpha}_{ew}, \hat{m}_Z, \hat{G}_F \}$ EW inputs, the $\mathcal{L}_{SM}$ Lagrangian parameters are fixed by the following definitions at tree level (with $\hct^2 = 1- \hst^2$)
\begin{equation}\label{alphainputs}
 \begin{aligned}
\hat e &=  \sqrt{4\pi\hat{\a}_{ew}}, & \quad  \hat g_1 &= \frac{\hat{e}}{\hct}, & \quad  \hat g_2 &= \frac{\hat{e}}{\hst},\\
   \hst^2 &= \frac{1}{2}\left[1-\sqrt{1-\frac{4\pi\hat{\a}_{ew}}{\sqrt2\hat G_F \hat m_Z^2}}\right], & \quad
 \hat v_T &= \frac{1}{2^{1/4}\sqrt{\hat G_F}},
 &
 \hat m_W^2 &= \hat m_Z^2\hct^2.
 \end{aligned}
\end{equation}
The Lagrangian parameters in the SMEFT differ from the SM Lagrangian terms due to $\mathcal{L}^{(6)}$ local operator corrections.
A generic parameter $\kappa$ receives a shift from its SM value due to $\mathcal{L}^{(6)}$ operators given by
\begin{equation}
 \d \kappa = \bar\kappa - \hat\kappa\,,
\end{equation}
and in the SM limit ($C_i\to 0$) one has $ \d \kappa \rightarrow 0$. We define the short hand notation\footnote{The $\rm U(3)^5$ limit used here treats the two flavour contractions
$(C_{\substack{ll}} \delta_{mn} \, \delta_{op} + C'_{\substack{ll}}  \delta_{mp} \, \delta_{no})(\bar l_m \gamma_\mu l_n)(\bar l_o \gamma^\mu l_p)$ as independent \cite{Cirigliano:2009wk}.}
\begin{align}\label{coreshifts}
\d m_h^2 &=  \frac{\hat m_h^2}{\sqrt2\hat G_F}\left(-\frac{3C_H }{2\lambda} + 2 \,\sqrt{2} \, \hat{G}_F \, \ckin \right), &
 \d G_F &= \frac{1}{\hat{G}_F} \left(C^{(3)}_{\substack{Hl}} - \frac{C_{\substack{ll}}+ C_{\substack{ll}}'}{4}\right),
\\
 \d m_Z^2 &= \frac{1}{2 \, \sqrt{2}} \, \frac{\hat{m}_Z^2}{\hat{G}_F} C_{HD} + \frac{2^{1/4} \sqrt{\pi \hat{\alpha}} \,
 \hat{m}_Z}{\hat{G}_F^{3/2}} C_{HWB}, & \quad
\d  m_W^2 &= \hat m_W^2 \left(\sqrt2 \d G_F + 2 \frac{\d g_2}{\hat g_2}\right),
 \end{align}
and using this notation, related results are \cite{Alonso:2013hga,Berthier:2015oma,Berthier:2015gja,Berthier:2016tkq,Brivio:2017bnu,Brivio:2017vri}
\begin{align}
 \d v_T^2 =\bar v_T^2 - \hat v_T^2&= \frac{\d G_F}{\hat G_F},\\
 \d g_1 =\bar g_1-\hat g_1&=\frac{\hat g_1}{2\hcdt}\left[\hst^2\left(\sqrt2 \d G_F+\frac{\d m_Z^2}{\hat m_Z^2}\right)+\hct^2\,\hsdt\bar v_T^2 C_{HWB}\right],\\
 \d g_2 =\bar g_2 - \hat g_2&=-\frac{\hat g_2}{2\hcdt}\left[
 \hct^2\left(\sqrt2 \d G_F+\frac{\d m_Z^2}{\hat m_Z^2}\right)+\hst^2\,\hsdt\bar v_T^2 C_{HWB}\right],\\
 \dsw^2 = s_{\bar \theta}^2- \hst^2 &= 2\hct^2\hst^2\left(\frac{\d g_1}{\hat g_1}-\frac{\d g_2}{\hat g_2}\right) +\bar v_T^2  \frac{\hsdt\hcdt}{2} C_{HWB}.
\end{align}
Using the $\{\hat{\alpha}_{ew}, \hat{m}_Z, \hat{G}_F \}$ EW input parameters one has $\{\bar{\alpha}_{ew}, \bar{m}_Z\}=\{\hat{\alpha}_{ew}, \hat{m}_Z\}$.
We still define a parameter $\d m_Z^2$ for latter convenience. Note that $\bar{G}_F$ for EW applications is defined as $\bar{G}_F = 1/\sqrt{2} \, \bar{v}_T^2$.
The input parameters are measured at different experimental scales $p^2 \simeq \{0, \hat{m}_Z^2, \hat{m}_{\mu}^2 \}$ and
are defined as follows.

\subsubsection{Extraction of \texorpdfstring{$\hat{\alpha}_{ew}$}{alpha\_ew}}
The extraction of $\hat{\alpha}_{ew}$  occurs in the measurement of the Coulomb potential
of a charged particle in the  Thomson limit ($p^2 \rightarrow 0$). The low scale extraction of $\hat{\alpha}_{ew}$ in the
SMEFT is given by
\bea
- i \, \left[\frac{4 \, \pi \, \hat{\alpha}_{ew}(q^2)}{q^2}\right]_{q^2 \rightarrow 0}
\equiv \frac{- i \, (\bar{e}_0)^2}{q^2} \left[1 + \frac{\Sigma^{AA}(q^2)}{q^2} \right]_{q^2 \rightarrow 0}.
\eea
In this expression, $\Sigma^{AA}$ is the two point function of the canonically normalized photon field in the SMEFT at a fixed order in perturbation theory.
Formally unknown finite terms in the low scale extraction are present due to the vacuum polarization of the photon
in the $q^2 \rightarrow 0$ limit, and in addition further related uncertainties are introduced using this input parameter in running this low scale parameter up through the
Hadronic resonance region (due to $\pi^\pm$ loops etc.). The expression for $\Sigma^{AA}$ is generally rearranged into the form
\bea
\left[\frac{\Sigma^{AA}(q^2)}{q^2} \right]_{q^2 \rightarrow 0} = {\rm Re} \frac{\Sigma^{AA}(m_Z^2)}{m_Z^2} - \left[\frac{{\rm Re} \Sigma^{AA}(m_Z^2)}{m_Z^2}  - \left[\frac{\Sigma^{AA}(q^2)}{q^2} \right]_{q^2 \rightarrow 0}\right],
\eea
where the last quantity in square brackets dominantly leads to an uncertainty that is far larger than the low scale measurement uncertainty in the limit
$q^2 \rightarrow 0$.
This introduces a significant numerical uncertainty as these nonperturbative contributions must be estimated. This is the dominant effect, leading to a parameter
$\tilde{\alpha}_{ew}(\hat{m}_Z^2)$ defined incorporating this correction and leading to the reduced
theoretical precision given by \cite{Olive:2016xmw,Mohr:2015ccw}
\bea
1/\tilde{\alpha}_{ew}(\hat{m}_Z) = 127.950 \pm 0.014, \quad {\rm while} \quad 1/\hat{\alpha}_{ew}(0) = 137.035 999139(31).
\eea
We use as a code input the value of $\hat{\alpha}_{ew}(\hat{m}_Z) = \tilde{\alpha}_{ew}(\hat{m}_Z)$ given in Refs.~\cite{Olive:2016xmw,Mohr:2012tt} which includes
an estimate of this numerical uncertainty. When $\hat{\alpha}_{ew}$ is used as an input parameter, the mapping of this expression
to Lagrangian parameters is given in Eqn.~\ref{alphainputs}.

\subsubsection{Extraction of \texorpdfstring{$\hat{G}_{F}$}{G\_F}}
The extraction of $\hat{G}_F$ defined by the Effective Lagrangian generated in the $p^2 \ll \hat{m}_W^2$ limit of the SM interactions (see Ref.~\cite{Jenkins:2017jig})
is defined at the scale $\mu^2 \sim m_\mu^2$
\begin{align}\label{GFdefinition}
\mathcal{L}_{G_F} &\equiv  -\frac{4\hat{G}_F}{\sqrt{2}} \, \left(\bar{\nu}_\mu \, \gamma^\mu P_L \mu \right) \left(\bar{e} \, \gamma_\mu P_L \nu_e\right),
\end{align}
with the measurement of $\mu^- \rightarrow e^- + \bar{\nu}_e + \nu_\mu$ defining the numerical value through
the measurement of the total muon lifetime ($\tau_\mu$). The extraction of $\hat{G}_F$ in this manner is subject to the condition that
the $\nu$ flavours are summed over experimentally.
This is required so that unitarity allows the neglect of contributions to the corresponding decay rate due to the
Pontecorvo-Maki-Nakagawa-Sakata (PMNS) matrix \cite{Pontecorvo:1957cp,Maki:1962mu} relating the weak and mass
eigenstates of the Neutrino's, which is suppressed in Eqn.~\ref{GFdefinition}.

When $\hat{G}_{F}$ is used as an input parameter in the SMEFT, the introduction of the shift $\delta G_F$ is still required for EW
studies at the LHC. This is because the predictions of observables for LHC do not follow from the lower scale ($\mu^2 \sim m_\mu^2 \ll \bar{v}_T^2$) Effective Lagrangian given in Eqn.~\ref{GFdefinition}.

\subsubsection{Extraction of \texorpdfstring{$\hat{m}_{Z}$}{m\_Z}}
The extraction of $\hat{m}_Z$ is defined in the simultaneous fit to the LEPI pseudo-observables $\{\hat{m}_Z,\hat{\Gamma}_Z, \hat{\sigma}_{had}^0,\hat{R}_\ell^0\}$
as defined in Ref.~\cite{ALEPH:2005ab} that occurs in the pole scan through the $Z$ mass that was preformed at LEP.\footnote{
For more discussion on the interpretation of these pseudo-observables (PO) in the SMEFT see Ref.~\cite{Brivio:2017vri}.}
The extraction of $\hat{m}_Z$ occurs with the subtraction of soft initial and final state QED radiation (captured in a QED radiator function denoted $R_{QED}$)
from the peak cross-section $\sigma^0$ as \cite{ALEPH:2005ab}
\bea
\sigma_{\bar{f}f}^0 = \frac{12 \, \pi}{\hat{m}_Z^2} \, \frac{\hat{\Gamma}_{ee} \, \hat{\Gamma}_{\bar{f}f}}{\hat{\Gamma}_{Z}^2} \, \frac{1}{R_{QED}}, \quad \sigma_{had}^0 =
\frac{12 \, \pi}{\hat{m}_Z^2} \, \frac{\hat{\Gamma}_{ee} \, \hat{\Gamma}_{had}}{\hat{\Gamma}_{Z}^2} \, \frac{1}{R_{QED}}, \quad R_\ell^0 = \frac{\hat{\Gamma}_{had}}{\hat{\Gamma}_{\ell}}.
\eea
The input $\hat{m}_{Z}$ is defined to be extracted from the simultaneous fit to these pseudo-observables.
When $\hat{m}_{Z}$ is used as an input the introduction of the shift $\delta m_Z^2$ given in Eqn.~\ref{coreshifts}
still appears  in some cases, when experimental predictions depend on the inferred values of $\hat{g}_1,\hat{g}_2$.

\subsection{\texorpdfstring{$\{\hat{m}_{W}, \hat{m}_Z, \hat{G}_F \}$}{\{m\_W, m\_Z, G\_f\}} input parameter scheme}
Using the $\{\hat{m}_{W}, \hat{m}_Z, \hat{G}_F \}$ input parameter scheme offers several theoretical advantages:
\begin{itemize}
\item{The use of $\hat{m}_{W}$ has been shown to have a subdominant
measurement bias to the quoted experimental error in the SMEFT \cite{Bjorn:2016zlr}. This has not been established for $\hat{\alpha}_{ew}$.
Furthermore, the impressive intrinsic precision of the low scale measurements of $\hat{\alpha}_{ew}$ is not relevant for the comparison of the two schemes,
due to the large error introduced running $\hat{\alpha}_{ew}$ up through the hadronic resonance region being dominant.
Due to this, the percentage errors of each input parameter are within a factor of two as far as
global constraint studies at EW scales are concerned.}
\item{The use of $\hat{m}_{W}$ as an input allows SMEFT studies to expand around the physical poles defining
scattering amplitudes such as $\bar{\psi} \, \psi \rightarrow \bar{\psi} \, \psi \, \bar{\psi} \, \psi$ through charged currents in a double pole
expansion, leading to more consistent global constraint studies, see Refs.~\cite{Berthier:2016tkq,Brivio:2017vri} for more discussion.}
\item{The use of an $\hat{m}_{W}$ input parameter has some advantages when developing one loop results in the SMEFT, see the discussion in Ref.~\cite{Ghezzi:2015vva,Gauld:2015lmb,Gauld:2016kuu,Hartmann:2016pil}. In addition,
the scales of the input parameters are closer together using $\{\hat{m}_{W}, \hat{m}_Z, \hat{G}_F \}$ reducing logarithmic enhancements
present in running $\hat{\alpha}_{ew}$ for LHC predictions, and related mixing effects with $\mathcal{L}^{(6)}$ operators.}
\end{itemize}
The precise extraction of $\hat{m}_{W}$ at the Tevatron occurred historically after LEPI-II operations. This fact largely explains the
current dominance of the $\{\hat{\alpha}_{ew}, \hat{m}_Z, \hat{G}_F \}$ input scheme. Initial investigations of
the input scheme dependence of the global constraint conclusions in the SMEFT framework
indicate that this scheme dependence is below experimental uncertainties \cite{Brivio:2017bnu}.
The advantages of the
$\{\hat{m}_{W}, \hat{m}_Z, \hat{G}_F \}$ input scheme are substantial enough that transitioning to this approach is theoretically favoured. We provide codes utilizing each input parameter scheme to aid in this transition.
Using the $\{\hat{m}_{W}, \hat{m}_Z, \hat{G}_F \}$ input scheme, the SM Lagrangian parameters are defined as
\begin{equation}\label{mwinputs}
 \begin{aligned}
\hat e &=  2\cdot 2^{1/4} \hat{m}_W\sqrt{\hat{G}_F} s_{\hat\theta}, & \quad  \hat g_1 &= 2\cdot 2^{1/4}\hat{m}_Z\sqrt{\hat{G}_F \left(1 -\frac{\hat{m}_W^2}{\hat{m}_Z^2}\right)},
 & \quad  \hat g_2 &= 2\cdot 2^{1/4}\hat{m}_W\sqrt{\hat{G}_F},\\
   \hst^2 &= 1-\frac{\hat{m}_W^2}{\hat{m}_Z^2}, & \quad
 \hat v_T &= \frac{1}{2^{1/4}\sqrt{\hat G_F}}.
 \end{aligned}
\end{equation}
In this scheme $\{\delta G_F, \delta m_h^2\}$ are unchanged and
\bea\label{coreshiftsmw}
 \frac{\delta \alpha}{2 \, \hat{\alpha}}  &=& -\frac{\delta G_F}{\sqrt{2}} + \dfrac{\d m_Z^2}{\hat{m}_Z^2} \frac{\hat{m}_W^2}{2 \, (\hat{m}_W^2 - \hat{m}_Z^2)}
  - \frac{C_{HWB}}{\sqrt{2} \, \hat{G}_F} \frac{\hat{m}_W}{\hat{m}_Z} \, s_{\hat\theta}, \\
  \delta s_{\bar \theta}^2 &=& 2\hct^2\hst^2\left(\frac{\d g_1}{\hat g_1}-\frac{\d g_2}{\hat g_2}\right) +\bar v_T^2  \frac{\hsdt\hcdt}{2} C_{HWB}, \\
  \dfrac{\d m_Z^2}{\hat{m}_Z^2} &=&
  \dfrac{1}{2\sqrt2 \hat{G}_F}C_{HD}+\dfrac{\sqrt{2}}{\hat{G}_F}\dfrac{\hat{m}_W}{\hat{m}_Z}\sqrt{1-\dfrac{\hat{m}_W^2}{\hat{m}_Z^2}}C_{HWB}, \quad  \quad \quad \quad \quad \dfrac{\d m_W^2}{\hat{m}_W^2} = 0, \\
    \d g_1 &=& - \frac{\hat{g}_1}{\sqrt{2}} \, \delta G_F  - \hat{g}_1 \, \dfrac{\d m_Z^2}{2 \,s_{\hat{\theta}}^2 \, \hat{m}_Z^2},\\
    \d g_2 &=& - \frac{\hat{g}_2}{\sqrt{2}} \, \delta G_F.
    \eea
    \subsubsection{Extraction of \texorpdfstring{$\hat{m}_W$}{m\_W}}\label{transversemass}
    An input parameter $\hat{m}_W$ can be extracted using a fit to
    the transverse mass $m_W^T$ at hadron colliders. Recall that when utilizing transverse variables (defined in the plane orthogonal to the collision axis) one defines a missing $E_T$ vector
    \bea
    {\vec{E}_T^{miss}} = - \sum_i {\vec{p}_{T}(i)},
    \eea
    summing over all visible final state particles $i$. ${\vec{E}_T^{miss}}$ is so reconstructed in the case of $W \rightarrow \ell \nu$ decays
    and this defines $|\vec{p}^{\, \nu}_T|$. Combined with a measured $p_T$ of an identified $\ell$ (the momentum of the lepton in the plane transverse to the collision axis), these variables are used to construct
    \bea
    (m_W^T)^2 = 2 |\vec{p}^{\,\ell}_T||\vec{p}^{\, \nu}_T| (1 - \cos \phi_{\ell \nu})
    \eea
where $\phi_{\ell \nu}$ is the angle between the leptons in the plane perpendicular
    to the collision axis. In the limit of $|\vec{p}_T^{\, W}| \rightarrow 0$ one has $m_T^2 = s \sin^2 \theta$
where $\theta$ is defined as the angle between the $W$ boson decay products and the beam axis in the $W$ boson rest frame
and $s$ is the partonic energy of the produced $W$. The Jacobian of transforming between the variable $\theta$ and $m_T$
is given in the $p_T^W = 0$ case by $m_T/ (s- m_T^2)$. This introduces a sharp Jacobian peak in the $m_T$ spectrum that allows
an extraction of the $W$ mass from the shape of the spectra and its endpoint. Precise extractions of $\hat{m}_W$ are strongly impacted by detector resolution effects and $p_T^W \neq 0$
requiring template fits to the derived spectra to fit for $\hat{m}_W$.
See Refs.~\cite{vanNeerven:1982mz,Bjorn:2016zlr,Abazov:2012bv,Aaltonen:2012bp,Aaboud:2017svj} for more details and the mapping to $\hat{m}_{W}$ from such spectra.

\subsection{Numerical values of inputs}
\begin{table}[t]\centering
  \begin{tabular}{ccc}
  Input parameters & Value & Ref. \\ \hline
  $\tilde{\alpha}_{ew}(\hat m_Z)$ & $1/(127.950 \pm 0.017) $ &  \cite{Olive:2016xmw,Mohr:2012tt} \\
  $\hat{m}_{W}$ & $ 80.365 \pm 0.016 \, \, {\rm[GeV]}$ &  \cite{Aaltonen:2013iut} \\\hline
  $\hat{m}_Z$ & $91.1876 \pm 0.0021 \, \, {\rm[GeV]}$ & \cite{Z-Pole,Olive:2016xmw,Mohr:2012tt} \\
  $\hat{G}_F$ & $1.1663787(6) \times 10^{-5} {\rm[GeV]^{-2}}$ &  \cite{Olive:2016xmw,Mohr:2012tt} \\

  $\hat{m}_h$ & $125.09 \pm 0.21 \pm 0.11 \, \, {\rm[GeV]}$ & \cite{Aad:2015zhl} \\
  $\hat{\alpha}_s(\hat m_Z)$&$0.1181 \pm 0.0011$&\cite{Olive:2016xmw}\\
  \hline
  $\hat{m}_e$& 	   $0.5109989461(31) \times 10^{-3} \, \, {\rm[GeV]}$ & \cite{Olive:2016xmw}\\
  $\hat{m}_\mu$&   $105.6583745(24)\times 10^{-3} \, \, {\rm[GeV]}$ & \cite{Olive:2016xmw}\\
  $\hat{m}_\tau$&  $1.77686 \pm 0.00012 \, \, {\rm[GeV]} $ & \cite{Olive:2016xmw}\\
  $\hat{m}_u$&     $2.2\substack{+0.6\\-0.4} \times 10^{-3} \, \, {\rm[GeV]}$ & \cite{Olive:2016xmw}\\
  $\hat{m}_c$&     $1.28 \pm 0.03 \, \, {\rm[GeV]}$ & \cite{Olive:2016xmw}\\
  $\hat{m}_t$&     $173.21 \pm 0.51 \pm 0.71 \, \, {\rm[GeV]}$ & \cite{Olive:2016xmw}\\
  $\hat{m}_d$&     $4.7\substack{+0.5\\-0.4} \times 10^{-3} \, \, {\rm[GeV]}$ & \cite{Olive:2016xmw}\\
  $\hat{m}_s$&     $0.096\substack{+0.008\\-0.004} \, \, {\rm[GeV]} $ & \cite{Olive:2016xmw}\\
  $\hat{m}_b$&     $4.18\substack{+0.04\\-0.03}  \, \, {\rm[GeV]}$ & \cite{Olive:2016xmw}\\
  \hline
  \end{tabular}
  \caption{Set of parameters used as inputs and corresponding numerical values. Only one parameter between $\tilde\alpha_{ew}(\hat m_Z)$ and $\hat m_W$ is retained, depending on the input scheme chosen.
  Note that the value of the $\hat m_W$ is the Tevatron extracted value, not the global average that includes LEP extractions of $\hat m_W$ that are
  harder to interpret in the SMEFT \cite{Bjorn:2016zlr}.} \label{tab:inputs}
\end{table}
The numerical values used to define the mass and coupling input parameters in each scheme are given in Table~\ref{tab:inputs}.
In addition, when including flavour violating effects, the Cabibbo Kobayashi Maskawa (CKM) matrix \cite{GellMann:1960np,Cabibbo:1963yz,Kobayashi:1973fv} is defined
through the Wolfenstein parameterization \cite{Wolfenstein:1983yz} as
\bea
V_{\rm CKM} &=& \left(\begin{array}{ccc}
c_{12} \, c_{13} & s_{12} \, c_{13} & s_{13}  \, e^{- i \delta}\\
- s_{12} \, c_{23}  - c_{12} \, s_{23} \, s_{13} \, e^{i \delta}& c_{12} \, c_{23} - s_{12} \, s_{23} \, s_{13} \, e^{i \delta}& s_{23} \, c_{13} \\
s_{12} \, s_{23} - c_{12} \, c_{23}  \, s_{13} \, e^{i \delta}& -c_{12} \, s_{23} - s_{12} \, c_{23}  \, s_{13} \, e^{i \delta}& c_{23} \, c_{13} \\
\end{array}\right), \\
&\approx&
\begin{pmatrix}
1-\lambda^2/2&  \lambda&  		A  \lambda^3 (\rho - i \eta)\\
-\lambda&       1-\lambda^2/2& 	A  \lambda^2\\
A \lambda^3(1-\rho-i \eta)& 	-A \lambda^2&	1
 \end{pmatrix}.
\eea
where the numerical parameters are input as ~\cite{Olive:2016xmw}:
\begin{center}
\begin{tabular}{>{$}c<{$}>{\tt}ll}
\text{parameter}& \rm name& value\\\hline
 \lambda&	CKMlambda& 0.22506 $\pm$ 0.00050\\
 A&		CKMA&	   0.811 $\pm$ 0.026\\
 \rho&		CKMrho&	   0.124 $\substack{+0.019\\-0.018}$\\
 \eta&		CKMeta&   0.356 $\pm$ 0.011
\end{tabular}
\end{center}
Here ``name" refers to the label of the parameter in the \FR \, codes. Note that the
Pontecorvo Maki Nakagawa Sakata (PMNS) \cite{Pontecorvo:1957qd,Maki:1962mu} matrix is not implemented in SMEFTsim as
neutrino masses are neglected, but it can be directly incorporated in an extension.

In the context of the SMEFT, experimental
extractions and fits to the CKM matrix elements get corrections due to $\mathcal{L}^{(6)}$ operators. Such corrections define
a difference between ``bar'' and ``hatted'' CKM quantities that are neglected here. The reason we have neglected these effects
on the CKM inputs is that to our knowledge, no complete analysis in the SMEFT defining such corrections to the global fit to Wolfenstein
parameters exists in the literature. Analyses that build up results central to the effort to determine such corrections
include Refs.~\cite{Cirigliano:2009wk,Cirigliano:2012ab,Gonzalez-Alonso:2016etj,Falkowski:2017pss,Jenkins:2017jig}.
When such results are available they will be included in the SMEFTsim package as an update.

\subsubsection{Fermion mass inputs}
The fermion mass inputs are given in Table \ref{tab:inputs}. The relation of these measured quantities and SMEFT Lagrangian
parameters is more subtle than in the SM. Following Ref.~\cite{Alonso:2013hga}
\bea
\mathcal{L}_{{\cal Y}} =- h\ \overline \psi_{R,r} \, \left[ {\cal Y}_\psi \right]_{rs} \, \psi_{L,s} + \ldots
\eea
and  the relation between the complex fermion mass matrix
$\left[ M_\psi \right]_{rs}$ and effective complex Yukawa coupling in the SMEFT is given by
\begin{align}
\left[ M_\psi \right]_{rs} &= \frac{\bar{v}_T}{\sqrt 2} \left( \left[Y_\psi \right]_{rs}   - \frac12 \bar{v}_T^2 C^*_{\substack{\psi H \\ sr}}   \right), \quad \quad
\left[ {\cal Y}_\psi \right]_{rs}
= \frac{1}{\bar{v}_T}\left[ M_\psi \right]_{rs} \left[ 1+ \ckin  \right]   - \frac{\bar{v}_T^2}{\sqrt{2}} C^*_{\substack{\psi H \\ sr}}.
\end{align}
These matrices are not simultaneously diagonalizable. Nevertheless the rotational freedom of the fermion kinetic terms (see Eqn.~\ref{weakmassrotations})
that leaves the kinetic terms invariant
allows the mass matrix to be taken to diagonal form so that
\bea
\mathcal{U}(\psi,R)^\dagger \left[ M_\psi \right]_{rs} \mathcal{U}(\psi,L) \equiv {\rm diag}\{\hat{m}^1_\psi,\hat{m}^2_\psi,\hat{m}^3_\psi\}.
\eea
The effective Yukawa matrices are then off diagonal in general and complex in the mass eigenstate basis
\bea
\mathcal{U}(\psi,R)^\dagger  \,  \left[ {\cal Y}_\psi \right]_{rs} \, \mathcal{U}(\psi,L) &=& 2^{1/4} \hat{G}_F \, {\rm diag}\{\hat{m}^1_\psi,\hat{m}^2_\psi,\hat{m}^3_\psi\} \left[ 1+ \ckin - \frac{\hat{G}_F}{\sqrt{2}} \delta v_T^2 \right], \\
&-& \frac{1}{2 \, \hat{G}_F} \left[\mathcal{U}(\psi,R)^\dagger C^*_{\substack{\psi H}} \, \mathcal{U}(\psi,L)\right]_{rs}, \nonumber
\eea
In taking $\left[ M_\psi \right]_{rs}$ to diagonal form, non-SM phases that are present  in the SMEFT are shifted into the general $3 \times 3$ $V_{CKM}$
matrix. The rephasing freedom of the SM fermion field kinetic terms still reduces $V_{CKM}$ to a unitary matrix with the same number of parameters as in the SM
(i.e. three real parameters and one phase). New non-SM relative phases do persist in the effective Yukawa couplings in general. In the case of ${\rm U}(3)^5$ symmetry
where $C^*_{\substack{\psi H \\ rs}} \rightarrow  C^*_{\substack{\psi H}} \left[Y_\psi \right]_{rs}$ with $C^*_{\substack{\psi H}} \in \mathbb{C}$ non-SM phases remain.
In the case of MFV being assumed
$C^*_{\substack{\psi H \\ rs}} \rightarrow  C^*_{\substack{\psi H}} \left[Y_\psi \right]_{rs}$ and $C^*_{\substack{\psi H}} \in \mathbb{R}$,
and non-SM phases are absent.
\subsubsection{Remaining inputs}
The remaining inputs in either scheme are the Higgs mass and the strong coupling given directly by
\begin{align}
\hat{m}_h^2 &\equiv \bar{m}_h^2 \left(1 + \frac{\delta \hat{m}_h^2}{\hat{m}_h^2} \right), &
\hat{g}_3 &\equiv  \sqrt{4 \pi \hat{\alpha}_s}.
\end{align}
The Higgs mass is directly extracted from the fitted reconstructed peaks in
$h \rightarrow \gamma \, \gamma$ and
$h \rightarrow Z^\star Z \rightarrow \ell^+ \, \ell^- \, \ell^+ \, \ell^-$, see Ref.~\cite{Aad:2015zhl}
for a discussion. The extraction of $\hat{g}_3$ is performed by multiple methods.
One of the most prominent in recent years is a joint fit to $e^+ e^-$ event shapes and leading non-perturbative
corrections, for a review see Ref.~\cite{Bethke:2011tr}.
The Lagrangian parameter $\bar{\lambda}$ is a derived quantity from the input of $\hat{m}_h$ and $\hat{G}_F$
which defines
\begin{align}
\hat{\lambda} &\equiv \frac{\hat{m}_h^2 \, \hat{G}_F}{\sqrt{2}},   & \bar{\lambda} = \hat{\lambda} \left(1 - \frac{\delta \hat{m}_h^2}{\hat{m}_h^2} - \sqrt{2} \, \delta G_F\right).
\end{align}

\section{Flavour symmetries in the operator basis}\label{Section:symassumptions}
\subsection{Flavour symmetry assumptions}
  A general SMEFT contains a large number of real parameters in $\mathcal{L}^{(6)}$, as listed in Table \ref{tab:parametercounts}.
There are 1350/53 $\rm CP$-even parameters and 1149/23 $\rm CP$-odd parameters in $\mathcal{L}^{(6)}$ for three/one generations \cite{Alonso:2013hga}.
Most of the parameters in the SMEFT are in
the $\psi^4$ operators due to flavour indices. This makes clear the importance of flavour symmetry assumptions and carefully utilizing
numerical suppressions of $\psi^4$ operators contributing to cross-sections.

We present codes that span several different flavour symmetry assumptions in the operator basis: a flavour
symmetric ${\rm U}(3)^5$ case that allows $\rm CP$ violating phases, a (linear) minimal flavour violating version where flavour change
follows the SM pattern and new $\rm CP$ violating phases are neglected,
and the general $\mathcal{L}^{(6)}$ case.  In this section, we summarize the required theoretical results for
each case.
\subsubsection{\texorpdfstring{${\rm U}(3)^5$}{U(3)\^{} 5} limit}
The ${\rm U}(3)^5$ limit refers to the limit of unbroken global flavour symmetry in the SM Lagrangian, restored in the limit $Y_{u,d,e} \rightarrow 0$.
To define this global symmetry group we define the relation between the weak (unprimed) basis and the mass (primed) basis as
\begin{align}\label{weakmassrotations}
u_L &= \mathcal{U}(u,L)\, u_L^\prime, & u_R &= \mathcal{U}(u,R)\, u_R^\prime,
& \nu_L &= \mathcal{U}(\nu,L)\, \nu_L^\prime, \\
d_L &= \mathcal{U}(d,L)\, d_L^\prime, & d_R &= \mathcal{U}(d,R)\, d_R^\prime, &
e_L &= \mathcal{U}(e,L)\, e_L^\prime, & e_R &= \mathcal{U}(e,R)\, e_R^\prime.
\end{align}
Each $\mathcal{U}$ rotation defines a ${\rm U}(3)$ flavour group. The ${\rm U}(3)^5$ group of the SM is defined as
\bea
{\rm U}(3)^5 = \mathcal{U}(u,R) \times \mathcal{U}(d,R) \times \mathcal{U}(Q,L)\times \mathcal{U}(\ell,L) \times \mathcal{U}(e,R).
\eea
The relative $\mathcal{U}$ rotations between components of the lepton and quark ${\rm SU}_L(2)$ doublet fields define the PMNS
and CKM matrices as
\bea
V_{\rm CKM} = \mathcal{U}(u,L)^\dagger \, \mathcal{U}(d,L), \quad \quad U_{\rm PMNS} = \mathcal{U}(e,L)^\dagger \, \mathcal{U}(\nu,L).
\eea
At times, it is useful to have defined the unbroken flavour groups of the SM (with the $U(1)$ global flavour number groups removed) as the quark and lepton subgroups
\begin{align}
\rm G_q &= \rm SU_{u_R}(3) \times SU_{d_R}(3) \times SU_{q_L}(3), & \quad \quad
\rm G_{\ell} &= \rm SU_{\ell_L}(3)  \times SU_{e_R}(3),
\end{align}
and a Yukawa matrix transforms as $Y_i \sim \{\rm G_q, G_\ell\}$ for these groups. Yukawa spurion
transformations are defined as
\begin{align}\label{spurion}
Y_u &\sim (3,1,\bar{3},1,1), & \quad \quad Y_d & \sim (1,3,\bar{3},1,1), & \quad \quad
Y_e &\sim (1,1,1,\bar{3},3),
\end{align}
so that one can restore the full $\{G_q, G_{\ell}\}$ flavour symmetry by inserting the Yukawa matrices in a manner
that makes flavour singlet structures manifest.
Furthermore, retaining only the top and bottom quark Yukawa coupling, defines a phenomenologically interesting
breaking of the ${\rm U}(3)^5$ limit
\begin{align}
\mathcal{U}(e,R)^\dagger \, Y_e \, \mathcal{U}(e,L) &\approx \left[ \begin{array}{ccc} 0 & 0 & 0 \\ 0 & 0 & 0 \\ 0 & 0 & 0 \end{array}\right],
& \mathcal{U}(d,R)^\dagger \, Y_d \, \mathcal{U}(d,L) & \approx \left[ \begin{array}{ccc} 0 & 0 & 0 \\ 0 & 0 & 0 \\ 0 & 0 & y_b \end{array}\right], \\
 \mathcal{U}(u,R)^\dagger \, Y_u \, \mathcal{U}(u,L)& \approx \left[ \begin{array}{ccc} 0 & 0 & 0 \\ 0 & 0 & 0 \\ 0 & 0 & y_t \end{array}\right],
\end{align}
where $y_i = \sqrt{2} m_i/v$ for a quark of mass $m_i$ in the SM. We supply a \FR \,  code implementing full diagonal Yukawa matrices
that directly simplify numerically to this ${\rm U}(3)^5$ breaking limit. Below we demonstrate the breaking of this limit that occurs numerically
in the $\rm U(3)^5$ codes, and only retain the leading breaking terms linear in $y_b,y_t$.

The $\mathcal{L}^{(6)}$ operators
are broken down to the Classes given in Table \ref{op59}. The Wilson coefficients of the operators
in Classes 1,2,3 and 4 are unchanged going to the  $\rm U(3)^5$ flavour symmetric limit and allowing complex Wilson coefficients.
The following Wilson coefficients are then defined in the $\rm U(3)^5$ limit. \\
\\
Class 5, ($\psi^2H^3$):
\bea\label{class5ops}
C_{\substack{e H \\ rs}} \, Q_{\substack{e H \\ rs}} &\rightarrow& C_{\substack{e H}} \, [Y_e^\dagger]_{rs} \, Q_{\substack{e H \\ rs}} \approx 0, \\
C_{\substack{d H\\ rs}} \, Q_{\substack{d H \\ rs}} &\rightarrow& C_{\substack{d H}} \, [Y_d^\dagger]_{rs} \, Q_{\substack{d H  \\ rs}} \approx  y_b^\star \, C_{\substack{b H}} \, Q'_{\substack{b H \\ 33}}, \\
C_{\substack{u H \\ rs}} \, Q_{\substack{u H \\ rs}} &\rightarrow& C_{\substack{u H}} \, [Y_u^\dagger]_{rs} \, Q_{\substack{u H \\ rs}}  \approx  y_t^\star \, C_{\substack{t H}} \, Q'_{\substack{t H \\ 33}},
\eea
In the last step we have further neglected all Yukawa's other than the top and bottom Yukawa's in the leading breaking of the $\rm U(3)^5$ limit
while rotating to the mass eigenstate basis. The operators where the fermion fields are taken to the mass eigenstate basis are indicated with a prime superscript.
Note that this is an IR limit defined in the SMEFT and $\{C_{\substack{b H}},C_{\substack{t H}}\} \in \mathbb{C}$ in general. The breaking of the $ \rm U(3)^5$
limit also follows from inserting spurions that are functions of $[Y_u Y_u^\dagger]$,$[Y_d Y_d^\dagger]$. In these terms, the expansion in the $y_{t}^2 \sim 1$ dependence
can be considered to be implicitly absorbed into an effective Wilson coefficient parameter for the
Class 5 and 6 operators.
\\
\\
Class 6 ($\psi^2 X H$) operators:
\begin{align}\label{class6ops}
C_{\substack{u G \\ rs}} \, Q_{\substack{u G \\ rs}} &\rightarrow C_{\substack{u G}} \, [Y_{u}^\dagger]_{rs} \, Q_{\substack{u G \\ rs}} \, \, \, \approx  y_t^\star C_{\substack{t G}} \, Q'_{\substack{t G \\ 33}}, \\
C_{\substack{d G \\ rs}} \, Q_{\substack{d G \\ rs}} &\rightarrow C_{\substack{d G}} \, [Y_{d}^\dagger]_{rs} \, Q_{\substack{d G \\ rs}} \, \, \, \approx  y_b^\star C_{\substack{b G}} \, Q'_{\substack{b G \\ 33}}, \\
C_{\substack{u W \\ rs}} \, Q_{\substack{u W \\ rs}} &\rightarrow C_{\substack{u W}} \, [Y_{u}^\dagger]_{rs} \, Q_{\substack{u W \\ rs}}\approx  y_t^\star C_{\substack{t W}} \, Q'_{\substack{t W \\ 33}}, \\
C_{\substack{d W \\ rs}} \, Q_{\substack{d W \\ rs}} &\rightarrow C_{\substack{d W}} \, [Y_{d}^\dagger]_{rs} \, Q_{\substack{d W \\ rs}}\, \approx  y_b^\star C_{\substack{b W}} \, Q'_{\substack{b W \\ 33}},\\
C_{\substack{u B \\ rs}} \, Q_{\substack{u B \\ rs}} &\rightarrow C_{\substack{u B}} \, [Y_{u}^\dagger]_{rs} \, Q_{\substack{u B \\ rs}} \,\,\, \approx  y_t^\star C_{\substack{t B}} \, Q'_{\substack{t B \\ 33}}, \\
C_{\substack{d B \\ rs}} \, Q_{\substack{d B \\ rs}} &\rightarrow C_{\substack{d B}} \, [Y_{d}^\dagger]_{rs} \, Q_{\substack{d B \\ rs}}  \,\,\, \approx  y_t^\star C_{\substack{b B}} \, Q'_{\substack{b B \\ 33}}, \\
C_{\substack{e W \\ rs}} \, Q_{\substack{e W \\ rs}} &\rightarrow C_{\substack{e W}} \, [Y_e^\dagger]_{rs} \, Q_{\substack{e W \\ rs}} \,  \approx  0, \\
C_{\substack{e B \\ rs}} \, Q_{\substack{e B \\ rs}} &\rightarrow C_{\substack{e B}} \, [Y_e^\dagger]_{rs} \, Q_{\substack{e B \\ rs}}\,\,\, \approx  0,
\end{align}
in the $\rm U(3)^5$ limit the Wilson coefficients of the $\psi^2H^3$ and $\psi^2 X H$ operators $\in \mathbb{C}$ as $Y_{e,u,d}$ are $3 \times 3$ complex matrices
in general.
In the last approximation, again,  all Yukawa's other than the top and bottom are neglected in the leading breaking of the $\rm U(3)^5$ limit
while rotating to the mass eigenstate basis.\\
\\
Class 7 ($\psi^2 H^2 D$):
\begin{align}
  C_{\substack{H l \\ rs}}^{(1)} \, Q_{\substack{H l \\ rs}}^{(1)} &\leadsto C_{\substack{H l}}^{(1)} \, Q_{\substack{H l \\ rr}}^{' (1)}, & \quad
  C_{\substack{H l \\ rs}}^{(3)} \, Q_{\substack{H l \\ rs}}^{(3)} &\leadsto C_{\substack{H l}}^{(3)} \,  Q_{\substack{H l \\ rr}}^{' (3)}, \\
  C_{\substack{H q \\ rs}}^{(1)} \, Q_{\substack{H q \\ rs}}^{(1)} &\leadsto C_{\substack{H q}}^{(1)} \, Q_{\substack{H q \\ rr}}^{' (1)}, & \quad
  C_{\substack{H q \\ rs}}^{(3)} \, Q_{\substack{H q \\ rs}}^{(3)} &\leadsto C_{\substack{H q}}^{(3)} \, Q_{\substack{H q \\ rr}}^{' (3)}, \\
  C_{\substack{H e \\ rs}} \, Q_{\substack{H e \\ rs}} &\leadsto C_{\substack{H e}} \, Q'_{\substack{H e \\ rr}},& \quad
  C_{\substack{H u \\ rs}} \, Q_{\substack{H u \\ rs}} &\leadsto C_{\substack{H u}} \, Q'_{\substack{H u \\ rr}}, \\
  C_{\substack{H d \\ rs}} \, Q_{\substack{H d \\ rs}} &\leadsto C_{\substack{H d}} \, Q'_{\substack{H d \\ rr}},
\end{align}
and $C_{\substack{Hud \\ rs}} \, Q_{\substack{Hud \\ rs}} \rightarrow C_{\substack{Hud}} [Y_u \, Y^\dagger_d]_{rs} \, Q_{\substack{Hud \\ rs}} \approx y_t \, y^\star_b \, (V_{CKM})_{33}\, C_{\substack{Hud}} \, Q'_{\substack{Hud \\ 33}}$
(with $C_{\substack{Hud}}\in \mathbb{C}$) in the leading breaking of the $\rm U(3)^5$ limit.  Rotating to the mass eigenstate basis has formally led to a redefinition of the initial Wilson coefficient of the form
$C \rightarrow \mathcal{U}(L/R)^\dagger C \, \mathcal{U}(L/R)$ and these rotations have been absorbed into a redefinition of the Wilson coefficients on the right hand side of the $\leadsto$.
An implicit sum over flavour indices $rr = \{11,22,33\}$ has been used here. As the operators $\{Q_{\substack{H l}}^{(1,3)},Q_{\substack{H q}}^{(1,3)},Q_{\substack{H e}},Q_{\substack{H u}},Q_{\substack{H d}}\}$
are self Hermitian, $\{C_{\substack{H l}}^{(1,3)},C_{\substack{H q}}^{(1,3)},C_{\substack{H e}},Q_{\substack{H u}},C_{\substack{H d}}\} \in \mathbb{R}$.

The $\rm U(3)^5$ limit of the four fermion operators is more subtle. The $(\bar{L}L)(\bar{L}L)$ operators simplify to \cite{Cirigliano:2009wk,Alonso:2013hga}
\begin{align}
C_{\substack{l l \\ prst}} Q_{\substack{l l \\ prst}} &\leadsto (C_{\substack{l l}} \, Q'_{\substack{l l \\ pptt}} + \mathcal{C}_{\substack{l l}} \, Q'_{\substack{l l \\ pttp}}), & \quad
C^{(1)}_{\substack{l q \\ prst}} Q^{(1)}_{\substack{l q \\ prst}} &\leadsto C^{(1)}_{\substack{l q}} \, Q^{' (1)}_{\substack{l q \\ pptt}}, \\
C_{\substack{qq \\ prst}}^{(1)} Q_{\substack{qq \\ prst}}^{(1)} &\leadsto (C^{(1)}_{\substack{qq}} \, Q_{\substack{qq \\ pptt}}^{' (1)} + \mathcal{C}^{(1)}_{\substack{qq}} \, Q_{\substack{qq \\ pttp}}^{' (1)}), & \quad
C_{\substack{l q \\ prst}}^{(3)} Q^{(3)}_{\substack{l q \\ prst}} &\leadsto C^{(3)}_{\substack{l q}} \, Q^{' (3)}_{\substack{l q \\ pptt}}, \\
C_{\substack{qq \\ prst}}^{(3)} Q_{\substack{qq \\ prst}}^{(3)} &\leadsto (C^{(3)}_{\substack{qq}} \, Q_{\substack{qq \\ pptt}}^{' (3)} + \mathcal{C}^{(3)}_{\substack{qq}} \, Q_{\substack{qq \\ pttp}}^{' (3)}),
\end{align}
with $\{C_{\substack{l l}}, \mathcal{C}_{\substack{l l}}, C^{(1)}_{\substack{qq}}, \mathcal{C}^{(1)}_{\substack{qq}}, C^{(3)}_{\substack{qq}}, \mathcal{C}^{(3)}_{\substack{qq}},C^{(1)}_{\substack{l q}},C^{(3)}_{\substack{l q}}\} \in \mathbb{R}$
due to the operators being self-Hermitian and not transforming under any external group. Furthermore,
The use of $C$ and $\mathcal{C}$ denote the two different flavour contractions and the operators with all fermion fields identical
also satisfy $Q_{ijkl} = Q_{klij}$ due to relabeling freedom of dummy indices
in all flavour symmetry cases, and in the flavour general case.
The $(\bar{R}R)(\bar{R}R)$ operators simplify in the $\rm U(3)^5$ limit to
\begin{align}
C_{\substack{ee \\ prst}} Q_{\substack{ee \\ prst}} &\leadsto C_{\substack{ee}} \, Q'_{\substack{ee \\ pptt}}, & \quad
C_{\substack{uu \\ prst}} Q_{\substack{uu \\ prst}} &\leadsto (C_{\substack{uu}} \, Q'_{\substack{uu \\ pptt}} + \mathcal{C}_{\substack{uu}} \, Q'_{\substack{uu \\ pttp}}),\\
C_{\substack{eu \\ prst}} Q_{\substack{eu \\ prst}} &\leadsto C_{\substack{eu}} \, Q'_{\substack{eu \\ pptt}}, & \quad
C_{\substack{dd \\ prst}} Q_{\substack{dd \\ prst}} &\leadsto (C_{\substack{dd}} \, Q'_{\substack{dd \\ pptt}} + \mathcal{C}_{\substack{dd}} \, Q'_{\substack{dd \\ pttp}}),\\
C_{\substack{ed \\ prst}} Q_{\substack{ed \\ prst}} &\leadsto C_{\substack{ed}} \, Q'_{\substack{ed \\ pptt}}, & \quad
C_{\substack{ud \\ prst}}^{(1)} Q_{\substack{ud \\ prst}}^{(1)} &\leadsto C_{\substack{ud}}^{(1)} \, Q_{\substack{ud \\ pptt}}^{'(1)}, \\
C_{\substack{ud \\ prst}}^{(8)} Q_{\substack{ud \\ prst}}^{(8)} &\leadsto C_{\substack{ud}}^{(8)} \, Q_{\substack{ud \\ pptt}}^{' (8)},
\end{align}
where the $C_{\substack{ee}}$ operator only allows one term due to the fact that the $e$ fields are singlets under $\rm{SU}_C(3) \times \rm{SU}_L(2) \times \rm{U}_Y(1)$,
and a Fierz identity \cite{Alonso:2013hga} reduces the number of effective parameters.
Again $\{C_{\substack{e e}}, C_{\substack{u u}}, \mathcal{C}_{\substack{u u}}, C_{\substack{d d}}, \mathcal{C}_{\substack{d d}},C_{\substack{e u}}, C_{\substack{e d}}, \mathcal{C}_{\substack{u d}}^{(1)}, C_{\substack{u d}}^{(8)}\} \in \mathbb{R}$
due to Hermitian operators that are not transforming under an external flavour group.

The $(\bar{L}L)(\bar{R}R)$ operators are trivial and have one real Wilson coefficient for each operator.
The chirality flipping $\psi^4$ operators have the $\rm U(3)^5$ limits
\begin{align}
C_{\substack{l e d q \\ prst}} Q_{\substack{l e d q \\ prst}} &\rightarrow C_{\substack{l e d q}} \, [Y_e^\dagger]_{rs} [Y_d]_{pt} \, Q_{\substack{l e d q \\ rspt}} \, \, \approx 0, \\
C_{\substack{q u q d \\ prst}}^{(1)} Q^{(1)}_{\substack{q u q d \\ prst}} &\rightarrow C^{(1)}_{\substack{q u q d}} \, [Y_u^\dagger]_{rs} [Y_d^\dagger]_{pt} \, Q^{(1)}_{\substack{q u q d \\ rspt}}
\approx C^{(1)}_{\substack{q u q d}} \, y_t^\star y_b^\star \, Q^{' (1)}_{\substack{q u q d \\ 3333}}, \\
C_{\substack{q u q d \\ prst}}^{(8)} Q^{(8)}_{\substack{q u q d \\ prst}} &\rightarrow C^{(8)}_{\substack{q u q d}} \, [Y_u^\dagger]_{rs} [Y_d^\dagger]_{pt} \, Q^{(8)}_{\substack{q u q d \\ rspt}}
\approx C^{(8)}_{\substack{q u q d}} \, y_t^\star y_b^\star \, Q^{' (8)}_{\substack{q u q d \\ 3333}}, \\
C_{\substack{l e q u \\ prst}}^{(1)} Q^{(1)}_{\substack{l e q u \\ prst}} &\rightarrow C^{(1)}_{\substack{l e q u}} \, [Y_e^\dagger]_{rs} [Y_u^\dagger]_{pt} \, Q^{(1)}_{\substack{l e q u \\ rspt}}
\, \, \approx 0, \\
C_{\substack{l e q u \\ prst}}^{(3)} Q^{(3)}_{\substack{l e q u \\ prst}} &\rightarrow C^{(3)}_{\substack{l e q u}} \, [Y_e^\dagger]_{rs} [Y_u^\dagger]_{pt} \, Q^{(3)}_{\substack{l e q u \\ rspt}}
\, \, \approx 0,
\end{align}
with $\{C_{\substack{l e d q}},C^{(1)}_{\substack{q u q d}},C^{(8)}_{\substack{q u q d}},C^{(1)}_{\substack{l e q u}},C^{(3)}_{\substack{l e q u}}\} \in \mathbb{C}$.
This limit does not forbid $\rm CP$ violation beyond the SM due to the presence of complex Wilson coefficients.
Again, in the last approximation all Yukawa's other than the top and bottom are neglected.\\
\subsubsection{MFV breaking}
Assuming that a $\rm CP$ violating phase only appears in $\mathcal{L}^{(6)}$ due to the
SM source of CP violation present in the CKM matrix, and that the breaking of flavour symmetry in the SMEFT
follows the breaking pattern in the SM, defines the MFV paradigm \cite{DAmbrosio:2002vsn}
(see also Refs.~\cite{Chivukula:1987py,Buras:2000dm,Cirigliano:2005ck,Kagan:2009bn}).

The reasons to adopt these strong symmetry
assumptions are basically twofold. First, the set of experimental constraints derived in the flavour physics program push a naive flavour violating suppression scale
in the SMEFT expansion $\Lambda \gg {\rm TeV}$, rendering SMEFT studies of LHC in this case unlikely to extract evidence of physics beyond the SM.\footnote{
See Refs.\cite{Buras:2000dm,DAmbrosio:2002vsn,Isidori:2010kg}
for discussion on flavour changing physics beyond the SM bounds.} The converse point also holds in that if SMEFT studies do uncover deviations in LHC data,
then an underlying sector must be consistent with strong lower energy flavour constraints. If this occurs due to a MFV symmetry breaking pattern then this symmetry
assumption acts to reduce tuning of parameters. Second, and equally important is that the introduction of such (IR) symmetry assumptions render systematic SMEFT constraint studies
feasible to practically carry out.\footnote{It is arguably possible that a completely flavour general constraint program can also be carried
out in the SMEFT,
see Refs.\cite{Efrati:2015eaa,Falkowski:2017pss} for analyses aiming at the flavour general case.}
\\

We introduce the Jarlskog invariant \cite{Jarlskog:1985ht,Jarlskog:1985cw,Kobayashi:1973fv,Wolfenstein:1983yz} as
\bea
J = c_{12} c_{13}^2 c_{23} s_{12} s_{13} s_{23} \sin (\delta) \simeq A^2 \lambda^6 \eta (1- \lambda^2/2) \simeq 3 \times 10^{-5},
\eea
in the MFV limit (at leading order in the MFV expansion)
\bea
\{C_{\tilde{G}},C_{\tilde{W}}, C_{H \tilde{G}}, C_{H \tilde{W}}, C_{H \tilde{B}}, C_{H \tilde{W} B}\} &\propto& J, \\
\{C_{eH},C_{uH},C_{dH},C_{eW},C_{eB},C_{uW},C_{uB},C_{uG},C_{dW},C_{dB},C_{dG},C_{Hud}\}  &\sim& \mathbb{R}, \\
\{C_{\substack{l e d q}},C^{(1)}_{\substack{q u q d}},C^{(8)}_{\substack{q u q d}},
C^{(1)}_{\substack{l e q u}}, C^{(3)}_{\substack{l e q u}}\}  &\in& \mathbb{R},
\eea
which renders $\{C_{\tilde{G}},C_{\tilde{W}}, C_{H \tilde{G}}, C_{H \tilde{W}}, C_{H \tilde{B}}, C_{H \tilde{W} B}\}$ and
\bea
{\rm Im} \{C_{eH},C_{uH},C_{dH},C_{eW},C_{eB},C_{uW},C_{uB},C_{uG},C_{dW},C_{dB},C_{dG},C_{Hud}\},
\eea
neglectable in studies that also neglect next to leading order corrections, such as the codes reported here, as loop suppressions are $\propto g^2_{SM}/16 \pi^2 \gg J $
are not systematically included.

MFV does not preserve flavour as in the  $\rm U(3)^5$ limit, but dictates that the flavour breaking pattern follows the SM.
 Following Ref.~\cite{DAmbrosio:2002vsn} an MFV expansion can be constructed by expanding in flavour invariants
determined using the spurion transformation properties in Eqn.~\ref{spurion}.
The Class $X^3$, $H^6$, $D^2 H^4$, $X^2 H^2$  Lagrangian terms are unchanged from the $\rm U(3)^5$ limit results given above.
The leading MFV breaking spurion's are given by
\bea\label{spurions}
S^q_{pr} &=& [Y_u^\dagger Y_u]_{pr} + \Delta [Y_d^\dagger Y_d]_{pr}  \approx y_t^2 [V_{3p} V_{3r}^\star] + \Delta \, y_b^2 [V_{3p} V_{3r}^\star], \\
S^u_{pr} &=& [Y_u Y_u^\dagger]_{pr} \approx y_t^2 \delta_{p3}\delta_{r3},\\
S^d_{pr} &=& [Y_d Y_d^\dagger]_{pr} \approx y_b^2 \delta_{p3}\delta_{r3}.
\eea
Here and below $\Delta$ indicates a relative normalization of terms resulting from the spurion insertions and the first line is
simplified with the implicit understanding that the $y_t^2/y_b^2$ spurion breakings leading to flavour change will be inserted for the $d_L/u_L$ fields respectively
expanding out the $Q_L$ doublet field.
The spurions transform as $\{\bf{8},\bf{1},\bf{1}\}$, $\{\bf{1},\bf{8},\bf{1}\}$, $\{\bf{1},\bf{1},\bf{8}\}$ for $S^u_{pr},S^d_{pr},S^q_{pr}$
under the global favour symmetries $\rm G_q$. The $S^q$ spurion that dictates flavour changing neutral currents, acts to absorb the rotation matrices
between the weak and mass eigenstate bases for the fermion fields. Similarly the $S^u,S^d$ spurions absorb the rotation matrices
going to the mass eigenstate basis fields. Incorporating the effects of all of these spurion breakings
leads to the following extra parameters in the supplied codes constructed in unitary gauge.

For the Class 5,6 ($\psi^2 H^3$,$\psi^2 X H$) operators, the effects of the spurions $S^{u,d}$ can be absorbed into a redefinition
of $C_{uH}$,$C_{dH}$ noted above, as only the third generation entry is retained.
The same point also holds for the $(\bar{L}R)(\bar{R}L)$ and $(\bar{L}R)(\bar{L}R)$ operators.
The Class 7 ($\psi^2 H^2 D$) operators have the following extra parameters in the linear MFV breaking limit:
\begin{align}
  C_{\substack{H q \\ rs}}^{(1)} \, Q_{\substack{H q \\ rs}}^{(1)} &\approx
  \left[ S^q_{rs} \, \Delta \, C_{\substack{H q}}^{(1)}\right] \, Q_{\substack{H q \\ rs}}^{' (1)}, \\
  C_{\substack{H q \\ rs}}^{(3)} \, Q_{\substack{H q \\ rs}}^{(3)} &\approx \left[S^q_{rs}  \, \Delta C_{\substack{H q}}^{(3)}\right] \, Q_{\substack{H q \\ rs}}^{' (3)},\\
  C_{\substack{H u \\ rs}} \, Q_{\substack{H u \\ rs}} &\approx \left[S^u_{rs} \Delta C_{\substack{H u}}\right] \, Q_{\substack{H u \\ rs}}^{'},\\
  C_{\substack{H d \\ rs}} \, Q_{\substack{H d \\ rs}} &\approx \left[S^d_{rs} \Delta C_{\substack{H d}}\right] \, Q_{\substack{H d \\ rs}}^{'}.
\end{align}
The $(\bar{L}L)(\bar{L}L)$ operators introduce the extra parameters
\begin{align}
C_{\substack{qq \\ prst}}^{(1)} Q_{\substack{qq \\ prst}}^{(1)} &\approx
\left[ \Delta_1 C^{(1)}_{\substack{qq}} \, S^q_{pr}  \, \delta_{st}
+ \Delta_2 C^{(1)}_{\substack{qq}} \, S^q_{st}  \, \delta_{pr} \right]  Q_{\substack{qq \\ prst}}^{' (1)}, \\
& \, \ + \left[\Delta_1 \mathcal{C}^{(1)}_{\substack{qq}} \, S^q_{pt}  \, \delta_{sr}
+ \Delta_2 \mathcal{C}^{(1)}_{\substack{qq}} \, S^q_{sr}  \, \delta_{pt} \right]  Q_{\substack{qq \\ prst}}^{' (1)}, \nonumber \\
C_{\substack{qq \\ prst}}^{(3)} Q_{\substack{qq \\ prst}}^{(3)} &\approx
\left[\Delta_1 C^{(3)}_{\substack{qq}} \, S^q_{pr}  \, \delta_{st}
+ \Delta_2 C^{(3)}_{\substack{qq}} \, S^q_{st}  \, \delta_{pr} \right]  Q_{\substack{qq \\ prst}}^{'(3)}, \\
& \, \ + \left[\Delta_1 \mathcal{C}^{(3)}_{\substack{qq}} \, S^q_{pt}  \, \delta_{sr}
+ \Delta_2 \mathcal{C}^{(3)}_{\substack{qq}} \, S^q_{sr}  \, \delta_{pt} \right]  Q_{\substack{qq \\ prst}}^{' (3)}, \nonumber \\
C^{(1,3)}_{\substack{l q \\ prst}} Q^{(1,3)}_{\substack{l q \\ prst}} &\approx \left[\Delta C^{(1,3)}_{\substack{l q}} \delta_{pr} S^q_{st}
\right] \,  Q^{' (1,3)}_{\substack{l q \\ prst}},
\end{align}
where the $\Delta C_i,\Delta \mathcal{C}_i$ parameters are normalizations that can differ from the one multiplying the spurion insertions.
The $(\bar{R}R)(\bar{R}R)$ operators have the extra parameters
\begin{align}
C_{\substack{uu \\ prst}}Q_{\substack{uu \\ prst}}  &\approx
\left[S^u_{pr}\, \delta_{st} \, \Delta_1 C_{uu} + S^u_{st}\, \delta_{pr} \, \Delta_2 C_{uu} + S^u_{pt}\, \delta_{sr} \, \Delta_3 C_{uu} + S^u_{sr}\, \delta_{pt} \, \Delta_4 C_{uu} \right]  Q_{\substack{uu \\ prst}}^{'}, \\
C_{\substack{dd \\ prst}}Q_{\substack{dd \\ prst}}  &\approx
\left[S^d_{pr}\, \delta_{st} \, \Delta_1 C_{dd} + S^d_{st}\, \delta_{pr} \, \Delta_2 C_{dd} + S^d_{pt}\, \delta_{sr} \, \Delta_3 C_{dd} + S^d_{sr}\, \delta_{pt} \, \Delta_4 C_{dd} \right]  Q_{\substack{dd \\ prst}}^{'}, \\
C_{\substack{eu \\ prst}} Q_{\substack{eu \\ prst}}  &\approx
\left[S^u_{st}\, \delta_{pr} \, \Delta C_{eu} \right]  Q_{\substack{eu \\ prst}}^{'}, \\
C_{\substack{ed \\ prst}} Q_{\substack{ed \\ prst}}  &\approx
\left[S^d_{st}\, \delta_{pr} \, \Delta C_{ed} \right]  Q_{\substack{ed \\ prst}}^{'}, \\
C_{\substack{ud \\ prst}}^{(1)} Q_{\substack{ud \\ prst}}^{(1)}  &\approx
\left[S^u_{pr}\, \delta_{st} \, \Delta_1 C_{ud}^{(1)} + S^d_{st}\, \delta_{pr} \, \Delta_2 C_{ud}^{(1)} \right]  Q_{\substack{ud \\ prst}}^{' (1)},\\
C_{\substack{ud \\ prst}}^{(8)} Q_{\substack{ud \\ prst}}^{(8)}  &\approx
\left[S^u_{pr}\, \delta_{st} \, \Delta_1 C_{ud}^{(8)} + S^d_{st}\, \delta_{pr} \, \Delta_2 C_{ud}^{(8)} \right]  Q_{\substack{ud \\ prst}}^{' (8)}.
\end{align}
The $(\bar{L}L)(\bar{R}R)$ operators have the extra parameters
\begin{align}
  C_{\substack{l u \\ prst}} Q_{\substack{l u \\ prst}}  &\approx
  \left[S^u_{st}\, \delta_{pr} \, \Delta C_{l u} \right]  Q_{\substack{l u \\ prst}}^{'},\\
  C_{\substack{l d \\ prst}} Q_{\substack{l d \\ prst}}  &\approx
  \left[S^d_{st}\, \delta_{pr} \, \Delta C_{l d} \right]  Q_{\substack{l d \\ prst}}^{'},\\
C_{\substack{q e \\ prst}} Q_{\substack{q e \\ prst}} &\approx
\left[S^q_{pr} \delta_{st} \, \Delta C_{\substack{q e }}
\right]Q'_{\substack{q e \\ prst}}, \\
C_{\substack{q u \\ prst}}^{(1,8)} Q_{\substack{q u \\ prst}}^{(1,8)} &\approx
\left[S^q_{pr} \delta_{st} \, \Delta_1 C_{\substack{q u}}^{(1,8)} + S^u_{st} \delta_{pr} \, \Delta_2 C_{\substack{q u }}^{(1,8)}
\right]Q_{\substack{q u \\ prst}}^{' (1,8)},\\
C_{\substack{q d \\ prst}}^{(1,8)} Q_{\substack{q d \\ prst}}^{(1,8)} &\approx
\left[S^q_{pr} \delta_{st} \, \Delta_1 C_{\substack{q d}}^{(1,8)} + S^d_{st} \delta_{pr} \, \Delta_2 C_{\substack{q d }}^{(1,8)}
\right]Q_{\substack{q d \\ prst}}^{' (1,8)}.
\end{align}
The remaining operators follow the pattern of the $\rm U(3)^5$ limit.

\section{Operator normalizations}\label{Section:normalization}

The normalization used in the SMEFTsim codes also differs from other codes, which should be noted in comparing results.
The HEL implementation \cite{Alloul:2013naa}, eHDECAY \cite{Contino:2014aaa}, Higgs Characterization \cite{Artoisenet:2013puc} and ROSETTA \cite{Falkowski:2015wza}
use a varying suppression scale $1/\hat{m}_W^2$ or $1/v^2$ for operators. Furthermore, these codes
normalize a subset of operators by powers of gauge couplings.

Following Weinberg \cite{Weinberg:1978kz} we take a different approach that conforms with a traditional
EFT construction.
We retain the general EFT with the most general interaction terms consistent with the assumed symmetries without extra UV specific dynamical content or assumptions.
The $\mathcal{L}^{(6)}$ operators are normalized in the SMEFTsim codes to a naive mass dimension suppression scale $\Lambda^2$.
Operators with field strengths are not normalized to be proportional to a corresponding SM gauge coupling, or suppressed by $16 \pi^2$.
The former normalization is not
required to respect $\rm SU_C(3)\times SU_L(2) \times U_Y(1)$ symmetry and the latter is not model
independent\footnote{A historically widespread approach of suppressing operators containing field strengths by loop factors was
shown to not be a model independent EFT statement in Ref.~\cite{Jenkins:2013fya}. See also the discussion in Ref.~\cite{Liu:2016idz}
agreeing with these developments.}.  No assignment of UV specific coupling factors can be made in $\mathcal{L}^{(6)}$ without introducing
further UV assumptions, so we do not include such factors in the \FR  \, codes.

Such normalizations can introduce a very problematic non-commutation with the
equations of motion when interested in EFT studies that seek to obtain basis independent results.
Furthermore, unusual arguments that imply some SMEFT operator bases are  preferred have also appeared in the literature
related to this challenge. These problems can be avoided if the corresponding
Wilson coefficients of the operators normalized differently are then varied sufficiently widely in experimental studies to cancel a chosen normalization.
By using a normalization by the naive mass dimension suppression scale $1/\Lambda^2$ we avoid placing this serious burden on a user of the SMEFTsim codes. We note that this standard EFT approach is also used in DsixTools~\cite{Celis:2017hod} and in Hto4l~\cite{Boselli:2017pef}. This makes it easy to interface with these two programs in the future.
We caution that it does not follow, when using a $1/\Lambda^2$ normalization,  that scan procedures assuming a homogeneous size for the Wilson coefficients is sufficient to cover all possible UV scenarios.

When comparing results with other codes, we caution that to our knowledge, the SMEFTsim codes, and the
implementations of Ref.~\cite{Dedes:2017zog,Celis:2017hod}, are the only
example of complete (public) codings of the $\mathcal{L}^{(6)}$ SMEFT available to date.\footnote{SMEFTsim and the implementation of
Ref~\cite{Dedes:2017zog} are different in scope. Ref~\cite{Dedes:2017zog} provides a \FR \, model formulated in $R_\xi$ gauge,
which is an important step towards NLO results being developed in time.
SMEFTsim includes a \FR\, and UFO implementation formulated in unitary gauge aimed at enabling consistent LO
SMEFT analyses. In particular, the model files generated by SMEFTsim, including input parameter corrections, can be directly employed for montecarlo event generation.}

Note also that (of these complete codes) only SMEFTsim incorporates input parameter corrections.
Missing operators can have non-intuitive consequences on the interpretation
of Wilson coefficients that are retained comparing two SMEFT codes, and make comparing complete operator basis results to incomplete results (that are also at times
ill-defined) challenging.
This is due to the equations of motion being extensively used to define the SMEFT in a minimal basis at $\mathcal{L}^{(6)}$, so that the resulting Wilson coefficients
in the reduced basis reflect many removed operator forms not retained. In short, when comparing SMEFTsim results to other codes {\it caveat emptor}.

\subsection{One loop functions}\label{Sec:oneloopfunctions}
The codes supplied are designed to enable numerical studies of the LO (tree-level) interference of the SMEFT with the SM,
while neglecting NLO corrections. This approach is phenomenologically insufficient if universally applied to all SM interactions.

The processes $h \rightarrow gg$, $h \rightarrow \gamma \, \gamma$, $h \rightarrow \gamma \, Z$
only occur at one loop in the SM due to renormalizability. To obtain a non-zero interference for these processes as a leading numerical correction, we implement the one loop functions for these
processes in the SM following the results in Refs~\cite{Ellis:1975ap,Shifman:1979eb,Bergstrom:1985hp,Dawson:1990zj,Manohar:2006gz}.
An explicit SM Lagrangian term {\tt LHSMloop} that is defined as
\begin{equation}
 \mathcal{L}_{{\rm SM\, loop}}=\frac{h}{\hat{v}_T} \left(g_{Hgg} \de_\mu G^a_\nu \de^\mu G^{a\nu} + g_{Haa} A_{\mu\nu} A^{\mu\nu} + g_{HZa}A_{\mu\nu} Z^{\mu\nu}\right),
\end{equation}
has been included with
\begin{align}
 g_{Hgg} &= \frac{g_s^2}{16\pi^2} I_f\left(\frac{m_h^2}{4m_t^2},0\right),\\
 g_{Haa} &= \frac{e^2}{8\pi^2}\left[I_w\left(\frac{m_h^2}{4 m_W^2}\right) + 3 \left(\frac23\right)^2 I_f\left(\frac{m_h^2}{4m_t^2},0\right)\right],  \\
 g_{Hza} &= \frac{e^2}{4\pi^2} \left[\frac{\hst}{\hct}I_w^Z\left(\frac{m_h^2}{4 m_W^2},\frac{m_Z^2}{4 m_W^2}\right) + 3 \frac23 \left(\frac12-\frac43 \hst^2\right)\frac{1}{2\hst\hct} I_f\left(\frac{m_h^2}{4m_t^2},\frac{m_Z^2}{4m_t^2}\right)\right] \,.
\end{align}
The loop functions are
\begin{align}
 I_f(a,b) &= \int_0^1\int_0^{1-x}\frac{1-4 x y}{1-4(a-b) x y -4b y (1-y)} dy dx,\\
 I_w(a) &=\int_0^1\int_0^{1-x}\frac{-4 + 6 x y + 4 a x y}{1-4 a x y } dy dx,\\
 I_w^Z(a,b) &= \frac{1}{t_{\hat{\theta}}^2}\int_0^1\int_0^{1-x}\frac{(5 - t_{\hat{\theta}}^2 + 2 a (1 - t_{\hat{\theta}}^2)) x y - (3 - t_{\hat{\theta}}^2)}{1-4(a-b) x y -4b y (1-y)} dy dx.
\end{align}
In the codes supplied they have been defined in a Taylor expansion up to cubic terms in the arguments
\begin{align}
 I_f(a,b) &= \frac13 + \frac{11 b}{90} + \frac{22 b^2}{315} + \frac{74 b^3}{1575} + \frac{7 a}{90} + \frac{16 b a}{315} + \frac{58 b^2 a}{1575} + \frac{2 a^2}{63} + \frac{2 b a^2}{75} + \frac{26 a^3}{1575},\\
 I_w(a,b) &=-\frac74 - \frac{11 a}{30} - \frac{19 a^2}{105} - \frac{58 a^3}{525},\\
 I_w^Z(a,b) &= \frac{11}{24} - \frac{31 \hct^2}{24 \hst^2} + \frac{11 a}{180} - \frac{11 \hct^2 a}{36 \hst^2} + \frac{19 a^2}{630} - \frac{19 \hct^2 a^2}{126 \hst^2}
      + \frac{29 a^3}{1575} - \frac{29 \hct^2 a^3}{315 \hst^2} + \frac{7 b}{45} - \frac{4 \hct^2 b}{9 \hst^2} + \frac{2 a b}{35} +\nonumber\\
      &- \frac{62 \hct^2 a b}{315 \hst^2}
      + \frac{16 a^2 b}{525} - \frac{4 \hct^2 a^2 b}{35 \hst^2} + \frac{53 b^2}{630} - \frac{17 \hct^2 b^2}{70 \hst^2} + \frac{67 a b^2}{1575}
      - \frac{43 \hct^2 a b^2}{315 \hst^2} + \frac{86 b^3}{1575} - \frac{10 \hct^2 b^3}{63 \hst^2}\,,
\end{align}
and they are called respectively {\tt Ifermion[x,y]}, {\tt Iw[x]}, {\tt IwZ[x,y]} in the SMEFTsim codes.
\section{SMEFTsim \FR ~packages}\label{codesections}
The SMEFTsim package is designed based on the theoretical outline of the previous sections and consists of several model files for the tree-level analysis of the $\mathcal{L}^{(6)}$ SMEFT corrections. It contains both model files
for \FR ~\cite{Alloul:2013bka} and pre-exported UFO files~\cite{Degrande:2011ua} to be interfaced e.g. with \MGNLO~\cite{Alwall:2014hca}.

Two independent models sets are supplied, called ``Models set A'' and ``Models set B'': each contains three different
theories:  a fully flavour general SMEFT, a $\rm U(3)^5-SMEFT$ with non-SM complex phases and $\rm MFV-SMEFT$.
In addition, each case has two different input schemes available $\{\hat\a_{\rm em}, \hat m_Z, \hat G_F\}$ and $\{\hat m_W, \hat m_Z, \hat G_F\}$.
The two models sets differ in the structure and in the technical implementation of $\mathcal{L}^{(6)}$,
but they produce equivalent results: the use of both sets is recommended for debugging and validation of the numerical results.

All the models are built upon the default SM implementation in \FR~\cite{SMFeynRules}, from which they inherit the SM fields, parameters and Lagrangian definitions.
The original file has been extended and modified to include the complete set of $\mathcal{L}^{(6)}$ baryon and lepton number conserving operators
of the Warsaw basis~\cite{Grzadkowski:2010es} and the input numbers have been updated according to Table~\ref{tab:inputs}.
The SM loop-induced effective couplings of the Higgs to $gg$, $\gamma\gamma$ and $Z\gamma$ have also been included, as detailed in the previous Section.
At this stage, the ghost Lagrangian has been left in its SM form. As a consequence the models give valid results only in unitary gauge, so {\tt \$FeynmanGauge = False} has been enforced in all cases.

The main purpose of the SMEFTsim package is to provide a complete tool for the analysis of the tree level interference terms between the  $\mathcal{L}^{(6)}$ dependent
amplitude and the SM amplitudes in a measured process. The implementation of the entire parameter space of the SMEFT and the automatic inclusion of the shifts
due to the choice of an input parameters set is a key feature. In this spirit, the models are not meant to be employed for the extraction
of accurate SM predictions and they are not equipped for NLO calculations in \MGNLO. The results obtained with SMEFTsim have a finite theoretical
uncertainty $\mathcal{O}(\%)$ for the interference term predicted due to
neglected higher orders in the SMEFT effective expansion ($\mathcal{L}^{(8)} + \cdots$) and radiative corrections that are not included.

 In this section we provide further details about the implementation of the package in \FR ~and \MG.

\subsection{Definition of the Wilson coefficients}
All the Wilson coefficients are assigned a specific interaction order called {\tt NP = 1.}
See Refs~\cite{Alwall:2014hca,Alloul:2013bka} for a definition of interaction order and other options relevant to the \MG~
implementation.
They are defined to be dimensionless, as the cutoff scale of the EFT has been defined as an independent external parameter called {\tt LambdaSMEFT} with a default value of 1~TeV, that can be modified by the user.
{\tt LambdaSMEFT} is defined with an interaction order {\tt QED = -1}, so that the ratio $\hv^2/\Lambda^2$ has overall {\tt QED = 0}.
The Wilson coefficients in the model files are free input parameters. For real Wilson coefficients, they are defined as external parameters
and can be assigned the values directly by the user. Due to the fact that \FR ~does not support complex external parameters,
complex Wilson coefficients are technically defined as internal parameters in the form of  {\tt cXX = cXXAbs Exp[I*cXXPh]} with two
independent external parameters: the absolute value {\tt cXXAbs} and the complex phase {\tt cXXPh} that are free to give numerical
values by the user\footnote{We note that this decomposition has the advantage of allowing to perform external scans on an $\mathbb{R}^n$ space easily.}
The assignment is applied via the attribute {\tt Value} rather than {\tt Definitions} in \FR ~so as to keep a compact notation
in the algebraic evaluation. All the real coefficients and the absolute values of the complex ones are assigned a default numerical value 1 while the phases are set to 0.
A restriction card called {\tt restrict\_SMlimit.dat}, that sets all the Wilson coefficients to zero, is supplied for each UFO model.

\subsection{Definition of the Lagrangian}
\begin{table}[t]
\begin{center}
\caption{Lagrangian terms defined in the SMEFTsim code.}
\label{tab:Lagsummary}
\vspace{2mm}
\begin{tabular}{|ll|}
\hline
\textsf{LSM} & The renormalizable SM Lagrangian. \\
\textsf{LSMlinear}& \multirow{2}{.8\textwidth}{\textsf{LSM} after performing the shifts due to redefinition of input parameters, linearized in the Wilson coefficients.}\\
& \\
\hline
\textsf{LSMloop} &  The effective Higgs couplings $gg$, $\gamma\gamma$ and $Z\gamma$. \\
\textsf{LSMincl} & \textsf{LSMlinear + LSMloop}. \\
\hline
\textsf{L6clN} & The dim-6 operators of the Class${\tt N}=1,2,\dots8$ classified as in Table~\ref{op59} \\
\textsf{L6} & The full dim-6 operators, $\sum_{N=1,\dots,8} \textsf{L6clN} $.\\
\hline
\textsf{LagSMEFT} & \textsf{LSMincl + L6}.\\
\hline
\end{tabular}
\end{center}
\end{table}

All the models contain the Lagrangian terms listed in
Table~\ref{tab:Lagsummary}. In particular, $\mathcal{L}^{(6)}$ has been split into 8 terms, one for each Class defined in Table~\ref{tab.nrparameters_fermion_operators}.
The Lagrangians are by definition Hermitian, while the individual $\mathcal{L}^{(6)}$ operators are not in general.

The Lagrangian is entirely written in the fermion mass eigenbasis, in which the Yukawa matrices are real and diagonal and the CKM matrix is consistently inserted in charged quark currents. By default all the Yukawa matrices have 3 non-zero diagonal entries and all fermion masses (except those of neutrinos) are non-vanishing. Restriction files are supplied for both the \FR ~and UFO models, that set to zero all the fermions' masses and Yukawas except those of the $t$ and $b$ quarks.
Analogously, the CKM matrix is defined as a $3\times3$ unitary matrix in the Wolfenstein parameterization~\cite{Wolfenstein:1983yz}, but it can be restricted to the $2\times 2$ Cabibbo rotation in \FR .

\subsection{Field redefinitions and shifts}
The field redefinitions required to have canonically normalized kinetic terms and the parameter shifts induced by the choice of a set of input parameters are automatically performed in the code, consistent
with Sections~\ref{Section:canonicalform} and~\ref{Section:inputschemes}. This means that all the parameters appearing in the output Lagrangian are ``hatted'' quantities.

The shift in $m_W$ induced in the alpha scheme is peculiar in that it does not suffice to have the shift reproduced correctly in the Lagrangian, but it is also necessary to embed it in the definition of the $W$ field for it to be read properly by \MG.
This is done defining {\tt MW} as an internal parameter that includes the shift {\tt dMW}.
This solution is ineffective for the \textsc{FeynArts/FormCalc/FeynCalc} interface~\cite{Hahn:2000kx,Hahn:1998yk,Shtabovenko:2016sxi} that defines mass parameters independently.
When employing the \{$\hat\a_{ew}\,\hat m_z,\,\hat G_F$\}-scheme models within either of these frameworks, the user needs to apply manually the replacement
\begin{equation}
\label{eq:Wmasscor}
 \mathtt{MW \to MW0(1 + dMW / MW0 )}\hspace*{2cm}
 \mathtt{MW2 \to MW0^2(1 + 2 dMW / MW0 )}
\end{equation}
in all the analytic expressions.

\subsection{Specifics of the implementation for different flavour structures}\label{flavour_differences}
\subsubsection{Flavor general models}
In the flavour general models,  the $\mathcal{L}^{(6)}$ operators constructed out of the fermion fields have free flavour indices that are contracted with those of the associated Wilson coefficients.
The latter $\in \mathbb{C}$ in the flavour space, and are therefore defined as internal tensorial parameters in \FR, with norms and phases given independently for all the complex entries.

Hermiticity and symmetry constraints require some entries to be real and enforce relations among different entries of a Wilson coefficient matrix, reducing the number of free parameters as detailed
in Section~\ref{Section:canonicalform}. This has been taken into account in the codes. For instance, the Wilson coefficient of an Hermitian 2-fermion operator
is specified by 9 real parameters (the 6 absolute values of the $(11),(22),(33),(12),(13),(23)$ entry and the 3 phases of the off-diagonal ones among these)
that can be assigned values in the model file.
The same method has been applied for 4-fermion operators.
The multi-dimensional flavour space makes the reduction of the parameter set more involved in this case.
We summarize the number of independent moduli and phases for each category of Wilson coefficients in Table~\ref{tab.nrparameters_fermion_operators}.

\begin{table}[t]\centering
 \caption{The number of independent parameters per Wilson coefficient for fermionic operators. The operators constructed out of 2 fermions and 4 fermions are divided into the upper and lower panels.}
 \label{tab.nrparameters_fermion_operators}
 \vspace*{2mm}
\begin{tabular}{|l|cc|cc|}
\hline
 \bf Classes & \bf Hermitian & \bf Sym. &\bf Moduli&\bf Phases\\\hline
 5, 6, $Q_{Hud}$& & & 9& 9\\
 7 excluding $Q_{Hud}$& $\surd$ &  & 6& 3\cr
 \hline
 8 -- $(\bar LR)(\bar RL)/(\bar LR)\bar LR)$& & & 81 & 81\\
 8 -- $(\bar LL)(\bar LL),\,(\bar RR)(\bar RR),\,(\bar LL)(\bar RR)$& $\surd$ & & 45& 36\\
  ~~~~~excluding the operators listed below& && & \\
 8 -- $Q_{ll},\,Q_{ee},\,Q_{uu},\,Q_{dd},\,Q_{qq}^{(1)},\,Q_{qq}^{(3)} $& $\surd$ & $\surd$ & 27& 18\\
 \quad\,\,\,\,$Q_{ee}$ & $\surd$ & $\surd$ & 21& 15\\\hline
 \end{tabular}
\end{table}

\subsubsection{\texorpdfstring{$\rm U(3)^5$}{U(3)\^{}5} flavour symmetric models}
In the $\rm U(3)^5$ flavour symmetric models all the Wilson coefficients are scalar parameters ($\in \mathbb{R}$ for Hermitian operators).
The Yukawa matrices used for internal flavour contractions in Classes 5, 6, 8 are diagonal, inclusive of the non-zero (1,1) and (2,2) entries.

\subsubsection{Linear MFV models}
The Wilson coefficients of the fermionic operators for the MFV models are defined so as to contain all the relevant spurions of flavour violation.
Although only the (3,3) Yukawa element is retained in the spurions, the (1,1) and (2,2) components are not set to zero in the leading order contributions. For this reason it is not possible to reabsorb flavour-diagonal spurions into a redefinition of the Wilson coefficients for the operators of Classes 5, 6 and 8 with $(\bar LR)(\bar RL)/(\bar LR)(\bar LR)$ contractions. All the spurions (including the diagonal ones) are therefore retained in the \FR ~model.
The restriction cards for massless light fermions consistently set to zero the flavour-diagonal spurions, as they become redundant in this limit.
The replacement of the Wilson coefficients in terms of the spurions is done explicitly in the Lagrangian (via the {\tt Definitions} attribute), so as to make the number of independent contractions manifest and to allow the reduction of symbolic CKM insertions (unitarity enforces cancellations in the product of CKM insertions stemming from field and spurions definitions).
The Jarlskog invariant is neglected and CP violating operators expected to be proportional to it are not implemented as they are significantly numerically suppressed.

\section{Models set A -- technical details}

\subsection{Code structure}
This models set contains one main file called \textsf{SMEFTsim\_A\_main.fr} that imports fields definitions from \textsf{SMEFTsim\_A\_fields.fr} and parameters definitions from \textsf{SMEFTsim\_A\_parameters.fr}. The latter contains switch commands that select the appropriate parameter definitions depending on the flavour framework and input parameters scheme selected: before importing the model in Mathematica,
the user should define the two flags {\tt Flavor} and {\tt Scheme} that take the values \{\texttt{general, U35, MFV }\} and \{\texttt{alphaScheme, MwScheme}\} respectively. The definitions of $\aem, m_W, \hst, \d g_1, \d g_2$ depend on the input scheme choice, while the flavour specification determines which Wilson coefficients set is imported among \textsf{d6\_parameters\_general.fr}, \textsf{d6\_parameters\_U35.fr}, \textsf{d6\_parameters\_MFV.fr} (see below) and, consequently, the form of $\d G_F$ and of the redefinition of the Yukawa couplings. The three files differ mainly in the implementation of the coefficients for fermionic operators, as described in Section~\ref{flavour_differences}.

The definitions of the $\mathcal{L}^{(6)}$ operators are also dependent on the flavour assumption adopted, and they are imported from one among the files \textsf{SMEFTsim\_A\_operators\_general.fr}, \textsf{SMEFTsim\_A\_operators\_U35.fr}, \textsf{SMEFTsim\_A\_operators\_MFV.fr}.

\subsection{Inputs and shifts}
The redefinitions of the Higgs and gauge fields required to bring the kinetic terms to their canonical form (see Section~\ref{Section:canonicalform}) are applied automatically
in the code. For the Higgs field, this takes place in the doublet field definition, while for the gauge bosons a field redefinition called {\tt rotateGaugeB} is applied on
the SM Lagrangian at the mass eigenstates level. This choice avoids performing unnecessary rotations on the gauge fields appearing in $\mathcal{L}^{(6)}$.
The redefinition of the vev and of the coupling constants due to fixing the input parameters set (see Section~\ref{Section:inputschemes}) is
 done applying the replacements {\tt redefConst} and {\tt redefVev} on the SM Lagrangian terms.

 The shifts  $\d G_F$, $\d m_Z^2$, $\d m_h^2$, $\d g_1$, $\d g_2$, $\d\hst^2$, $\d m_W$ are left explicitly in the Lagrangian and they are defined in \FR  ~as
 internal parameters depending on the Wilson coefficients, so that they are automatically assigned the correct numerical value in \MG.
 For instance the $\rm U(3)^5$ symmetric model produces directly the following Feynman rule for the $Z$ coupling to a pair of neutrinos:
 \begin{equation}
 -\frac{i\hat g_2}{2\hct}\delta_{rs} \gamma^\mu P_L\left[1+\hst^2 \d g_1 + \hct^2 \d g_2+\frac{\hv^2}{\Lambda^2}\hst\hct C_{HWB}+\frac{\hv^2}{\Lambda^2}(\cHlt-\cHls)\right]\,.
 \end{equation}

 It is worth noting that the Lagrangian expressions containing these quantities have the same form irrespectively of the input scheme chosen. What distinguishes the $\{\hat{\alpha}_{ew}, \hat{m}_Z, \hat{G}_F\}$ from the $\{\hat{m}_W, \hat{m}_Z, \hat{G}_F\}$ choice is the exact dependence of the shifts on of Wilson coefficients.
 The models contain a replacement list called either {\tt alphaShifts} or {\tt MwShifts} that allows one to make explicit the Wilson coefficient dependence in algebraic expressions. The replacements should be applied via {\tt ReplaceRepeated} in Mathematica.

 Finally, all the models contain the definition of the functions {\tt LinearWC} and {\tt SMlimit}. The former linearizes analytical expressions in the $\mathcal{L}^{(6)}$ corrections, while the latter sets them to zero, recovering the unshifted SM expression.

\subsection{Comments on the implementation for different flavour structures}
The definition of the $\mathcal{L}^{(6)}$ operators and associated Wilson coefficients has been optimized for each of the three flavour setups considered.
In the flavour general model the fermionic operators have free flavour indices and the corresponding Wilson coefficients are defined as tensorial parameters.
In the $\rm U(3)^5$ symmetric models the flavour contractions for all the fermionic operators are incorporated in the definition of the
operators themselves. This allows a reduction in the number of diagrams in the UFO model and consequently the computation time.
All the Wilson coefficients are therefore scalar numbers ($\in \mathbb{R}$ for Hermitian operators).

In the MFV case, only the Wilson coefficients of quark operators carry flavour indices and they are defined as the appropriate combination of flavour invariants.
Because the model is written in the fermion mass basis and the Yukawa matrices are real and diagonal, it is sufficient to define three spurions
$$S_u= Y_u Y_u^\dag =Y_u^\dag Y_u\approx y_t^2, \quad
S_d= Y_d Y_d^\dag =Y_d^\dag Y_d\approx y_b^2, \quad
S_{Vd} = V_{\rm CKM} \,S_d\, V_{\rm CKM}^\dag\,.$$
to implement the spurion breaking given in Eqn.\ref{spurions}.
In this way, for instance, $\cHqs$ can be introduced as
$$(\cHqs)_{rs} \mapsto (\cHqs)_0\, \delta_{rs} + \Delta_u\cHqs \,(S_u)_{rs} + \Delta_d\cHqs\, (S_{Vd})_{rs}$$
where the two components of the $S^q$ spurion have been assigned independent coefficients $\Delta_u \cHqs$ and $\Delta_d \cHqs$ and $q=(u_L, V_{CKM} d_L)$.
One can immediately verify that expanding the $\rm SU_L(2)$ components gives the correct expression where CKM insertions accompany $S_d$ in the $(\bar u_L u_L)$ current and $S_u$ in the $(\bar d_L d_L)$ current. Four fermion operators and their coefficients are defined in an analogous way.

The following notation is adopted for spurion insertions: the coefficients of the identity contractions are denoted with a final 0 (e.g. {\tt ceW0} ). The coefficients accompanying spurion insertions have names starting with {\tt Delta}: for operators that allow only one spurion insertion the associated coefficient is called {\tt DeltacXX} (where {\tt XX} ~stands for the operator name). Wherever both the $S_u$ and $S_d$ spurions are allowed we assign them coefficients called {\tt DeltaucXX} and {\tt  DeltadcXX} respectively. For four fermion operators that admit spurion insertions in both currents, those in the first current have a coefficient {\tt Delta1cXX} (or {\tt Delta1ucXX , Delta1dcXX} ) and those in the second one have coefficients {\tt Delta2cXX} (or {\tt Delta2ucXX , Delta2dcXX} ). All the parameters appearing here are real, as the only phases allowed from the MFV ansatz are those stemming from the CKM matrix.

\section{Models set B -- technical details}

\subsection{The structure of the model file}

\begin{figure}[t]
\begin{center}
 \includegraphics[width=0.9\textwidth]{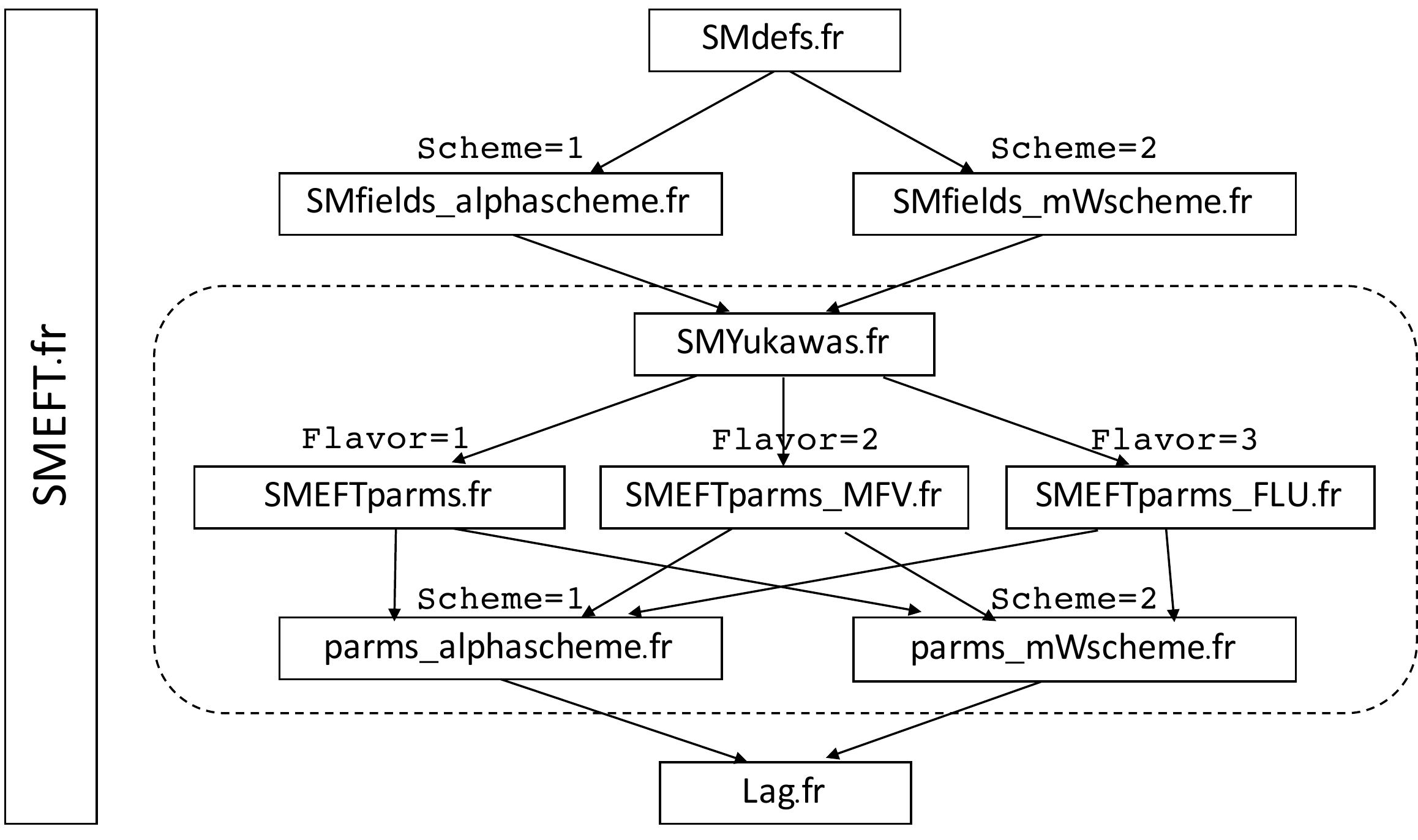}
\caption{Illustrative structure of the SMEFT model set B files.}
\label{fig:flowchart}
\end{center}
\end{figure}

This model file contain a single master code \textsf{SMEFT.fr} and a number of subroutines, along with several restriction files.
The internal structure of this model file is depicted in Fig.~\ref{fig:flowchart}.
The model file can be loaded in \FR ~using the notebook program \textsf{SMEFTsim.nb}, with the product of the UFOs (Universal  \FR ~Outputs).
In the master code two flags: {\tt Scheme} and {\tt Flavor} are established, which are used to identify the input scheme and flavour symmetry being adopted in loading the model.  For example, \\
\\
\texttt{Scheme=X; (* 1: alpha scheme; 2: mW scheme *)}\\
\texttt{Flavor=X; (* 1: flavour general; 2: MFV; 3: U(3)\^{}5 *)}\\
\\
This setup allows one to have different subroutines in different levels (see details in Fig.~\ref{fig:flowchart}), resulting 6 versions of UFOs obtained.
In the master code the \texttt{InteractionOrderHierarchy} is defined but the \texttt{InteractionOrderLimit} is not specified.\footnote{For the definitions of
interaction order and other attributes defined in MadGraph, we recommend the users to consult the \FR\, and \MG\, manuals~\cite{Alwall:2014hca,Alloul:2013bka}.}
In addition to the {\tt QCD} and {\tt QED}, we specify the {\tt NP} orders for the interactions that arise from dimension-6 operators.

\subsection{SM inputs}
The \textsf{SMdefs.fr} is a universal subroutine consisting of the definition of the gauge groups ({\tt U1Y, SU2L, SU3C}) and the indices associated with these groups. As the gauge group is not enlarged in the SMEFT, this subroutine are retained the same as the SM default implementation.

The description of the SM fields is contained in a separate subroutine. Even if no new field is introduced in the SMEFT, we make modifications for this subroutine, offering two versions for
$\{\hat{\alpha}_{ew}, \hat{m}_Z, \hat{G}_F \}$
or $\{\hat{m}_{W}, \hat{m}_Z, \hat{G}_F\}$ input schemes separately, as explained in Section~\ref{Section:inputschemes}. The difference occurs on the $W^{\pm}$ boson mass {\tt MW}. It is set {\tt Internal} in the \textsf{SMfields\_alphascheme.fr} file, while externally given a numerical value in the \textsf{SMfields\_mWscheme.fr} file.

The second part of the model file is the definition of the model parameters, which include the SM parameters and the Wilson coefficients of $\mathcal{L}^{(6)}$ operators. Due to the fact that Wilson coefficients carrying flavour indices are proportional to Yukawas in the flavour symmetric limits, the Yukawas and CKM must be defined before the Wilson coefficients. For this purpose, we divide the SM parameters into two blocks: {\tt YUKAWA} and {\tt SMINPUTS}.

In the subroutine \textsf{SMYukawas.fr}, the mass of all fermion (mass eigenstates) belonging to {\tt YUKAWA} block are externally given numerical values.

The remaining SM parameters are exclusively defined in \textsf{parms\_alphascheme.fr} and in
\textsf{parms\_mWscheme.fr} for the $\{\hat{\alpha}_{ew}, \hat{m}_Z, \hat{G}_F \}$
or $\{\hat{m}_{W}, \hat{m}_Z, \hat{G}_F\}$ input schemes, respectively.
The other block {\tt SMINPUTS} includes three external parameters: {\tt aEWM1}, {\tt Gf} and {\tt aS} in
the \textsf{parms\_alphascheme.fr}, and with the substitution of {\tt aEWM1} by {\tt MW} in the
subroutine  labeled as \textsf{parms\_mWscheme.fr}.

\subsection{Wilson coefficients}
 In general, the full list of Wilson coefficients contain two types of parameters: scalar parameters for flavour-singlet Wilson coefficients and tensorial parameters when the Wilson coefficients carry flavour indices. The dependence on the flavour
space can be reduced as shown in Section~\ref{Section:symassumptions}. In the model file, there are three versions of subroutine \textsf{SMEFTparms.fr},  \textsf{SMEFTparms\_MFV.fr} and \textsf{SMEFTparms\_FLU.fr}
provided corresponding to the cases of flavour general SMEFT, MFV-SMEFT and $\rm U(3)^5 -SMEFT$, respectively.  In each file, all the flavour-singlet Wilson coefficient parameters are stored in the {\tt NEWCOUP} block. The coefficients associated with spurion breaking in the MFV limit are individually stored in a block named {\tt MFVCOUP}.
We stress that the attributes {\tt BlockName} and {\tt OrderBlock} cannot be specified for tensorial parameters.
By default, their block name are taken as {\tt FRBlock\#X} in sequence.
In addition, the shifts on the Higgs vev and coupling, gauge boson mass and gauge couplings
{\tt dGf, dvev, dlam, dW, dM2Z, dgw, dg1, dsw2}
due to the $\mathcal{L}^{(6)}$ contribution are constructed in the subroutine \textsf{parms\_alphascheme.fr} and \textsf{parms\_mWscheme.fr}. Among them, the $W$ mass shift \verb"dW" presented in the $\{\hat{\alpha}_{ew}, \hat{m}_Z, \hat{G}_F \}$ input scheme follows Eqn.~\ref{eq:Wmasscor}.
and the analytic forms for the remaining ones are summarized in Section~\ref{Section:inputschemes}.

 \subsection{Lagrangian construction}
The SMEFT Lagrangian is constructed in the subroutine \textsf{Lag.fr}.
The SM Lagrangian {\tt LSM} is implemented by default.
As already shown, in the presence of $\mathcal{L}^{(6)}$ a shift at the leading {\tt NP} order is induced on the Higgs vev and SM gauge couplings compared to the SM values,
and meanwhile the field redefinition for the SM fields are also demanded, leading to a conversion from the initial SM Lagrangian {\tt LSM}
to a new defined SM Lagrangian {\tt LSMshift}. This important step is accomplished in the {\tt LSMshift} function by employing a series of substitution rules named {\tt redefXXX}:\\
\\
{\tt lagtmp=LSM;}\\
{\tt  Return[lagtmp/.redefHiggs/.redefYuk/.redefSMfield/.redefWeakcoup}\\
{\tt  /.Conjugate[CKM[a\_, b\_]]*CKM[a\_, c\_]->IndexDelta[b, c]//OptimizeIndex];}\\
\\
Here a series of substitution rules ({\tt redefXXX}) employed encode the above mentioned shifts and redefinitions.
It is clear that the terms of higher order in Wilson coefficients are included in the {\tt LSMshift}. This not only brings the inconsistency in the
perturbative expansion but also increases the difficulty in the numerical computation. In fact, for the purpose of the phenomenological study,
one may be interested in a SMEFT Lagrangian linearized in Wilson coefficients. To this end, we practically introduce an auxiliary variable {\tt WC} and multiply
it in front of each Wilson coefficient parameter in the {\tt redefXXX}. The usage of {\tt WC} multiplier allows us to expand {\tt LSMshift} in a Taylor series in {\tt WC} to
linear order by means of the following command:\\
\\
{\tt LSMlinear := Normal[Series[LSMshift, {WC, 0, 1}]]/.WC->1}\\
\\
For completeness, we add to the the inclusive SM Lagrangian the dimension-5 effective couplings of the SM Higgs to $gg$, $\gamma\gamma$ and $Z\gamma$ which is defined in the {\tt LhSMloop} function given in Section \ref{Sec:oneloopfunctions}..\\
\\
{\tt LSMincl := LSMlinear + LhSMloop; }\\
\\
On the other hand, the effect of shifts and redefinitions on the dimension-six terms $\mathcal{L}^{(6)}$ are higher {\tt NP} order and can
be safely dropped in the SMEFT. As a result, the SMEFT Lagrangian is $\mathcal{L}_{SM}+ \mathcal{L}^{(6)}$ denoted in the code as\\
\\
{\tt LagSMEFT := LSMincl+L6}.
\section{General recommendations for the use of the UFO models in \textsc{MadGraph5}}

When generating a process in the SMEFT it is always necessary to specify the order {\tt NP=1} to make sure that all and only the diagrams giving linear $\mathcal{L}^{(6)}$
contributions are included. Notice that the {\tt InteractionOrderLimit} is not specified by default in the
model files. In order to extract the tree level interference contribution between $\mathcal{L}^{(6)}$ and the SM
amplitudes, we suggest generating the process with the syntax {\tt NP\^{}2==1} in \MG.

In general, due to the fact that SM Lagrangian parameters ($g_i$, $Y_i$, $\hv$\dots) can multiply the Wilson coefficients in the Lagrangian, a given
interaction vertex can have multiple interaction orders. For instance, the $Z\bar ee$ coupling stemming from $Q_{He}$ is proportional to $\hv^2 g_1 C_{He}/\Lambda^2$ and has therefore interaction order {\tt \{NP=1, QED=1\}}. There is one coupling that has {\emph{negative}} {\tt QED} order, namely the contribution to the trilinear Higgs coupling $h^3$ stemming from $Q_H$, which is proportional to $C_H \hv^3/\Lambda^2$ and has therefore order {\tt \{NP=1, QED=-1\}}.
One should be careful when generating processes that include this coupling, as this may alter the intuitive interaction order hierarchies among diagrams.

Before generating events or calling a survey in MadEvent it is preferable to set all the relevant widths to {\tt auto} in the {\tt param\_card}. This is because  the value of the particle width is used to compute some cross-sections in a narrow-width approximation in \MG .
The values assigned by default to the widths in the model are those computed in the SM (sometimes including radiative corrections) and they are often inconsistent with a tree-level SMEFT prediction.\footnote{This has a particularly large impact in the
Higgs case: the default value assigned to its width in \FR   \, is obtained in the SM with the inclusion of radiative corrections, and it is significantly smaller than the tree level value, mainly due to a large negative loop contribution to the $h\to b\bar b$ partial width. Using the default width for Higgs-mediated processes may give unphysical results with branching fractions apparently larger than 1.}
Note for the general SMEFT, it takes  a few minutes to load the model, and at most two hours to generate the UFO files.

\section{A pole parameter global SMEFT fit}\label{poleparam}

The SMEFTsim codes enable LHC SMEFT studies to be carried out combined with lower energy data reported at LEP and
other experiments, while all parameters in $\mathcal{L}^{(6)}$ are retained. Such a global SMEFT physics program
is of interest long term, due to its importance for the development of model independent constraints.
This approach enables hints of new physics that could emerge in the data in time to be understood and decoded
systematically by combining measurements of deviations in a well defined field theory setting.
This approach is also valuable as it is a way to record the data in a field theory interpretation that
allows the SM to break down at higher energies for the long term.

Developing simplified fits as an intermediate step towards the general fit case is also important.
This can be done minimizing UV assumptions and exploiting
the kinematics of the relevant collider scattering events, in addition to the SMEFT power counting and flavour symmetries. This approach can
be followed when defining a `WHZ pole parameter' program to constrain an interesting subset of SMEFT parameters.
We consider one of the main applications of the SMEFTsim package is to directly enable this effort to be undertaken in the LHC experimental collaborations.
\begin{figure}[t]
\begin{center}
\includegraphics[width=0.9\textwidth]{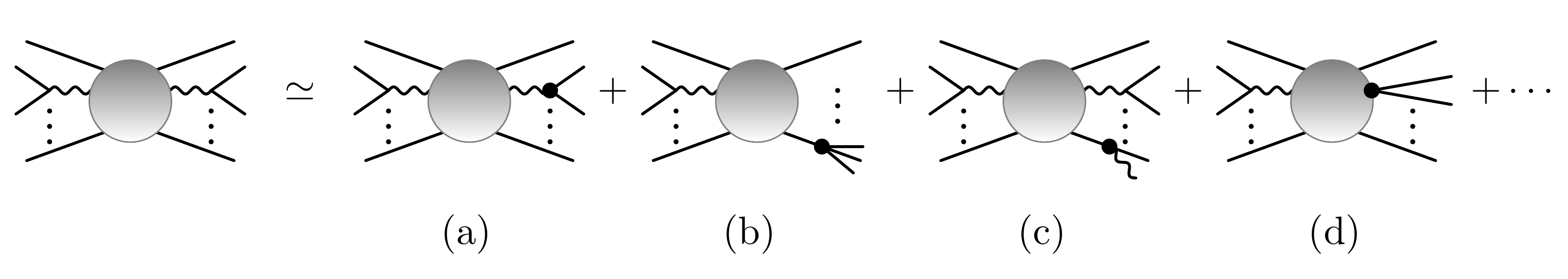}
\end{center}
\caption{Decomposing a general scattering amplitude in the SMEFT into pole and non-pole parameters. Here the black dot indicates
a possible insertion of $\mathcal{L}^{(6)}$ and shifts are only shown on the final states as an illustrative choice, but also appear
on the initial states in the suppressed $+ \cdots $ contributions. A WHZ pole parameter is shown in case a).}  \label{fig1:poleparam}
\end{figure}
The idea is to use the fact that $\mathcal{L}^{(6)}$ operator forms interfere with a SM process for the numerically leading correction, to supplement
the power counting of the SMEFT systematically by using:
\begin{itemize}
\item{Resonance domination of signal events enforced with optimized phase space cuts to further suppress (primarily) Class 8 $\psi^4$ operators.
This is discussed in Section~\ref{section:resonanance}.}
\item{Numerical suppressions in interference due to the presence of small symmetry breaking effects in the SM due to light quark masses, helicity configurations,  $\rm CP$ violation, CKM suppressions
and the GIM mechanism \cite{Glashow:1970gm} in SM amplitudes.}
\end{itemize}
Exploiting these IR physics effects in the SMEFT, in addition to the usual power counting of the theory, significantly reduces the number of WHZ pole parameters to a manageable set.
Many studies have been performed that utilize one or more of these IR effects, but we believe the systematic approach laid out here goes beyond past literature.
It is important to note that the parameter reduction that makes a LO SMEFT effort feasible using such IR physics occurs if flavour symmetry assumptions
are explicitly adopted or not, see Table~\ref{tab:parametercounts}. Some processes that are consistent with the discussion given here are
$\bar{\psi} \psi \rightarrow Z \rightarrow \bar{\psi} \psi $, $\bar{\psi} \psi \rightarrow W \rightarrow \bar{\psi} \psi $,
$\bar{\psi} \psi \rightarrow BB \rightarrow \bar{\psi} \psi \bar{\psi} \psi $, and phase space restricted $pp \rightarrow h \rightarrow Z Z^\star \rightarrow \bar{\psi} \psi \bar{\psi} \psi$ and
$pp \rightarrow h \rightarrow W W^\star \rightarrow \bar{\psi} \psi \bar{\psi} \psi$ when selecting for intermediate near on-shell massive bosons.

\subsection{Resonance domination and numerically suppressed interference}\label{section:resonanance}
The interference with the SM leads to a relative kinematic suppression of $\psi^4$ operators, compared to the parameters
retained in a  `WHZ pole parameter' program in several experimental signals of interest.\footnote{These arguments are
the logical extension of the reasoning used to justify the neglect of $\psi^4$ operators in studying LEPI data, as discussed in Refs.~\cite{Han:2004az,Berthier:2015oma}.}
This occurs so long as scattering events in a measurement are dominantly proceeding through a
near on-shell phase space pole (i.e. $\sqrt{p_{i}^2} - m_B \lesssim \Gamma_B$) of the narrow bosons of the SM ($B = \{W,Z,h\}$).

The `WHZ pole parameters' are generally defined by allowing a non-SM three point interaction of the SM narrow bosons $B = \{W,Z,h\}$ which allows a
contribution to the leading set of poles in the SM prediction. This largely limits the pole parameters to those
parameterizing the product $\langle H |\mathcal{L}_{SM}| H \rangle$ reduced into a minimal operator basis,
such as operators in Classes $2,3,4,5,7$ in the Warsaw basis. Operators of Class one contribute to anomalous massive vector boson and gluon scattering,
the $\rm CP$ even operators of the form $\mathcal{Q}_{W}$ are retained in the WHZ pole parameters.
Operators of Class 2 are not relevant for the near term
at LHC  -- until double Higgs production can be probed. Furthermore, only a small subset of Class 8 $(\bar{L} L)(\bar{L} L)$ parameters are retained due to
the redefinition of the vev in the SMEFT.

To understand the relevance of the `WHZ pole parameters' for hadron collider measurements, consider a general scattering amplitude depicted in Fig.~\ref{fig1:poleparam}.
The total amplitude can be decomposed around the physical poles of the narrow bosons of the SM in the process
\bea
\mathcal{A} &=& \frac{\mathcal{A}_a(p_1^2, \cdots p_M^2)}{(p_1^2 - m_{B_1}^2 + i \Gamma_{B_1} m_{B_1}) \cdots (p_N^2 - m_{B_N}^2 + i \Gamma_{B_N} m_{B_N})}, \nn
 &+& \frac{\mathcal{A}_b(p_1^2, \cdots p_M^2)}{(p_1^2 - m_{B_1}^2 + i \Gamma_{B_1} m_{B_1}) \cdots (p_{N-1}^2 - m_{B_{N-1}}^2 + i \Gamma_{B_{N-1}} m_{B_{N-1}})},\nn
 &+& \cdots + \mathcal{A}_j(p_1^2, \cdots p_M^2).
\eea
Fig.~\ref{fig1:poleparam} illustrates that $\mathcal{L}^{(6)}$ corrections that are the same order in the power counting can
modify a resonant process in the SM (as in $\mathcal{A}_a$ and  Fig.~\ref{fig1:poleparam} a), lead to a contribution to the scattering amplitude with fewer
poles than in the SM process (as in $\mathcal{A}_b, \cdots \mathcal{A}_{j-1}$ and Fig.~\ref{fig1:poleparam} b), or lead to a contribution to the process with no internal SM poles
(as in $\mathcal{A}_j$) from narrow SM bosons. Here the $p_i^2$ factors stand for general Lorentz invariants of dimension two.
Assume that selection cuts are made so that
the process is numerically dominated by a set of leading pole contributions of $\leq N$ narrow $B$ bosons.
Then the leading SMEFT cross-section corrections in this phase space volume  $\Omega$  are expected to be
\bea
\left(\frac{d \sigma}{\d \Omega}\right)_{pole}  &\simeq& \left(\frac{d \sigma_{SM}}{\d \Omega}\right)^1 \, \left[1+ \mathcal{O}\left(\frac{C_i \, \bar{v}_T^2}{g_{SM}\Lambda^2}\right)
+ \mathcal{O}\left(\frac{C_j \, \bar{v}_T^2 \, m_B}{\Lambda^2 \, \Gamma_B}\right)\right], \\
&+& \left(\frac{d \sigma_{SM}}{\d \Omega}\right)^2 \,  \left[1+ \mathcal{O}\left(\frac{C_k \, p_i^2}{g_{SM}\Lambda^2}\right)\right]. \nonumber
\eea
The differential cross-sections $\left(d \sigma_{SM}/\d \Omega\right)^{1,2}$ are distinct in each case and $C_i$ correspond to WHZ pole parameters
that are of the form $\langle H |\mathcal{L}_{SM} | H \rangle$, the $C_j$ correspond to scheme dependent corrections to the intermediate propagators\footnote{The mass shift of this form for $W$ propagators are present in the $\{\hat{\alpha}_{ew}, \hat{m}_Z, \hat{G}_F \}$ input scheme,
and known to be numerically small.} and the $C_k$ correspond to a subset of operators that lead to three point interactions with more than one derivative.

Additional corrections to the measured processes and relevant backgrounds also exist in the SMEFT, but they can be relatively numerically suppressed in a SMEFT oriented experimental analysis.
Consider the interference with a complex Wilson coefficient in $\mathcal{L}^{(6)}$, denoted $C$,
that occurs when a resonance exchange is not present compared to the leading resonant SM signal result (shown in Fig.~\ref{fig1:poleparam} d)).
The interference terms in the corresponding observable then scale as
\bea\label{intargument}
|\mathcal{A}|^2 &\propto& \left(\frac{g_{SM}^2}{(p_i^2 - m_B^2 + i \, \Gamma(p) m_B)} + \frac{C}{\Lambda^2} \right) \left(\frac{g_{SM}^2}{(p_i^2 - m_B^2 + i \, \Gamma(p) m_B)} + \frac{C}{\Lambda^2} \right)^\star \cdots \\
&\propto& \left[\frac{g_{SM}^2}{(p_i^2 - m_B^2)^2 + \Gamma_B^2 \, m_B^2} + \frac{(p_i^2 - m_B^2) (C/\Lambda^2 + C^\star/\Lambda^2) - i \Gamma_B \, m_B (C^\star/\Lambda^2 - C/\Lambda^2)}{(p_i^2 - m_B^2)^2 + \Gamma_B^2 \, m_B^2}\right] \cdots \nonumber
\eea
In the near on-shell region of phase space $(\sqrt{p_i^2} - m_B \sim \Gamma_B)$ for the narrow boson $B$, the SMEFT then has the additional numerically subleading corrections
\bea
&\,&\left(\frac{d \sigma_{SM}}{\d \Omega}\right)^1 \,  \mathcal{O}\left(\frac{\Gamma_B \, m_B \, \{{\rm Re}(C),{\rm Im}(C)\}}{g_{SM}^2\Lambda^2}\right)
+ \left(\frac{d \sigma_{SM}}{\d \Omega}\right)^2 \, \mathcal{O}\left(\frac{\Gamma_B \, m_B \, \{{\rm Re}(C),{\rm Im}(C)\}}{g_{SM}^2\Lambda^2}\right)
\cdots
\eea
For this reason, the numerical effect of the non-pole parameters are relatively suppressed by a factor
of
\bea\label{scalingrule}
\left(\frac{\Gamma_B \, m_B}{\bar{v}_T^2}\right) \frac{\{{\rm Re}(C),{\rm Im}(C)\}}{g_{SM} \, C_i}, \quad \quad \left(\frac{\Gamma_B \, m_B}{p_i^2}\right) \frac{\{{\rm Re}(C),{\rm Im}(C)\}}{g_{SM} \,  C_k},
\eea
compared to a Wilson coefficient that is a (scheme independent) pole parameter. Such a suppression factor appears for each missing resonance selected for with selection cuts.
This relative numerical suppression occurs in addition to the power counting in the SMEFT. It is the combination of these
two suppressions that is experimentally and theoretically relevant.

\begin{figure}[t]\centering
 \includegraphics[width=1\textwidth]{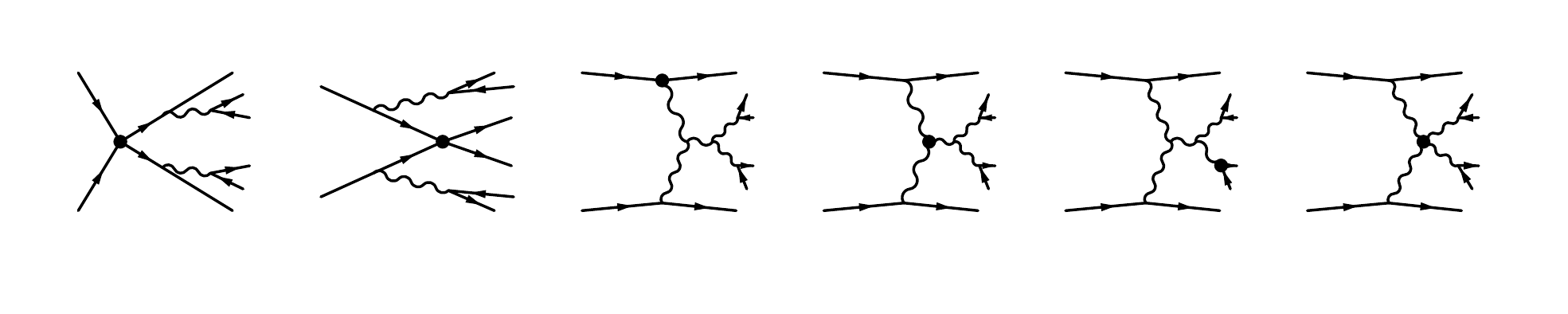}
 \caption{Contributions to $\bar{\psi} \psi \rightarrow \bar{\psi} \psi \bar{\psi} \psi \bar{\psi} \psi $ due to $\mathcal{L}^{(6)}$ which is indicated with
 a black dot.}\label{VBSexample}
\end{figure}

As experimental selection cuts for narrow SM $B$ bosons do not isolate all of the poles in a process in general,
in some cases $\psi^4$ operators can be classified as leading parameters numerically, and should be retained in a global SMEFT analysis examining such a process.
For example, consider $\bar{\psi} \psi \rightarrow \bar{\psi} \psi \bar{\psi} \psi \bar{\psi} \psi $ which can occur through many Feynman diagrams, including the diagrams in
Fig. \ref{VBSexample}.  In the left two diagrams of Fig. \ref{VBSexample}, isolation cuts to identify and reconstruct only the two bosons that
decay into pairs of final state fermions do not suppress these contributions. A $\psi^4$ operator can then be classified
as a leading contribution to be retained in an alternate interesting subset of SMEFT parameters for this process, if the remaining phase space selection cuts
did not further suppress these contributions.\footnote{This particularly occurs in the case of $\psi^4$ operators leading to top final states which
themselves emit $W^\pm$ when decaying.} We use the nomenclature `WHZ pole parameters' which is intended to signal that we exclude such cases by definition
in this parameter set.

Other parameters neglected from the `WHZ pole parameters' set are
numerically suppressed due to the lack of tree level flavour changing neutral currents in the SM, see Section \ref{interferenesuppress} for details.
We reiterate, these additional numerical suppressions come about due to IR kinematics and symmetries in near on-shell regions of phase space for
the narrow $B$ bosons selected for, not UV assumptions.
As NLO corrections are neglected in the SMEFTsim package anyway, these numerically suppressed effects can be consistently neglected - so long as an
appropriate theoretical error is assigned for this approximation.

Pole parameter SMEFT studies are an important step towards a
global SMEFT analysis. This approach has some similarity to a pseudo-observable (PO) approach to LHC data \cite{Passarino:2010qk,Isidori:2013cla,Isidori:2013cga,Gonzalez-Alonso:2014eva,David:2015waa,Gonzalez-Alonso:2015bha,Bordone:2015nqa,Greljo:2015sla},
and can be considered to extend and improve it by embedding this approach it in a well defined field theory. Some minor differences in the approaches do exist, and follow from the use of
a consistent $\mathcal{L}_{SMEFT}$ construction, as we discuss below.

\subsection{Interference of anomalous interactions in the SMEFT}\label{interferenesuppress}

\begin{figure}[t]\centering
 \includegraphics[width=0.75\textwidth]{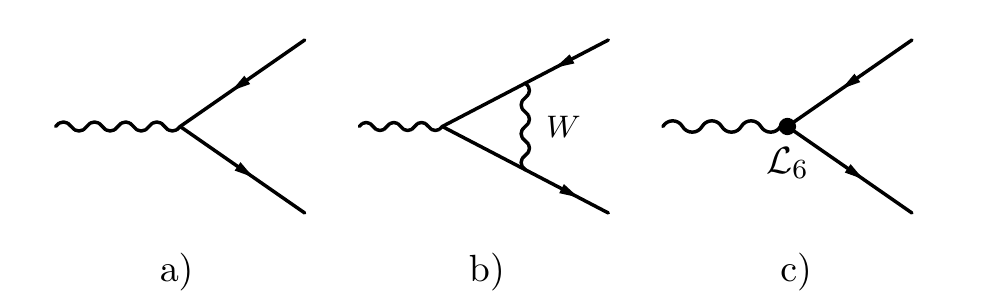}
 \caption{Interference diagrams for anomalous three point interactions in the SMEFT.}\label{interferencethreepoint}
\end{figure}

Retaining a general flavour conserving anomalous dipole interaction
\bea
\mathcal{L}_{\rm dipole} = C_{\psi X} \, \bar{\psi} \, \sigma_{\mu \, \nu} \, P_R \, \psi H X^{\mu \, \nu}+ h.c.
\eea
due to the arguments of the Appendix on the experimental constraints on dipole operators, leads to off-shell interference with the SM Lagrangian in Fig.~\ref{interferencethreepoint} a) and  c)
that give
\bea
|\mathcal{A}|^2 \propto m_\psi \, \epsilon_\alpha  \, \epsilon_\beta^\star \left(i \, 8 {\rm Re}[C_{\psi X}] \, (g_L + g_R) g^{\alpha \, \beta} p_1 \cdot p_2
- 8 {\rm Re}[C_{\psi X}] \, (g_R - g_L) \epsilon^{\alpha \, \beta \, p_1 \, p_2} , \nonumber \right.\\
\left.- i \, 4  \,  \left[C_{\psi X} \, g_L + C^\star_{\psi X} \, g_R \right] p^\alpha \, p_2^\beta
- i \, 4   \,  \left[C^\star_{\psi X} \, g_L + C_{\psi X} \, g_R \right] p^\alpha \, p_1^\beta\right).
\eea
Here a general chiral interaction in the SM is parameterized by $g_{R/L}$.
As the interference is suppressed by $m_\psi$, it follows that $C_{\psi X}$ insertions can be initially neglected in LO global SMEFT studies involving light fermions ($\psi \neq \{t, b\}$),
not due to experimental constraints, or a UV model assumption, but as a numerical suppression due to the IR physics of the SMEFT.
This can be done until experimental precision advances to overcome such additional numerical suppressions.
This is in agreement with Ref.~\cite{Gonzalez-Alonso:2014eva,Gonzalez-Alonso:2015bha,Bordone:2015nqa,Greljo:2015sla} argued in the context of a PO framework.
Nevertheless, the results of the Appendix argue for retaining dipole operators when $\psi = \{t, b\}$, and we note that the inclusion of top dipole interactions has been shown to have
an important effect on Higgs phenomenology in many works, including Refs.~\cite{Degrande:2012gr,Zhang:2013xya,Durieux:2014xla,Buckley:2015nca}.
Note that if such $\mathcal{L}^{(6)}$ terms are not arbitrarily neglected, then contributions such as shown in Fig.~\ref{fig1:poleparam} c) require a deconvolution of
possible non-SM soft emissions in the LHC collider environment to extract model independent PO. For more discussion see
Refs.~\cite{Passarino:2010qk,Bordone:2015nqa,Brivio:2017vri}.

IR SM-$\mathcal{L}^{(6)}$ interference effects can also justify the neglect of flavour off diagonal three point interactions. Such interactions can be present
due to Class 5, 6 and 7 operators.
Flavour changing neutral currents vanish at tree level in the SM,
and the one loop contributions for the three point vertices $Z \bar{\psi}_i \, \psi_k$, come about due to the interference
of Fig.~\ref{interferencethreepoint} b) and  c). The one loop flavour changing three point interaction in the SM for the $Z$ (Fig.~\ref{interferencethreepoint} b)
scales as \cite{Gaillard:1975ds,Ma:1979px,Clements:1982mk,Ganapathi:1982xy}
\bea\label{gimsuppressionZ}
\mathcal{A}^{Z}_{ik} &\simeq & \, -\frac{3 \sqrt{\bar{g}_1^2 + \bar{g}_2^2} \, \bar{g}_2^2 \, V_{jk}^\star \, V_{ji}}{32 \, \pi^2} \frac{m_j^2}{m_W^2} \bar{\psi}_k \, \gamma^\mu \, P_L \, \psi_i \, \epsilon^Z_{\mu} + \cdots,
\eea
due to the presence of a GIM mechanism \cite{Glashow:1970gm}, with $m_j$ the mass of the internal quarks summed over. Similarly the amplitude following form
the effective one loop coupling $h \bar{\psi}_i \, \psi_k$ also has a GIM suppression \cite{Botella:1986hs,Eilam:1989zm,Dedes:2003kp}
\bea\label{gimsuppressionH}
\mathcal{A}^{h}_{ij} &\simeq& \frac{3 \bar{v}_T \, \bar{g}^3_2}{16^2 \, \pi^2 \, \hat{m}_W} \, \bar{\psi}_i \, \left[y_i \, V_{ik}^\dagger \, V_{kj} \frac{m_k^2}{\hat{m}_W^2} P_L
+ y_j \, V_{kj}^\dagger \, V_{ik} \frac{m_k^2}{\hat{m}_W^2} P_R \right] \, \psi_j, + \cdots
\eea
Interference with flavour off diagonal corrections then experiences
an additional numerical suppression of this form in both cases. This can be used to justify neglecting such effects in LO SMEFT analyses
when neutral currents are present. No such extra GIM suppression is present in the case of charged currents.\footnote{Note that a one loop result for a
flavour changing neutral current in the SM can be compared to the one loop improvement of an effective three point interaction in the SMEFT due to the insertion of $\mathcal{L}^{(6)}$
in the loop diagram. The latter does not in general experience the extra suppression from the GIM mechanism on top of the one loop suppression, and can introduce
a number of $\mathcal{L}^{(6)}$ parameters not present in a tree level analysis. This is another reason that one loop SMEFT results are of interest when incorporating precise
experimental constraints such as LEP data into a global SMEFT fit, see Refs.~\cite{Berthier:2015oma,Berthier:2015gja,deFlorian:2016spz,Passarino:2016pzb,Hartmann:2015aia,Hartmann:2016pil}.}
For this reason, flavour off diagonal entries in the Wilson coefficient matrices $C_{\substack{H l}}^{(3)},C_{\substack{H q}}^{(3)}$
can be practically retained in a LO analysis, while neglecting the remaining flavour off diagonal Wilson coefficients of the Class 5 and Class 7 operators.
Even so, the numerical size of the first-third generation charged currents in the SM is smaller than neglected one loop corrections as a result of the CKM parameter suppression,
and such SMEFT parameters are thus neglected in the WHZ pole parameter counts.

A similar argument holds for the Wilson coefficient matrix $C_{\substack{H u d}}$, however in this case, the neglect of this set of parameters
follows from the corresponding right handed currents not interfering at leading order with the SM interactions. Such interference
first comes about proportional to two insertions of light quark masses \cite{Alioli:2017ces}.

The interference of $\rm CP$ violating phase stemming from operators of Classes 1,4,5,6,7  with the corresponding effective operators generated in the SMEFT
are also numerically suppressed to the level of neglected loop corrections. These operators are neglected in the pole parameter set
but we note that we provide a fully general SMEFT code, and a $\rm U(3)^5$-SMEFT code with all phases so that $\rm CP$ violating effects can be studied as desired
using the SMEFTsim package.

Neglecting numerically suppressed contributions from $\mathcal{L}^{(6)}$ operators in a LO `WHZ pole parameter' program can be justified in this manner.
Developing the SMEFT in time to the level of NLO corrections is required for the interpretation of the most precise experimental data,
see Refs.~\cite{deFlorian:2016spz,Passarino:2016pzb,Brivio:2017vri,Ghezzi:2015vva,Gauld:2015lmb,Gauld:2016kuu,Hartmann:2016pil} for discussion and results developing this effort.
In the mean time LO fits in a pole program can and should be pursued. A theory error metric must be chosen in this effort to make such simplifying LO approximations.
Some theory error metrics were proposed in Refs.~\cite{Berthier:2015oma,Berthier:2015gja,deFlorian:2016spz,Passarino:2016pzb}.
In addition, we note the exact number of parameters in a pole constraint program is weakly basis dependent as exchanging $\psi^4$ operators for operators
without fermion fields is largely blocked as the latter do not carry sufficient flavour indices.
As the vast majority of the parameters of the SMEFT reside in the operators with a maximal set of flavour
indices the simplification in the number of parameters present is dramatic.

The SMEFTsim codes do not contain anomalous one loop flavour changing neutral current interactions for the SM. The loop level generation of $\rm CP$ odd
operators in the SM proportional to the Jarlskog invariant are also absent. As such,
a restricted set of parameters is a natural result of numerical simulations using SMEFTsim. The restricted set of parameters
comes about when the leading order interference terms with the SM are calculated, which is the purpose of the SMEFTsim package.
These arguments on numerically suppressed interference also lead to the corresponding second order terms
in a constructed $\chi^2$ to fit experimental data also being suppressed.
As such retaining such parameters in a fit is subject to large theoretical uncertainties. In addition $\mathcal{L}^{(8)}$ is of the same order as such terms
in a constructed $\chi^2$ and neglected and only retaining a subset of $\mathcal{O}(1/\Lambda^4)$ corrections is not basis independent.

Restricting to a  `WHZ pole parameter'  program the number of parameters in a LO global constraint program in the Warsaw basis are estimated in Table~\ref{tab:parametercounts}.

\subsection{`WHZ pole parameter' counts}
\begin{table}[t]\centering
  \begin{tabular}{c|c|c|c}
Class & Parameters  & $n_f = 1$ & $n_f = 3$\\
\hline \hline
1 & $C_{W} \in \mathbb{R}$ & 1 & 1\\
\hline
3 & $\{C_{HD}, C_{H\Box}\} \in \mathbb{R}$ & 2 & 2\\
\hline
4 & $\{C_{HG}, C_{HW}, C_{HB}, C_{HWB}\} \in \mathbb{R}$ & 4 & 4\\
\hline
5 & $\{C_{\substack{uH \\ 3 3}},C_{\substack{dH \\ 3 3}}\}  \in \mathbb{R}$ & 2 & 2\\
\hline
6 & $\{C_{\substack{uW \\ 3 r}},C_{\substack{uB \\ 3 3}},C_{\substack{uG \\ 3 3}},C_{\substack{dW \\ 3 r}},C_{\substack{dB \\ 3 3}},C_{\substack{dG \\ 3 3}}\}_{r \neq 1}  \in \mathbb{R}$ & 6 & 10\\
\hline
7 &  $\{C^{(3)}_{\substack{H l \\ p r}},C^{(3)}_{\substack{H q \\ p r}}, C^{(1)}_{\substack{H l \\ r r}},C^{(1)}_{\substack{H q \\ r r}},C_{\substack{H e \\ r r}},C_{\substack{H u \\ r r}},C_{\substack{H d \\ r r}}\}_{pr \neq \{(1,3),(3,1)\}}  \in \mathbb{R}$
& 7 & 26\\
 \hline
8 $(\bar{L} L)(\bar{L} L)$ &  $\{C_{\substack{l l \\ \mu e e \mu}}, C_{\substack{l l \\ e \mu \mu e}} \} \in \mathbb{R}$ & 1 & 1\\
\hline
 & {\rm \hspace{7cm} Total Count} & 23 & 46 \\
  \hline \hline
    \end{tabular}
  \caption{LO parameter counts in the general SMEFT flavour cases for $n_f$ generations for a `WHZ pole parameter' program. The parameters retained are
  those that lead to contributions to near on-shell regions of phase space, do not experience suppressions by light quark mases or
  GIM suppression when interfering with the SM, or violate $\rm CP$ and carry a resonant enhancement in this region of phase space.} \label{tab:poleprogramSMEFTG}
\end{table}
The parameter counts in the case of the $n_f = 1,3$ SMEFT are given in Table~\ref{tab:poleprogramSMEFTG}.
The only subtlety is in the counting of the $(\bar{L} L)(\bar{L} L)$ operators. The expression for the shift in the
extracted value of the Fermi constant in the general SMEFT is
\begin{align}
-\frac{4 \, G_F}{\sqrt{2}} &=  -\frac{2}{v_T^2} +  \left(C_{\substack{ll \\ \mu ee \mu}} +  C_{\substack{ll \\ e \mu\mu e}}\right) - 2 \left(C^{(3)}_{\substack{Hl \\ ee }} +  C^{(3)}_{\substack{Hl \\ \mu\mu }}\right).
\label{gfermi}
\end{align}
However, due to the self-Hermitian nature of the operator $Q_{l l}$, if follows that $C_{\substack{ll \\ \mu ee \mu}} =C_{\substack{ll \\ e \mu\mu e}}$.
Further,the diagonal entries of the self Hermitian operator $C^{(3)}_{\substack{Hl}} \in \mathbb{R}$. This leads to the parameter counts given in Table~\ref{tab:poleprogramSMEFTG}.
\begin{table}[t]\centering
  \begin{tabular}{c|c|c}
Class & Parameters  & \\
\hline \hline
1 & $C_{W} \in \mathbb{R}$ & 1 \\
\hline
3 & $\{C_{HD}, C_{H\Box}\} \in \mathbb{R}$ & 2 \\
\hline
4 & $\{C_{HG}, C_{HW}, C_{HB}, C_{HWB}\} \in \mathbb{R}$ & 4 \\
\hline
5 & $\{C_{\substack{uH}},C_{\substack{dH}}\}  \in \mathbb{R}$ & $\sim 2$ \\
\hline
6 & $\{C_{\substack{uW}},C_{\substack{uB}},C_{\substack{uG}},C_{\substack{dW}},C_{\substack{dB}},C_{\substack{dG}}\}  \in \mathbb{R}$ & $\sim 6$ \\
\hline
7 &  $\{C^{(1)}_{\substack{H l}},C^{(3)}_{\substack{H l}},C^{(1)}_{\substack{H q}},C^{(3)}_{\substack{H q}},C_{\substack{H e}},C_{\substack{H u}},C_{\substack{H d}}\}  \in \mathbb{R}, $& $\sim 7$ \\
 \hline
8 $(\bar{L} L)(\bar{L} L)$ &  $\{C_{\substack{l l}}, \mathcal{C}_{\substack{l l}}\} \in \mathbb{R}$ & 2 \\
\hline
 & {\rm \hspace{7cm} Total Count} & $\sim 24$ \\
  \hline \hline
    \end{tabular}
  \caption{Parameter counts in the $\rm U(3)^5$ SMEFT for a `WHZ pole parameter' program. The parameter counts that are approximate also rely on expanding numerically in the Yukawa eigenvalues.} \label{tab:poleprogramU3}
\end{table}
The parameter counts for the $\rm U(3)^5$-SMEFT and MFV-SMEFT in each operator Class are given in Table~\ref{tab:poleprogramU3} and Table~\ref{tab:poleprogramMFV} respectively.
\begin{table}[t]\centering
  \begin{tabular}{c|c|c}
Class & Parameters  & \\
\hline \hline
1 & $C_{W} \in \mathbb{R}$ & 1 \\
\hline
3 & $\{C_{HD}, C_{H\Box}\} \in \mathbb{R}$ & 2 \\
\hline
4 & $\{C_{HG}, C_{HW}, C_{HB}, C_{HWB}\} \in \mathbb{R}$ & 4 \\
\hline
5 & $\{C_{\substack{uH}},C_{\substack{dH}}\}  \in \mathbb{R}$ & $\sim 2$ \\
\hline
6 & $\{C_{\substack{uW}},C_{\substack{uB}},C_{\substack{uG}},C_{\substack{dW}},C_{\substack{dB}},C_{\substack{dG}}\}  \in \mathbb{R}$ & $\sim 6$ \\
\hline
7 &  $\{C^{(1)}_{\substack{H l}},C^{(3)}_{\substack{H l}},C^{(1)}_{\substack{H q}},C^{(3)}_{\substack{H q}},C_{\substack{H e}},C_{\substack{H u}},C_{\substack{H d}}\}  \in \mathbb{R},$& $\sim 13$ \\
&  $\{\Delta C^{(1)}_{\substack{H q}},\Delta C^{(3)}_{\substack{H q}},\Delta C_{\substack{H u}},\Delta C_{\substack{H d}}\}  \in \mathbb{R}$&  \\
 \hline
8 $(\bar{L} L)(\bar{L} L)$ &  $\{C_{\substack{l l}}, \mathcal{C}_{\substack{l l}}\} \in \mathbb{R}$ & 2 \\
\hline
 & {\rm \hspace{7cm} Total Count} & $\sim 30$ \\
  \hline \hline
    \end{tabular}
  \caption{Parameter counts in the MFV SMEFT for a `WHZ pole parameter' program. The parameter counts that are approximate rely on expanding numerically in the Yukawa eigenvalues.
  For $\Delta C^{(1)}_{H q},\Delta C^{(3)}_{H q}$ the notation corresponds to two spurion insertions as defined in Eqn.~\ref{spurions}.} \label{tab:poleprogramMFV}
\end{table}

\subsection{Selection/identification cuts in a pole program}\label{selectioncuts}
A detailed study of efficient experimental cuts to pursue a WHZ pole parameter program is beyond the scope of this work.
In this section, we discuss the plausibility of enforcing stronger narrow width selection cuts at LHC
to enable this effort. We then illustrate numerically
the effect of such selections on $\mathcal{L}^{(6)}$ parameter dependence.  Our purpose is to demonstrate
an application of the Model files, and to numerically illustrate how the scaling rules in Eqn.~\ref{scalingrule} translate into simulated results.

\subsubsection{Narrow width phase space selection cuts for Z}
The general prospects for isolating the near on-shell region of phase space for the $W,Z$ to enable a SMEFT program are strong at LHC, and consistent with
standard particle identification strategies. For example, enforcing a quasi-narrow $\Gamma_Z$ selection cut using $\ell^+ \ell^-$ or $J \, J$ final states is standard
in many LHC searches with an identified $Z$.  Historically $Z$ identification used a dilepton invariant mass cut \cite{Aad:2010yt}
of $\sim  10 \,\Gamma_Z$ in ATLAS \cite{Aad:2010yt} and $\sim  12 \,\Gamma_Z$ in CMS \cite{Khachatryan:2010xn}.
Recent studies at higher operating energies have used tighter cuts
$\sim \pm \{ 3 \,\Gamma_Z, (3 + 0.01 \,p_T^{\ell \ell})\}$ for selecting $\{ e^+ e^-,\mu^+ \mu^-\}$ final states
and $\sim \pm 7 \,\Gamma_Z$ selection for $JJ$ final states in a search at $\sqrt{s} = 13 \, {\rm TeV}$ for heavy resonances decaying into
diboson pairs reported by ATLAS ~\cite{Aaboud:2017itg}. Similar results from CMS \cite{Sirunyan:2017nrt} use comparable criteria.
Although wider selection criteria have also been utilized at ATLAS and CMS, for example in Refs~\cite{Aad:2011dm,CMS:2011aa,Aaboud:2016btc,Chatrchyan:2013tia}
we strongly encourage the development of analyses to enable cleaner SMEFT interpretations of LHC data in a pole parameter program, through optimized narrow width selection cuts.

The effect of further select cuts is non-trivial and requires dedicated experimental studies. For example, high $p_T$ selection cuts on the width of a reconstructed $Z$
is important, with a narrow peak persisting for a boosted on-shell $Z$ and a broader peak from off-shell production of a $Z$ dictated by the $p_T$ cut selected.
Detailed numerical studies are called for with realistic detector simulations and all selection criteria imposed. We note that preliminary studies
indicate that the narrow boosted on-shell peak is a subdominant, but non-negligible, source of $\sim 20\%$ of total events in $Z \rightarrow e^+ e^-$.
when $p_T(e) > 100 \, {\rm GeV}$ is enforced on each electron.

The near resonance numerical suppression of interest in collider SMEFT studies degrades linearly with the width of this cut and we advocate
enforcing even stronger selection criteria when a narrow resonance is present. This is an important alternative experimental strategy
to systematically develop that is currently understudied. To our knowledge, intrinsic detector energy resolution is not a barrier
to significantly tighter selection cuts of this form, as it is $< \Gamma_Z$ for identified leptons.

\subsubsection{Narrow width phase space  selection cuts for W}
Enforcing selection criteria for a narrow width region of phase space for $W \rightarrow \bar{q} \, q$ decays is also feasible. An example is the use of
$\sim \pm 8 \,\Gamma_Z$ selection for $JJ$ final states at $\sqrt{s} = 13 \, {\rm TeV}$ in Ref.~\cite{Aaboud:2017itg}.
For $\{W, Z\}$ using $JJ$ selection cuts of $\sim \pm 7 \,\Gamma_B$ is chosen due to the degradation of the reconstructed boson mass as a function of jet $p_T$
\cite{ATL-PHYS-PUB-2015-033}. Tighter selection criteria for $JJ$ invariant mass cuts can be imposed at the cost of rejecting high $p_T$ jet events.
Inspection of Ref.~\cite{ATL-PHYS-PUB-2015-033} (Fig. 7) indicates that a tighter cut of $\sim \pm 5 \,\Gamma_B$ is feasible when vetoing
high $p_T$ jet events $> 1 {\rm TeV}$.

More subtle is $W \rightarrow \ell \nu $ where transverse variables are used. The scaling that follows from Eqn.~\ref{scalingrule}
on resonance domination of the populated phase space still applies when transverse variables
are used, even though no resonance peak is reconstructed experimentally. Enforcing a narrow width condition using transverse variables\footnote{
See Section \ref{transversemass} for the variable definitions.}
considering the uncertainties involved in $\vec{E}_T^{miss}$ reconstruction and pileup is a challenge.
Define the final state $\ell$ and $\nu$ in the boosted $W$ frame in a plane orthogonal to the $z$ collision axis as
\bea
p_\ell^\mu = \frac{\sqrt{p_i^2}}{2}\left(\gamma - \alpha \sin \theta \cos \theta, - \alpha + \gamma \sin \theta \cos \phi, \sin \theta \sin \phi, \cos \phi \right)^T, \\
p_\nu^\mu = \frac{\sqrt{p_i^2}}{2}\left(\gamma + \alpha \sin \theta \cos \theta, - \alpha - \gamma \sin \theta \cos \phi, -\sin \theta \sin \phi, -\cos \phi \right)^T
\eea
where $\{\theta,\phi\}$ are defined relative to the $z$ collision axis and the boost factor is defined as $\alpha = \sqrt{1- \gamma^2} = |p_T^W|/\sqrt{(p_\ell+p_\nu)^2}$.
It follows that a narrow width region of phase space condition $(\sqrt{(p_\ell+p_\nu)^2}-m_W)^2 \simeq \Gamma_W^2$ corresponds to
\bea
m_T^2 + 2 |p_T^\ell|^2 \left(1 + \frac{2 \, |p_T^\nu|}{|p_T^\ell|} \, \cos \theta_{\ell \nu}\right) \simeq m_W^2 \left( 1 + \frac{\Gamma_W}{m_W} - 2 \, \frac{|p_T^\nu|^2}{m_W^2}\right)
\eea
Typical selection criteria for $W$ leptonic decays are $|p_T^\nu| > 25 \, {\rm GeV}$ and $m_T > 40 \, {\rm GeV}$
\cite{Aad:2010yt,Khachatryan:2010xn,Aaboud:2017itg,Sirunyan:2017nrt}. The lower bound on $m_T$ is limited by QCD backgrounds.
As such, to enforce a near on-shell region of phase space $|p_T^\ell| \lesssim 40 \, {\rm GeV}$ leptons
with a minimized $|p_T^\nu|$ is preferred. The uncertainty on the missing energy reconstruction $\Delta |p_T^\nu| \sim 15  \, {\rm GeV} \sim 7 \, \Gamma_W$ is the basic limiting factor.
Reducing this
uncertainty is limited by pile up at $\sqrt{s} = 13 \, {\rm TeV}$ energies in the RunII collision environment, see Ref.~\cite{Khachatryan:2014gga,Schoefbeck:2016ctt,ATLAS:2014nea}.
Dedicated studies to optimize selecting for near on-shell $W$ boson decays in $W \rightarrow \ell \nu$ decays are warranted, but pessimism on strongly enforcing
these selection cuts is reasonable. Due to these challenges, the numerical illustrations below utilize $JJ$ final states for isolating the $W$ boson
resonance region of phase space.

\subsection{Numerical illustration}\label{sec:numerics}
\begin{figure}[t]
\hspace*{1mm}(i)~~\parbox{\textwidth}{\includegraphics[height=3cm]{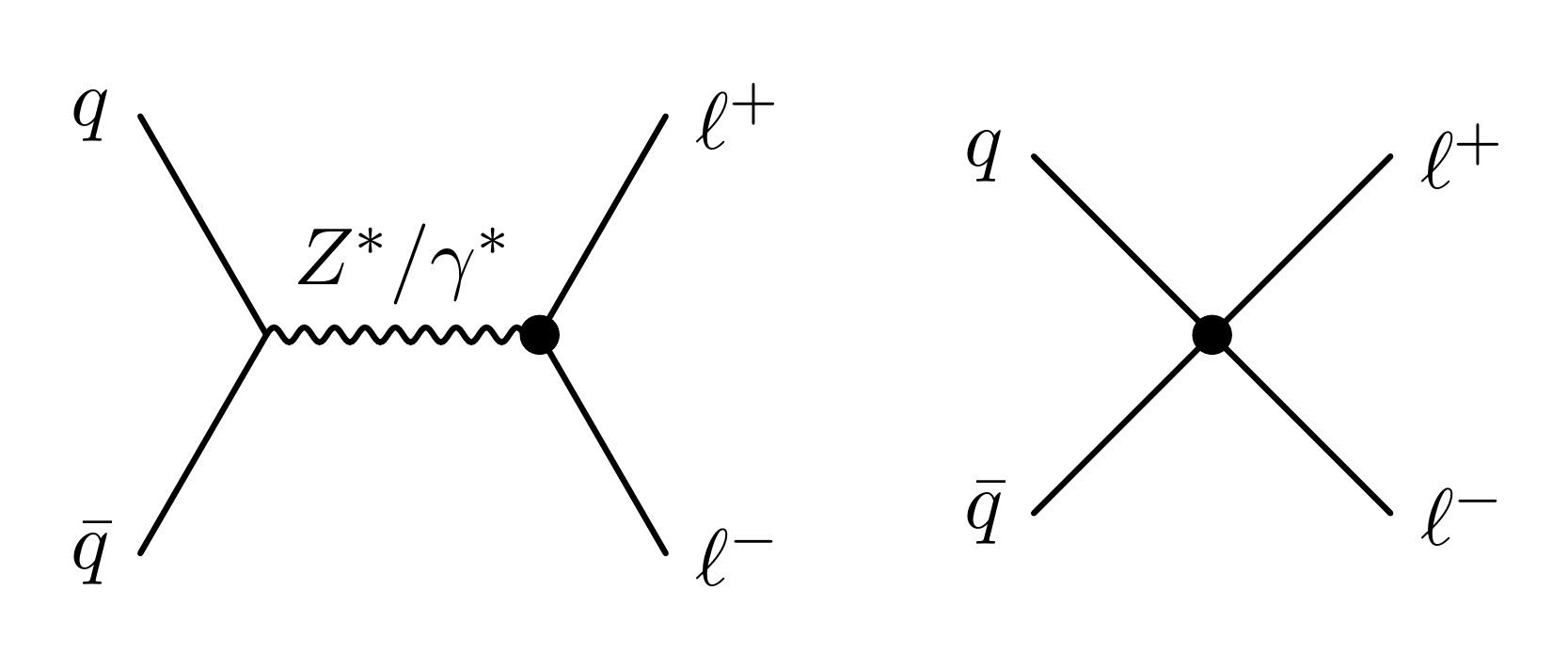}}\\
\hspace*{1mm}(ii)~~\parbox{\textwidth}{\includegraphics[height=2.5cm]{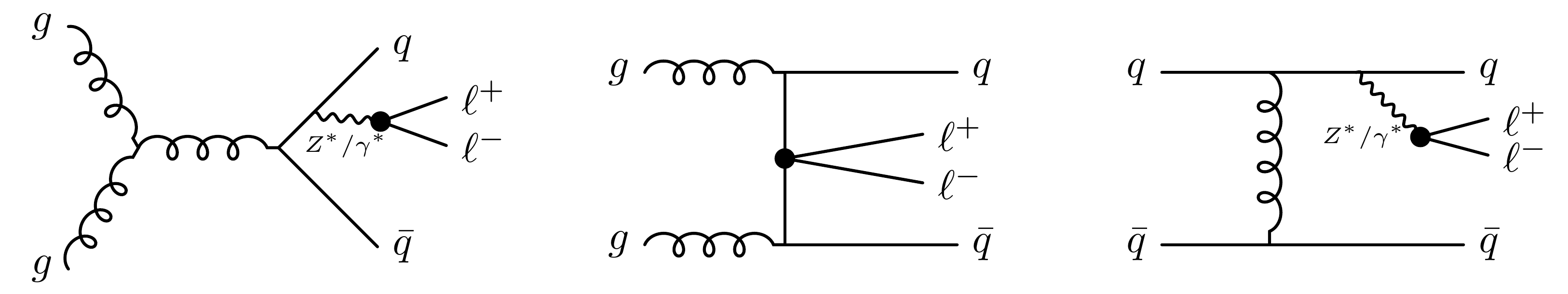}}\\
\hspace*{1mm}(iii)~~\parbox{\textwidth}{\includegraphics[height=2.8cm]{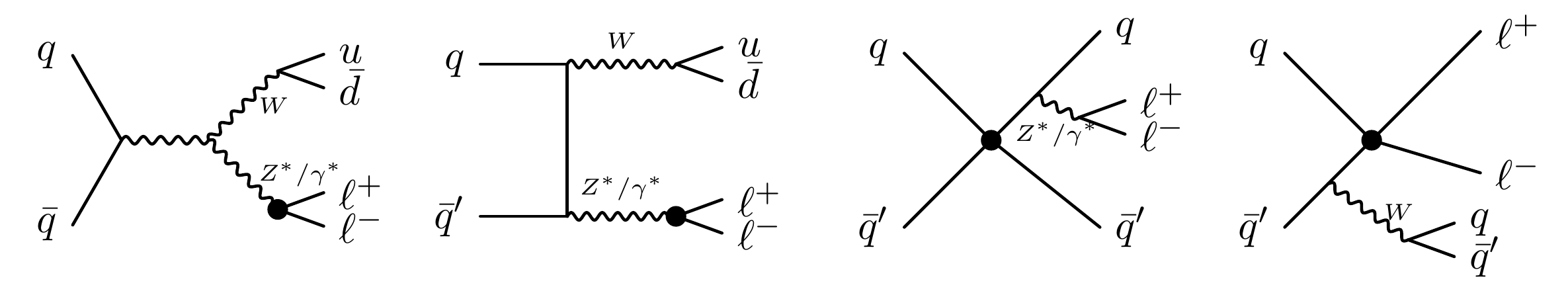}}
 \caption{Illustrative subset of the diagrams contributing to the three processes studied in Section~\ref{sec:numerics}. The dots indicate the (possible)
 insertion of one of the three operators considered in the analysis. }\label{fig:diagrams}
\end{figure}

This section provides a simple numerical demonstration of the arguments in Section \ref{poleparam}, employing the SMEFTsim package and \MG. We present
a basic analysis of the impact of pole vs. non-pole parameters for three LHC processes that receive significant resonant contributions in the SM when narrow width regions of phase
space are selected for:
\begin{enumerate}
 \item[(i)] $p p\to \ell^+\ell^-$, $\quad \ell=\{e,\mu\}$,
 \item[(ii)] $p p \to \bar q q \; \ell^+ \ell^-$, $\quad q\neq t,b,\;$ with the quark pair QCD produced (non-resonant),
 \item[(iii)] $p p \to q_u \,\bar{q}_d\; \ell^+\ell^-$, $\quad q_u=\{u,c\},\, q_d=\{d,s\},\;$ with the quark pair EW produced.
\end{enumerate}
Fig.~\ref{fig:diagrams} shows a sample of the diagrams that contribute to these processes in our study: process (i) is Drell-Yan production that
is resonant for $m(\ell^+\ell^-)\simeq m_Z$. In process (ii) the EW production of the quark pair is subdominant compared to the gluon-mediated diagrams
and, for the sake of cleanliness in the analysis, it is forbidden at the generation level, setting the interaction order limit {\tt QED<=2} in \MG. As a consequence,
this process has only one resonant structure corresponding to the $Z$ peak in the invariant mass distribution of the lepton pair.
Finally, process (iii) serves an illustration for processes with two resonances, around $m_W$ in the $(q_u\bar q_d)$ mass spectrum and around $m_Z$ in the
dilepton. The QCD contributions are conveniently removed in this case when generating events in \MG, requiring {\tt QCD=0}.

We choose 3 representative operators of the Warsaw basis: $\Q_{Hl}^{(1)}$, whose Wilson coefficient belongs to the category of pole parameters
and the two four-fermion operators $\Q_{qe}$, $\Q_{qq}^{(1)}$, that give $(\bar q q)(\bar q q)$ and $(\bar q q)(\bar \ell\ell)$ contact interactions respectively.
We adopt the $\rm U(3)^5$-SMEFT (for $\Q_{qq}^{(1)}$ so that only the flavour contraction $\Q_{\substack{qq\\pptt}}^{(1)}$ is retained)
and use the $\{\hat\a_{\rm em}, \hat m_Z,\hat G_F\}$ input scheme.

For each process we generate one event sample for the SM production and one for each interference term with one of the three effective operators considered,
included one by one.\footnote{The authors are well aware that one at a time operator analyses are generally not representative
of consistent IR limits in the SMEFT \cite{Jiang:2016czg}. For our numerical illustration, we reluctantly consider a one at a time operator analysis acceptable
to examine the numerical result of the scaling arguments underlying the pole parameter program.} The operator $\Q_{qq}^{(1)}$ is included only in process (iii).

The processes are generated in \MG ~using the UFO model from set~A for the $\rm U(3)^5$ symmetric case and with the $\{\hat\a_{\rm em},\hat m_Z,\hat G_F\}$ scheme. Restriction cards are employed to set all the fermion masses and Yukawa couplings to zero (with the exception of those of the top and bottom quarks) and to fix the values of the Wilson coefficients. In the SM limit all of them are vanished, while for the operator insertions the corresponding coefficient is set to $C_i=1$ with $\Lambda= 1$~TeV.
Interaction order limitations are used both to produce cleaner signals and to isolate the interference term in the case of $\mathcal{L}^{(6)}$ insertions. Specifications of the first class are {\tt QED<=2} for process (ii) and {\tt QCD=0} for process (iii), that reduce the number of relevant SM diagrams as explained above.
The interference is instead isolated\footnote{The estimate of the interference term obtained with this procedure is more accurate and numerically stable than the estimate obtained e.g. generating the full process with $\mathcal{L}_{SM}+\mathcal{L}^{(6)}$ and subsequently subtracting the pure SM contribution.} using the recently introduced \MG~syntax {\tt NP\^{}2==1}.

Each event sample contains $10^4$ events and it is produced setting the widths of the $W,Z,h$ bosons for automatic evaluation in \MG ~and restricting the phase space with broad kinematic cuts on the invariant masses of fermion pairs, as summarized in Table~\ref{tab:generation_details}. The invariant mass of the lepton pair in the final state is always required to be in the region between 40 and 120~GeV and the invariant mass of the $(q_u\bar q_d)$ pair is required to be in the same window for process (iii). This selection allows a more efficient scan of the near-resonant regions of the parameter space.

The invariant mass spectra of the relevant fermion pairs are extracted analyzing the event samples with ROOT~\cite{Brun:1997pa}. In the case of the SM--$\mathcal{L}^{(6)}$ interference terms, the histograms are further rescaled by $|\s(C_i,{\rm int.})| / \s(\rm SM)$ so that their bin content can be directly compared to that of the SM distributions.
In this way all the histograms have the same (arbitrary) normalization, which is such that the SM production has $10^4$ events in the kinematic region for which the events were generated.

The resulting distributions for the interference terms are shown in the lower panels of Figs.~\ref{fig:DY_qqZ_ee_mass}-\ref{fig:ZW_ll_mass} (colored lines) and can be easily compared. The figures also show, for reference, the error band on the complete distribution for the SM case (grey band). The latter is estimated bin-by-bin as\footnote{In our analysis the SM distribution has two sources of uncertainty: a statistical uncertainty due to analyzing a finite event sample and an error on the overall normalization of the distribution, stemming from the uncertainty in the determination of the total cross-section in \MG. The latter, however, is an effect of order {$\sim$~8~\textperthousand} and can be safely neglected.} $\Delta N_k = \sqrt{N_k}$ where $N_k$ is the number of events in the bin $k$.
Finally, the top panels show the total predictions for the SM and for the SM + one operator, obtained as the sum of the SM and interference histograms. Note that, while in the lower panel we plot the absolute size of the interference terms, in the upper panel its sign is kept into account. In particular, the interference is always negative for $C_{qe}$ and positive in all the other cases. The normalization of the histograms is arbitrary and such that the SM production has $10^4$ events in the kinematic region for which the events were generated.

\begin{table}[t]\centering
\renewcommand{\arraystretch}{1}
\begin{tabular}{c|*4{m{2.9cm}}}
 \hline
\bf Process& \multicolumn{4}{c}{\bf Specifications in \MG}\\\hline
& SM int. order &{\tt wz,ww,wh = auto}&{\tt mmll = 40 mmllmax = 120}& {\tt mmjj = 40 mmjjmax = 120}\\
(i)&  & $\surd$& $\surd$& \\
(ii)& {\tt QED <= 2}& $\surd$& $\surd$& \\
(iii)& {\tt QCD = 0}& $\surd$& $\surd$& $\surd$\\
 \hline
\end{tabular}
 \caption{Details of the options specified in \MG~ for the generation of events for each of the processes considered (see text for details).}\label{tab:generation_details}
\end{table}

\subsubsection[Results for \texorpdfstring{$p p \rightarrow \ell^+ \ell^-$}{pp -> l+ l-} and \texorpdfstring{$p p \rightarrow q q \; \ell^+ \ell^-$}{pp -> l+ l- q qbar}]{Results for processes (i) and (ii)}
Processes (i) and (ii) are both singly-resonant in the SM, and they show a particularly clean enhancement/suppression effect of the $\mathcal{L}^{(6)}$ contributions
in the distributions in near the resonant region of phase space.

\begin{figure}[t]\hspace*{-1.3cm}
\includegraphics[height=7.8cm]{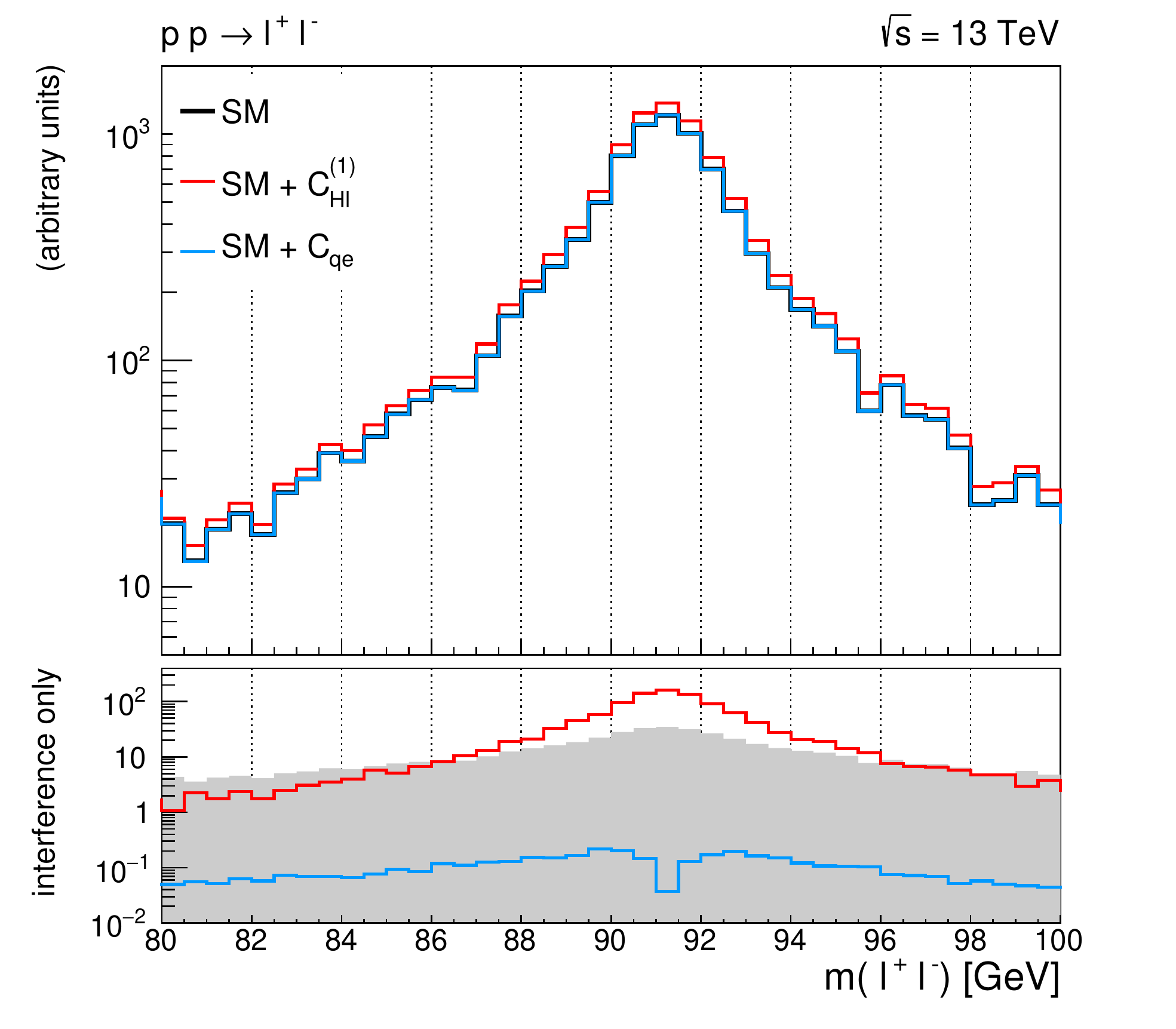} \hspace*{-8mm}
\includegraphics[height=7.8cm,trim={1.5cm 0cm 0cm 0cm},clip]{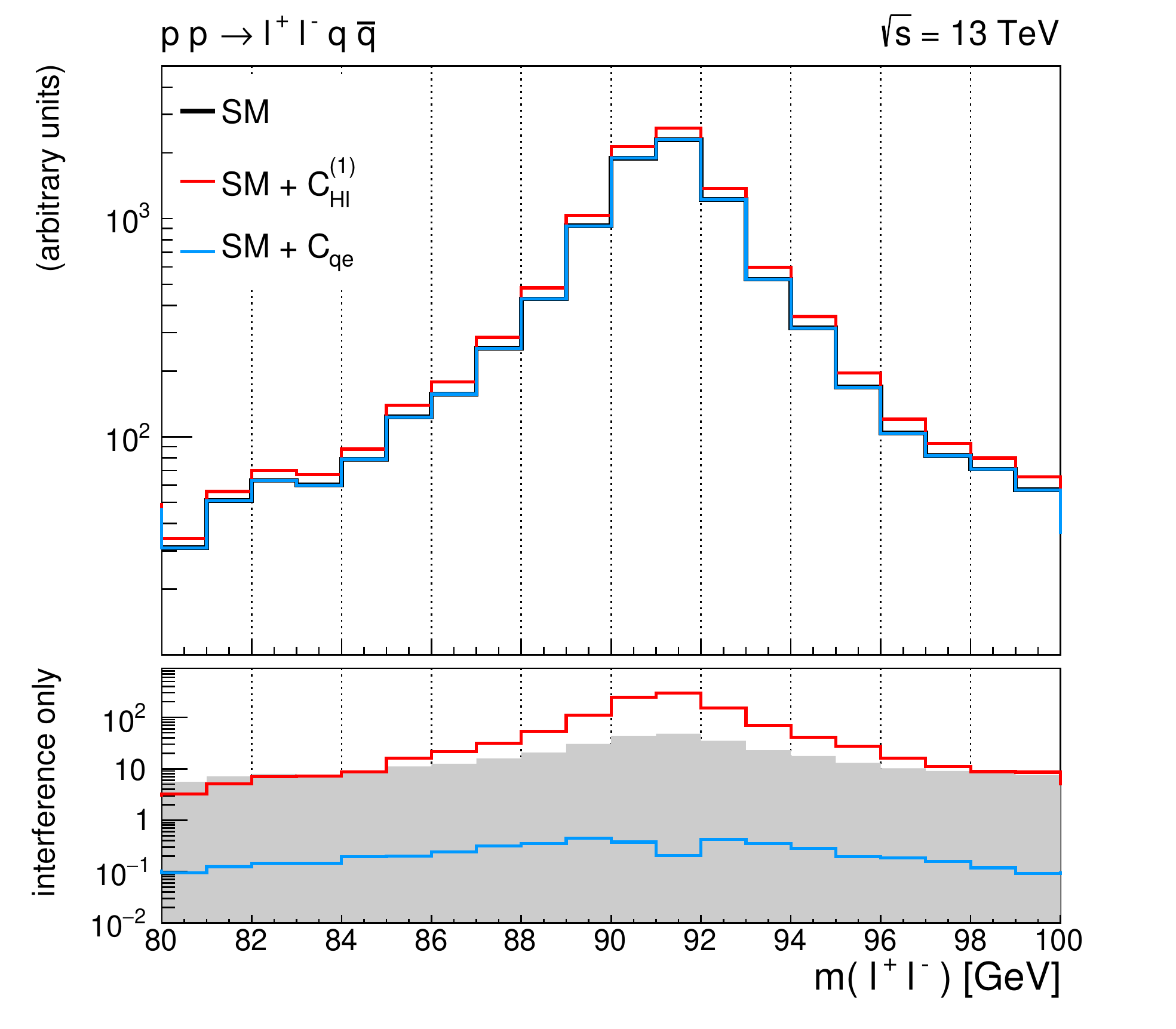}
\caption{Left: invariant mass distribution of the $\ell^+\ell^-$ pair from Drell-Yan production (process~(i)). Right:  invariant mass distribution of the $\ell^+\ell^-$ pair from process (ii). The top panels show the complete spectra obtained in the SM limit and in the presence of one effective operator with $C_i/\Lambda^2=1$~TeV$^{-2}$ (only the interference term is retained in these cases). The black and blue lines are overlapping in the right figure. The lower panels show the absolute size of the pure SM--$\mathcal{L}^{(6)}$ interference term for each operator (colored lines) compared to the statistical uncertainty on the SM distribution (grey band).
}\label{fig:DY_qqZ_ee_mass}
\end{figure}

As shown in Fig.~\ref{fig:DY_qqZ_ee_mass}, in both cases the impact of the ``pole operator'' $\Q_{Hl}^{(1)}$ is enhanced around the $Z$ peak.
On the other hand, the four-fermion operator $\Q_{qe}$ has a very small impact overall, which, with the statistics presented here, is always smaller than the
statistical uncertainty on the SM production alone. It is worth noting that,
around the $Z$ resonance, it undergoes a further suppression, that can be appreciated as a dip in the curve in the low panels of Fig~\ref{fig:DY_qqZ_ee_mass}.

The relative suppression of the $\Q_{qe}$ vs. $\Q_{Hl}^{(1)}$ operator emerging around a $B$ boson resonance can be
quantified as $N(C_{qe}) / N(C_{Hl}^{(1)})$, where $N(C_i)$ is the number of events in the region $|m(\bar ff)-m_B|\leq \Gamma_B$
for the interference spectrum of a the operator $\Q_i$. In this case, $N(C_i)$ is
given by the number of entries for the bins between 89 and 93~GeV of the interference histograms.
The numbers obtained in this way are summarized in Table~\ref{tab:integral_ratios}.
This gives $\sim 1/620$ and $\sim 1/570$ for processes (i) and (ii) respectively (see Table~\ref{tab:integral_ratios}),
which is consistent with the estimate of Eq.~\ref{scalingrule}.

\subsubsection[Results for \texorpdfstring{$p p \rightarrow \ell^+ \ell^- q_u \bar q_d$}{pp -> l+ l- q_u qbar_d}]{Results for process (iii)}
\begin{figure}[t]\hspace*{-1.5cm}
\includegraphics[height=8.5cm]{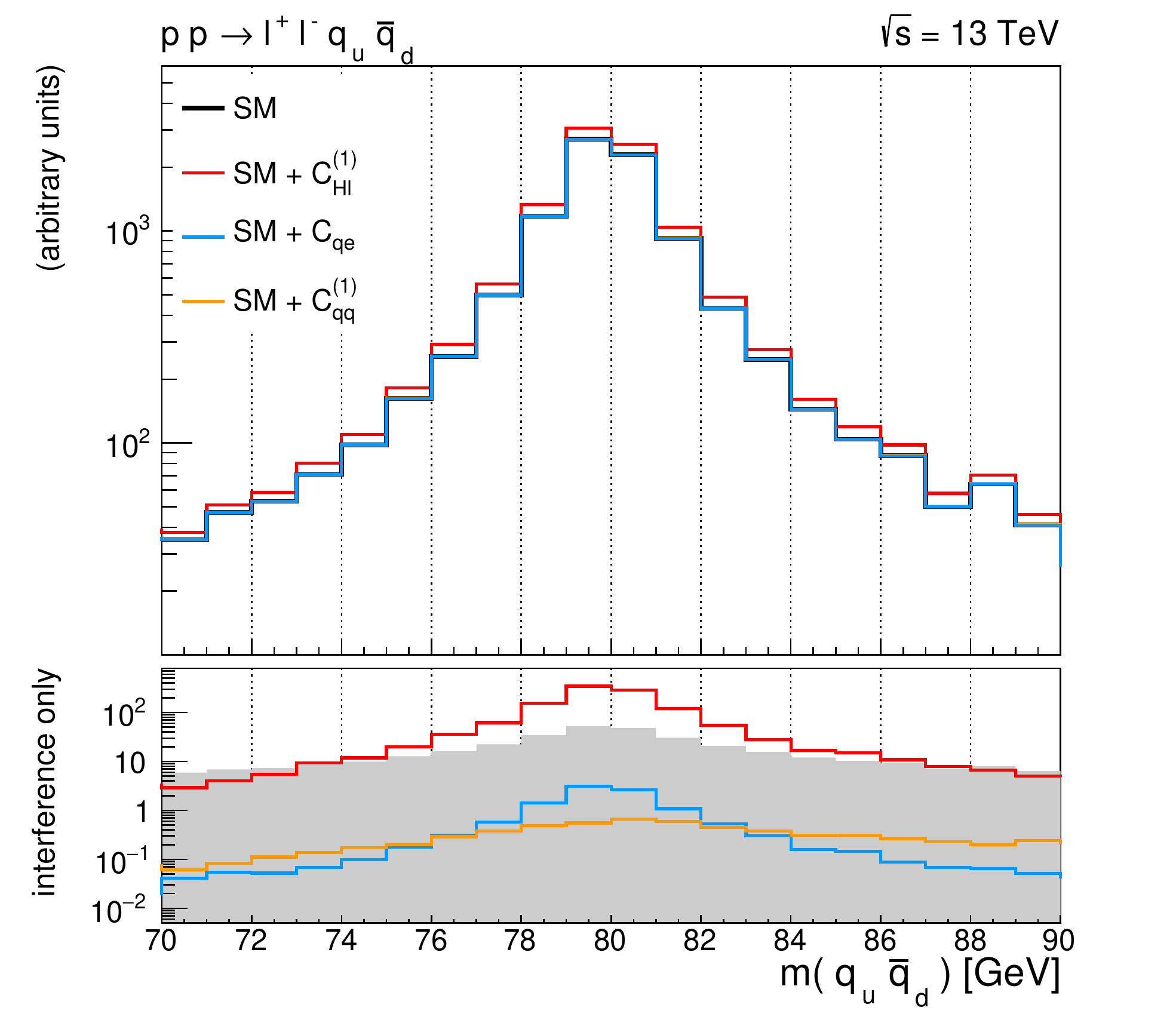}\hspace*{-8mm}
\includegraphics[height=8.5cm,trim={1.5cm 0cm 0cm 0cm},clip]{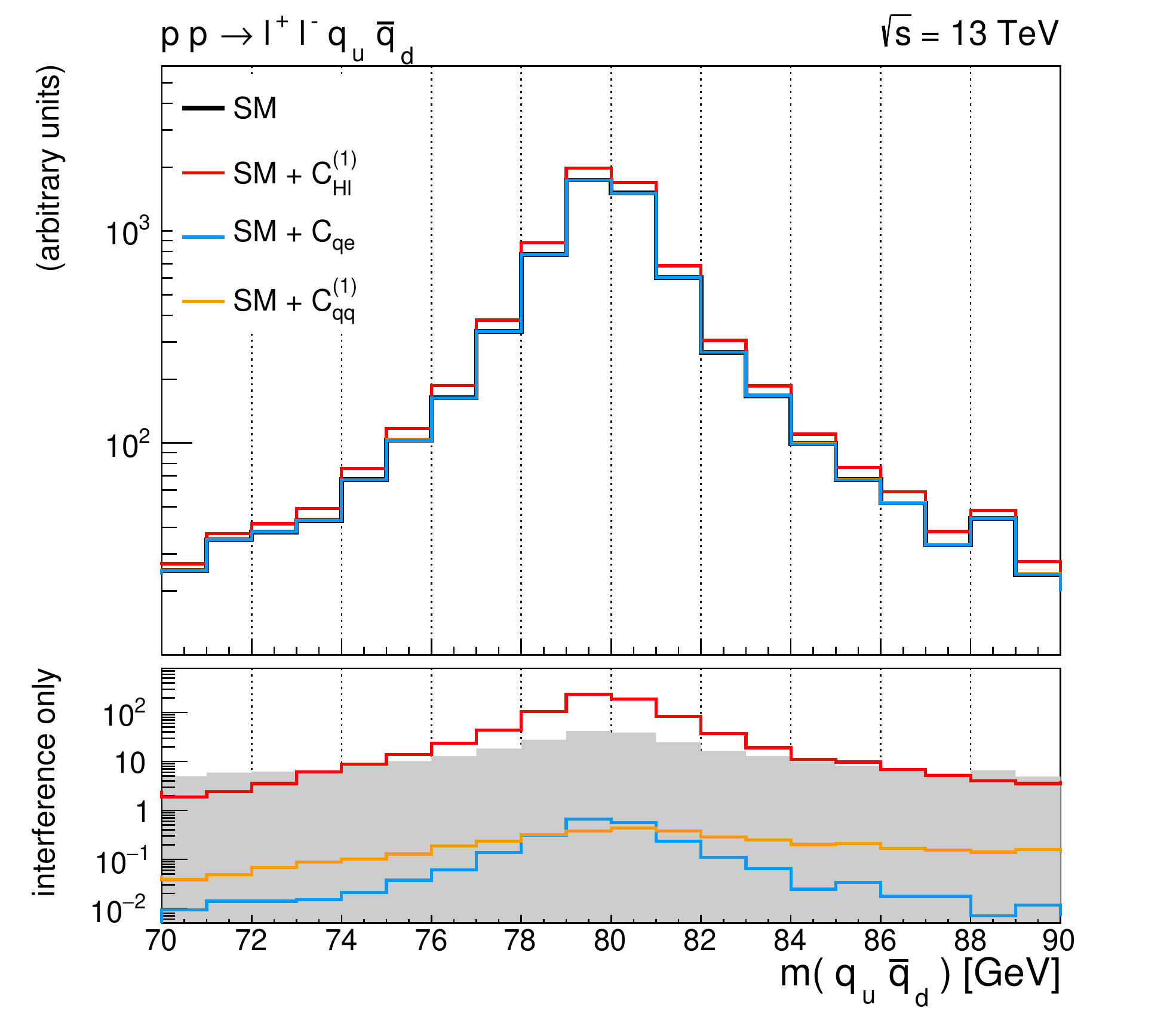}
\caption{Invariant mass spectrum of the $q_u\bar q_d$ pair in process (iii) before (left) and after (right) applying a narrow
selection cut on $m(\ell^+\ell^-)$ to isolate the $Z$ peak. The top panels show the complete distributions obtained in the SM limit and in the presence of one effective operator with {$C_i/\Lambda^2=1$~TeV$^{-2}$} (only the interference term is retained in these cases). The black, yellow and blue lines are overlapping in both figures. The lower panels show the absolute size of the pure SM--$\mathcal{L}^{(6)}$ interference term for each operator (colored lines) compared to the statistical uncertainty on the SM distribution (grey band). }\label{fig:ZW_qq_mass}
\end{figure}

\begin{figure}[t]\hspace*{-1.5cm}
\includegraphics[height=8.5cm]{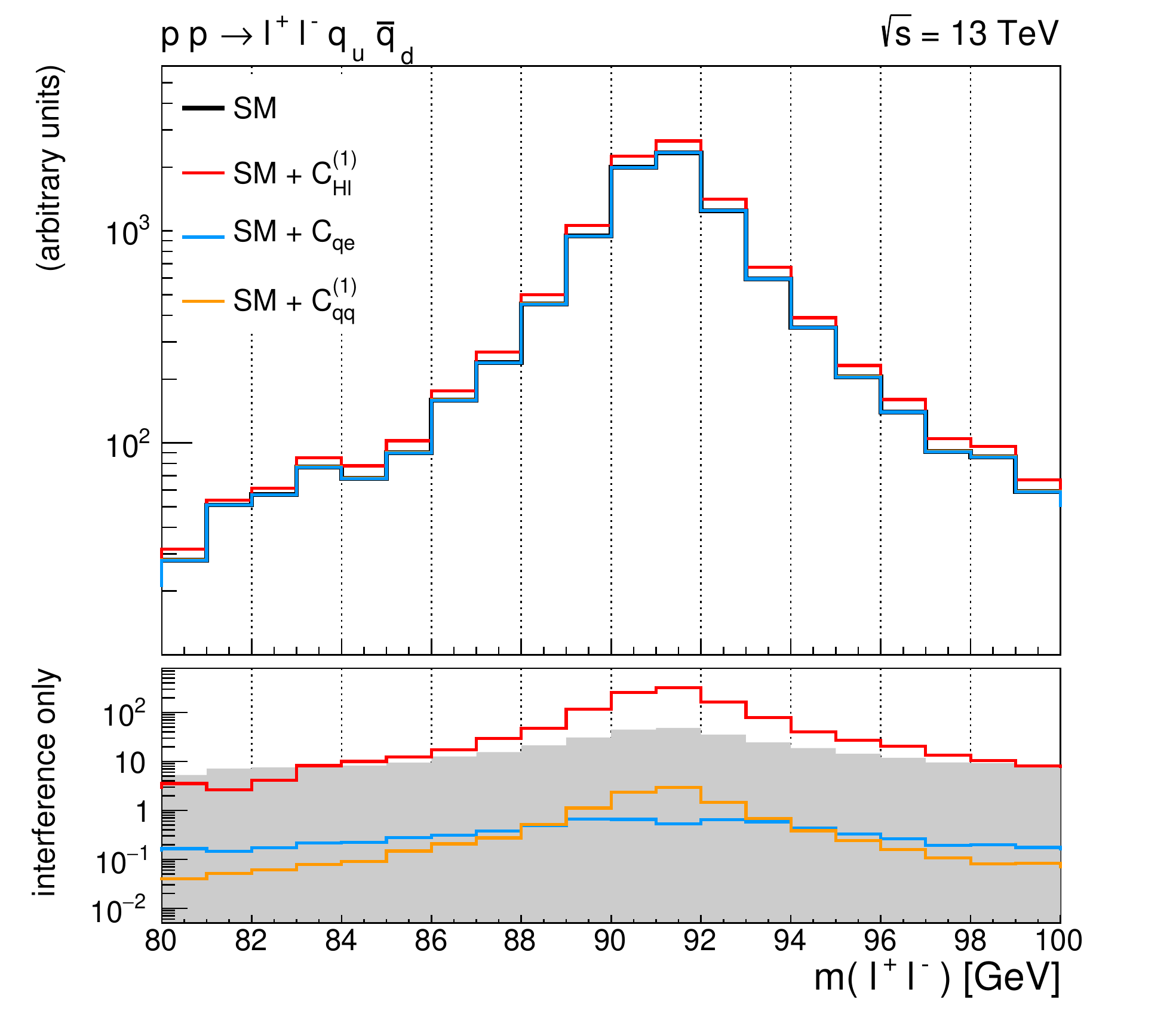}\hspace*{-8mm}
\includegraphics[height=8.5cm,trim={1.5cm 0cm 0cm 0cm},clip]{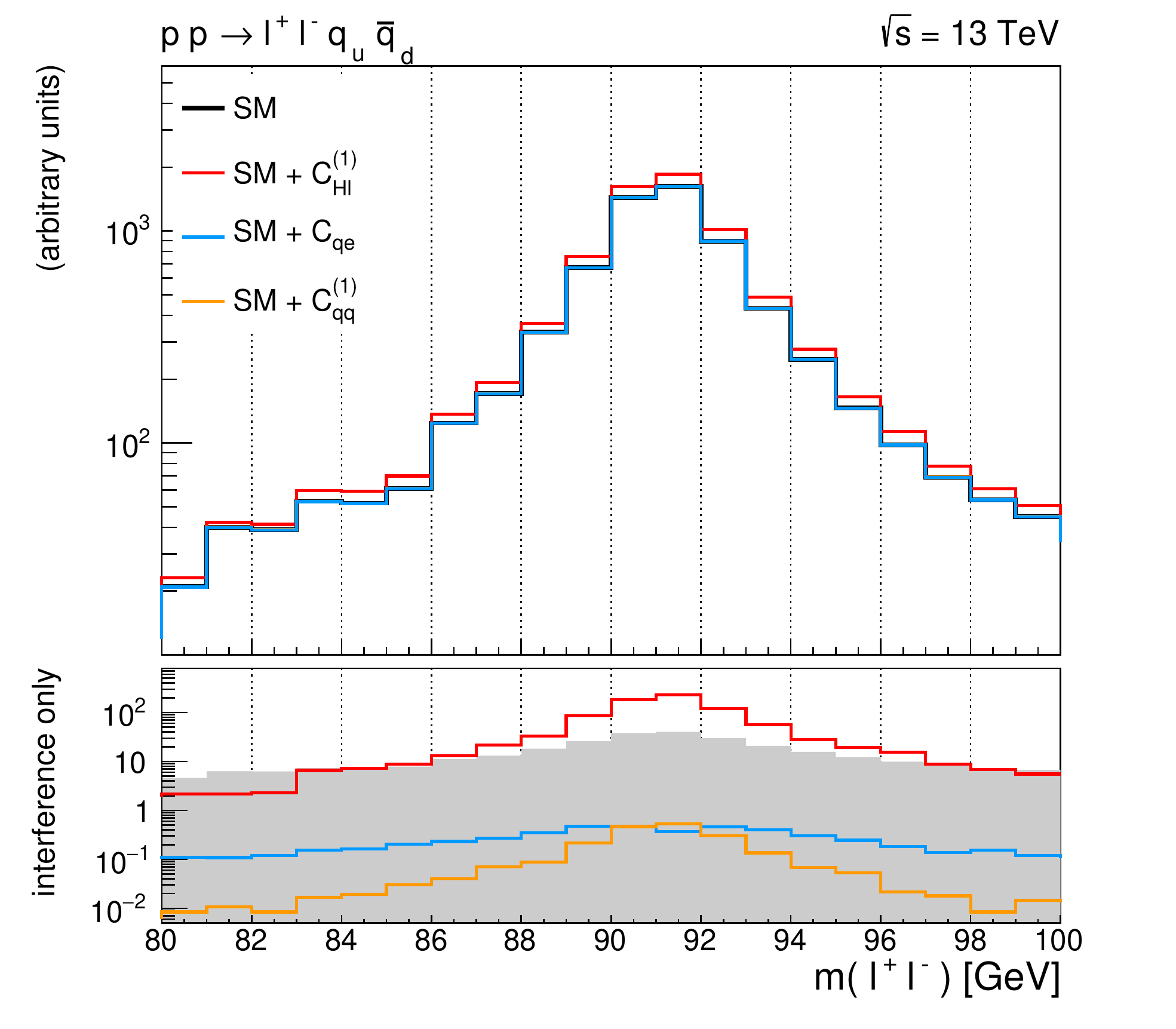}
\caption{Invariant mass spectrum of the $\ell^+\ell^-$ pair in process (iii) before (left) and after (right) applying a narrow selection cut on $m(q_u\bar q_d)$ to isolate the $W^+$ peak. The top panels show the complete distributions obtained in the SM limit and in the presence of one effective operator with {$C_i/\Lambda^2=1$}~TeV$^{-2}$ (only the interference term is retained in these cases). The black, yellow and blue lines are overlapping in both figures. The lower panels show the absolute size of the pure SM--$\mathcal{L}^{(6)}$ interference term for each operator (colored lines) compared to the statistical uncertainty on the SM distribution (grey band). }\label{fig:ZW_ll_mass}
\end{figure}

Because process (iii) can present two resonances, both the invariant mass spectra of the $q_u\bar q_d$ (Fig.~\ref{fig:ZW_qq_mass}) and dilepton (Fig.~\ref{fig:ZW_ll_mass})  pairs are analyzed.
For each pair we show the results obtained directly from the generated sample vs. after applying an additional narrow cut on the invariant mass of the complementary fermion pair:
in one case (Fig.~\ref{fig:ZW_qq_mass}, right) we select {89~GeV~$ < m(\ell^+\ell^-)<93$~GeV} to isolate the $Z$ peak and observe its impact on the $(q_u \bar q_d)$ spectrum, while in the other (Fig.~\ref{fig:ZW_ll_mass}, right) we select {78~GeV~$ < m(q_u\bar q_d)< 93$~GeV} to isolate the $W^+$ peak and observe its impact on the $m(\ell^+\ell^-)$ distribution.

While these cuts do not have a visible impact on the total spectra, they significantly reduce the size of the pure interference term for one four-fermion operator: selecting resonant $\ell^+\ell^-$ pairs suppresses the contribution of $\Q_{qe}$, while selecting resonant $q_u\bar q_d$ pairs removes that of $\Q_{qq}$. This effect is clearly visible in the lower panel of Fig.~\ref{fig:ZW_qq_mass} (\ref{fig:ZW_ll_mass}), comparing the blue (yellow) curves in the left and right plots.

As for processes (i) and (ii), the enhancement of the interference term for $C_{Hl}^{(1)}$ close to the resonance is clearly visible in the lower panels of Figs.~\ref{fig:ZW_qq_mass},~\ref{fig:ZW_ll_mass} and the impact of four fermion operators is smaller than the error band of the SM distribution. The extra suppression of the four-fermion operators, instead, is less evident and can only be seen as a tiny dip in the central bins of the $m(\ell^+\ell^-)$ distributions for $C_{qe}$ (Fig.~\ref{fig:ZW_ll_mass}, lower panels).

Finally, the relative suppression of the $\Q_{qe}$ and $\Q_{qq}^{(1)}$ vs. $\Q_{Hl}^{(1)}$ operators is quantified by the ratios $N(C_{qe}) / N(C_{Hl}^{(1)})$, $N(C_{qq}^{(1)}) / N(C_{Hl}^{(1)})$ summarized in Table~\ref{tab:integral_ratios}. The ratios found for this process are of the same order of magnitude as those for processes (i) and (ii) and consistent with the estimate
of Section~\ref{poleparam}.
The values $\sim 1/110$ obtained for the $C_{qq}^{(1)}/C_{Hl}^{(1)}$ ratio with cuts on $m(\ell^+\ell^-)$ and for the  $C_{qe}/C_{Hl}^{(1)}$ ratio with cuts on $m(q_u\bar q_d)$ essentially reflect the ratio of the cross-sections obtained for the interference terms, as the selection is ineffective in these cases. Restricting to the relevant resonant region induces an additional suppression of a factor 3--4. This effect is also observed in processes (i) and (ii).

\begin{table}[ht]\centering
\renewcommand{\arraystretch}{1.2}
\begin{tabular}{cl|cc}
\hline
\bf Process&\bf Resonant region& \boldmath $N(C_{qe})/N(C_{Hl}^{(1)})$ & \boldmath $N(C_{qq}^1)/N(C_{Hl}^{(1)})$\\\hline
(i)& $89.\leq m(\ell^+\ell^-)\leq 93$~GeV & 1/620 & -\\
(ii)& $89.\leq m(\ell^+\ell^-)\leq 93$~GeV& 1/566 & -\\
(iii)& $89.\leq m(\ell^+\ell^-)\leq 93$~GeV& 1/344 & 1/109\\
& $78.\leq m(q_u \bar q_d)\leq 82$~GeV& 1/109& 1/392\\
& both resonances & 1/333 & 1/388\\
\hline
\end{tabular}
\caption{Approximate ratios of the number of events contained in the central peak regions of the relevant invariant mass spectra determined by the pure SM--$\mathcal{L}_{d=6}$ interference of non-pole ($C_{qe},\,C_{qq}^{(1)}$) vs pole ($C_{Hl}^{(1)}$)
parameters in the three processes considered. The values can be compared with the order of magnitude estimate
of Eq.~\ref{scalingrule}, that gives ${\Gamma_B m_B/\bar g_{2} \bar v_T^2\approx 1/220 (250)}$ for $B=Z (W)$.
}\label{tab:integral_ratios}
\end{table}

\section{Conclusions}
In this paper we have advanced the SMEFT physics program on multiple fronts. We have developed and reported the \href{https://feynrules.irmp.ucl.ac.be/wiki/SMEFT}{SMEFTsim} package,
a set of \FR \,  implementations of the general SMEFT, the $\rm U(3)^5$-SMEFT and the MFV-SMEFT theories as defined in Sections~\ref{Section:canonicalform},\ref{Section:inputschemes},\ref{Section:symassumptions}.
We have provided these results in two input parameter schemes, $\{\hat{\alpha}_{ew}, \hat{m}_Z, \hat{G}_F \}$
or $\{\hat{m}_{W}, \hat{m}_Z, \hat{G}_F\}$ in all three cases. We have supplied two code sets based on this theoretical outline
for validation purposes.

We have also systematically developed a theoretical framework of a WHZ pole parameter program in the SMEFT as a key application to pursue with the SMEFTsim package.
The idea is to maximally exploit numerical suppressions in the scattering and interference of the SMEFT
corrections to the SM in addition to the power counting of the EFT. These IR effects come about due to approximate symmetries and the use of near on shell regions of phase space
enforced with selection cuts at hadron colliders. Rather ironically, a key complaint against EFT methods at hadron colliders
- an excess of parameters, can be arguably overcome using stronger versions of the very selection cuts that underlie standard particle identification of narrow bosons of the SM
in such environments.
We have advanced this argument and numerically demonstrated the impact of the pole parameter scaling using the SMEFTsim package interfaced with \MG.
Although our results do not rise to the level of realistic selection cuts at LHC with full detector effects included, we believe they are sufficient
and promising enough to strongly motivate the initiation of a systematic
pole parameter approach to data analysis at LHC.

The enormous data rate at LHC in RunII and in the high luminosity run is such that stronger selection cuts sacrificing pure rate in favour
of cleaner SMEFT motivated measurements, at lower energies, are of interest and reasonable to consider and develop. Such cuts can enable a
systematic program of constraining physics beyond the SM using powerful EFT techniques,
that are already argued to be relevant by the lack of beyond the SM resonances discovered to date at LHC. The number of parameters present in such WHZ pole parameter
efforts is manageable, and the SMEFTsim package allows the systematic study and optimization of selection cuts to develop this program, examining quantitatively
how neglected terms are suppressed with tight or weak selection cuts.
We strongly encourage the LHC experimental collaborations to study and develop a pole parameter SMEFT approach to LHC measurements using the tools provided, in addition to standard
searches already in place.

\acknowledgments
We thank the Villum Foundation, NBIA, the Discovery Centre at Copenhagen University and the Danish National Research Foundation
(DNRF91) for support. MT also thanks past collaborators and members of the Higgs Cross Section Working Group
for discussions. We thank Fabio Maltoni, Marco Zaro and Olivier Mattelær for assistance with MadGraph,
Chris Hays and Gabija Zemaityte for useful suggestions to optimize the code implementation and Will Shepherd for reporting a significant, bizarre bug in the first version of the code.
We thank  Ben Fuks for feedback and continual help with Feynrules in all aspects.
We thank Andr\'e David, Tilman Plehn and Zoltan Ligeti for helpful discussions and feedback. We thank Kristin Lohwasser and Troels Petersen for helpful
feedback on Section \ref{selectioncuts}.

\appendix

\section{Parameter tuning and experimental constraints on Class 5, 6 operators}

\begin{figure}[t]\centering
 \includegraphics[width=0.75\textwidth]{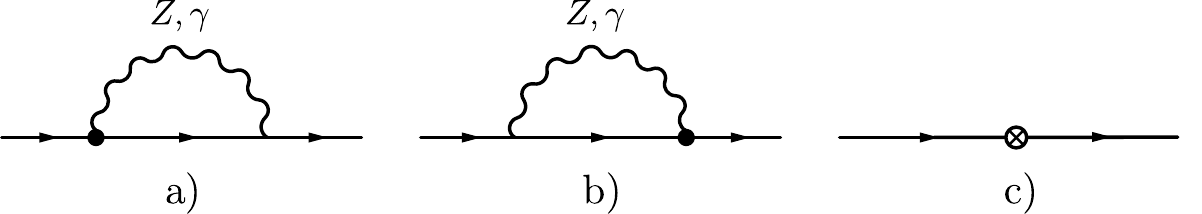}
 \caption{One loop corrections due to $\psi^2 X H$ dipole operators on the fermion two point function. The insertion of the $\mathcal{L}^{(6)}$
 operators is indicated with a black dot, and the counterterm matrix with a ``x".}\label{twopoint}
\end{figure}
Operators of Class 6 ($\psi^2 X H$) in the Warsaw basis can also contribute anomalous three point interactions,
in addition to operators of the form $\langle H |\mathcal{L}_{SM} | H \rangle$.
In this appendix we examine the constraints on these operators due to naturalness concerns and some experimental considerations.
In Section \ref{Section:symassumptions} the
insertion of a SM Yukawa matrix in this operator Class is made to formally restore $\rm U(3)^5$ flavour symmetry. One can utilize one loop corrections
in the $\mathcal{L}_{SMEFT}$ to examine the theoretical support of this approach and establish if  flavour diagonal interactions due to $\psi^2 X H$ operators with $C_{\psi X} \not \propto y_i$,
with $y_i$ a Yukawa coupling, introduce a significant tuning of parameters in the SMEFT. Consider the one loop diagrams shown in Fig.~\ref{twopoint} calculated in dimensional regularization
with $d= 4- 2 \epsilon$ in $\rm \overline{MS}$. The divergence structure of these
diagrams for the operator $C_{eB}$ is given by
\bea\label{dipoleexample}
i \mathcal{A}_\epsilon &=& - \bar{\psi}^{(0)}_L i \slashed{D} \, \psi^{(0)}_L \left(\frac{1}{16 \, \pi^2 \epsilon} \frac{3 \, v \, m_\ell \, \bar{g}_1 \, C_{eB}}{2 \, \sqrt{2}} \right)
- \bar{\psi}^{(0)}_R i \slashed{D} \, \psi^{(0)}_R \left(\frac{1}{16 \, \pi^2 \epsilon} \frac{3 \, v \, m_\ell \, \bar{g}_1 \, C_{eB}}{2 \, \sqrt{2}} \right), \nonumber \\
&-& i \bar{\psi}^{(0)}_L \, \psi^{(0)}_R \, \left(\frac{m_\ell^2 \, C_{eB} \, v}{16 \, \pi^2 \, \epsilon}\right) \left(\frac{9 \, \bar g_1}{2 \sqrt{2}}\right)
- i \bar{\psi}^{(0)}_L \, \psi^{(0)}_R \, \left(\frac{v^3 \, C_{eB}}{16 \, \pi^2 \epsilon} \right) \frac{(9 \bar{g}_1^3 - 3 \, \bar{g}_2^2 \, \bar{g}_1)}{4 \, \sqrt{2}}
\eea
The counterterms are introduced as
\bea
\psi^{(0)}_{L/R} = \sqrt{Z_{\psi_{L/R}}} \, \psi^{(r)}_{L/R}, \quad \quad Q_i^{(0)} = Z_{ij} \, Q_j^{(r)}.
\eea
The contribution to the chiral wavefunction renormalization factors cancels the divergences in the first line of Eqn.~\ref{dipoleexample},
while the counterterm for $C_{eH}$ reported in Ref.~\cite{Alonso:2013hga} (see   Fig.~\ref{twopoint} c) exactly cancels the divergences in the second line once
the SMEFT wavefunction renormalization is taken into account.
The finite terms follow the same pattern and a shift to the light fermion mass is present that is not proportional to $y_i$
\bea
\left(i \mathcal{A}_{finite}\right)_{m_\ell \rightarrow 0} = - i \, \frac{v^3 \, C_{eB}}{4 \, \sqrt{2}} \,  (3 \bar{g}_1^3 - \bar{g}_2^2 \, \bar{g}_1)
 \left(\frac{1}{16 \,  \pi^2} \right) \left(1 + 3\log \frac{\mu^2}{m_Z^2}\right) \bar{\psi}_L \, \psi_R +  \cdots
\eea
Similar corrections are present for all the fermion masses in the SMEFT,
generated by the dipole operators $Q_{eW},Q_{uW},Q_{uB},Q_{dW},Q_{dB}$.
It follows that if
\bea
- \frac{v^2 \, C_{eB}}{4} \,  (3 \bar{g}_1^3 - \bar{g}_2^2 \, \bar{g}_1)
 \left(\frac{1}{16 \,  \pi^2} \right) \left(1 + 3 \,\log \frac{\mu^2}{m_Z^2}\right) \gg y_\ell,
\eea
then a tuning of parameters would be present to obtain the Yukawa coupling $y_\ell$
inferred from the observed fermion masses. Taking $\mu \sim {\rm TeV}$ one finds
\bea
\delta y_\ell ({\rm TeV}) \simeq 3 \times 10^{-5} \,  C_{eB} \, \frac{[\rm TeV]^2}{\Lambda^2}
\eea
as a correction to the effective Yukawa coupling. Numerically as
\bea
y_e(\mu < {\rm TeV}) &\simeq 3 \! \times 10^{-6}, \quad
y_\mu(\mu  <  {\rm TeV}) &\simeq 6 \times 10^{-4}, \quad
y_\tau(\mu  < {\rm TeV}) \simeq 0.01, \\
y_u(\mu  < {\rm TeV}) &\simeq 6 \times 10^{-4}, \quad
y_c(\mu  < {\rm TeV}) &\simeq 4 \times 10^{-5}, \quad
y_t(\mu  < {\rm TeV}) \simeq 1, \\
y_d(\mu  < {\rm TeV}) &\simeq 3 \times 10^{-5}, \quad
y_s(\mu  < {\rm TeV}) &\simeq 6 \times 10^{-4}, \quad
y_b(\mu  < {\rm TeV}) \simeq 0.02,
\eea
dipole operators do not have to be $\propto y_i$ to avoid this tuning of parameters.

Operators of Class 5 directly lead to a tree level contribution to an effective Yukawa interaction for $\Lambda \sim {\rm TeV}$ that are numerically
\bea
\frac{3 \, C_{eH} v^2}{2 \, \sqrt{2}\, \Lambda^2} \sim 0.06 \, C_{eH} \frac{[{\rm TeV}]^2}{\Lambda^2} \gg y_{e,\mu}, \\
\frac{3 \, C_{dH} v^2}{2 \, \sqrt{2}\, \Lambda^2} \sim 0.06 \, C_{dH} \frac{[{\rm TeV}]^2}{\Lambda^2} \gg y_{d,s}, \\
\frac{3 \, C_{uH} v^2}{2 \, \sqrt{2}\, \Lambda^2} \sim 0.06 \, C_{uH} \frac{[{\rm TeV}]^2}{\Lambda^2} \gg y_{u,c},
\eea
which implies expanding around a $\rm U(3)^5$ symmetric limit in the SMEFT by inserting a SM Yukawa matrix for Class 5 operators reduces parameter tuning.
This then requires a $\rm U(3)^5$ limit of the SMEFT be taken consistently in the counterterm matrices, which requires the insertion of a SM Yukawa matrix for Class 6
operators in the $\rm U(3)^5$-SMEFT and MFV-SMEFT.

Finally, utilizing the SMEFT Lagrangian formalism, dipole operators can be directly related to a shift to a measured anomalous magnetic moment
\bea
\delta a_i = - \frac{4 \, m_i \, v}{\sqrt{2}} \, {\rm Re} \left[\frac{C_{\substack{eB \\ ii}}}{\bar{g}_1} - \frac{C_{\substack{eW \\ ii}}}{\bar{g}_2} \right] + \cdots
\eea
The muon anomalous magnetic moment is highly constrained \cite{Olive:2016xmw} $\delta a_\mu \lesssim 288(63)(49) \times 10^{-11}$ which argues for the neglect of the flavour symmetric component
of this dipole interaction operator Wilson coefficient
when $i = \mu$, but this conclusion does not hold for all $\mathcal{L}^{(6)}$ dipole interaction terms. Experimental constraints on $C_{\substack{\psi X \\rs}}$, with $r \neq s$
are significant in some cases, but not for all possible flavour transitions, see the recent discussion in Refs.~\cite{Raidal:2008jk,Crivellin:2017rmk,Pruna:2014asa,Pruna:2015jhf}.

Furthermore, the separation of flavour diagonal and flavour off diagonal interactions of the dipole operators is a scale dependent distinction,
flavour mixing is extensive in the SMEFT RGE (see Refs.~\cite{Jenkins:2013wua,Alonso:2013hga}), and the scales of the experimental constraints
are separated from the scales used to probe these interactions at the LHC.
The general neglect of all Class 6 operators does not seem to be supported due to experimental constraints in the SMEFT at this time.
%
\begin{table}[h!]
\begin{center}
\small
\begin{minipage}[t]{4.45cm}
\renewcommand{\arraystretch}{1.5}
\begin{tabular}[t]{c|c}
\multicolumn{2}{c}{$1:X^3$} \\
\hline
$Q_G$                & $f^{ABC} G_\mu^{A\nu} G_\nu^{B\rho} G_\rho^{C\mu} $ \\
$Q_{\widetilde G}$          & $f^{ABC} \widetilde G_\mu^{A\nu} G_\nu^{B\rho} G_\rho^{C\mu} $ \\
$Q_W$                & $\epsilon^{IJK} W_\mu^{I\nu} W_\nu^{J\rho} W_\rho^{K\mu}$ \\
$Q_{\widetilde W}$          & $\epsilon^{IJK} \widetilde W_\mu^{I\nu} W_\nu^{J\rho} W_\rho^{K\mu}$ \\
\end{tabular}
\end{minipage}
\begin{minipage}[t]{2.7cm}
\renewcommand{\arraystretch}{1.5}
\begin{tabular}[t]{c|c}
\multicolumn{2}{c}{$2:H^6$} \\
\hline
$Q_H$       & $(H^\dag H)^3$
\end{tabular}
\end{minipage}
\begin{minipage}[t]{5.1cm}
\renewcommand{\arraystretch}{1.5}
\begin{tabular}[t]{c|c}
\multicolumn{2}{c}{$3:H^4 D^2$} \\
\hline
$Q_{H\Box}$ & $(H^\dag H)\Box(H^\dag H)$ \\
$Q_{H D}$   & $\ \left(H^\dag D^\mu H\right)^* \left(H^\dag D_\mu H\right)$
\end{tabular}
\end{minipage}
\begin{minipage}[t]{2.7cm}

\renewcommand{\arraystretch}{1.5}
\begin{tabular}[t]{c|c}
\multicolumn{2}{c}{$5: \psi^2H^3 + \hbox{h.c.}$} \\
\hline
$Q_{eH}$           & $(H^\dag H)(\bar l_p e_r H)$ \\
$Q_{uH}$          & $(H^\dag H)(\bar q_p u_r \widetilde H )$ \\
$Q_{dH}$           & $(H^\dag H)(\bar q_p d_r H)$\\
\end{tabular}
\end{minipage}

\vspace{0.25cm}

\begin{minipage}[t]{4.7cm}
\renewcommand{\arraystretch}{1.5}
\begin{tabular}[t]{c|c}
\multicolumn{2}{c}{$4:X^2H^2$} \\
\hline
$Q_{H G}$     & $H^\dag H\, G^A_{\mu\nu} G^{A\mu\nu}$ \\
$Q_{H\widetilde G}$         & $H^\dag H\, \widetilde G^A_{\mu\nu} G^{A\mu\nu}$ \\
$Q_{H W}$     & $H^\dag H\, W^I_{\mu\nu} W^{I\mu\nu}$ \\
$Q_{H\widetilde W}$         & $H^\dag H\, \widetilde W^I_{\mu\nu} W^{I\mu\nu}$ \\
$Q_{H B}$     & $ H^\dag H\, B_{\mu\nu} B^{\mu\nu}$ \\
$Q_{H\widetilde B}$         & $H^\dag H\, \widetilde B_{\mu\nu} B^{\mu\nu}$ \\
$Q_{H WB}$     & $ H^\dag \tau^I H\, W^I_{\mu\nu} B^{\mu\nu}$ \\
$Q_{H\widetilde W B}$         & $H^\dag \tau^I H\, \widetilde W^I_{\mu\nu} B^{\mu\nu}$
\end{tabular}
\end{minipage}
\begin{minipage}[t]{5.2cm}
\renewcommand{\arraystretch}{1.5}
\begin{tabular}[t]{c|c}
\multicolumn{2}{c}{$6:\psi^2 XH+\hbox{h.c.}$} \\
\hline
$Q_{eW}$      & $(\bar l_p \sigma^{\mu\nu} e_r) \tau^I H W_{\mu\nu}^I$ \\
$Q_{eB}$        & $(\bar l_p \sigma^{\mu\nu} e_r) H B_{\mu\nu}$ \\
$Q_{uG}$        & $(\bar q_p \sigma^{\mu\nu} T^A u_r) \widetilde H \, G_{\mu\nu}^A$ \\
$Q_{uW}$        & $(\bar q_p \sigma^{\mu\nu} u_r) \tau^I \widetilde H \, W_{\mu\nu}^I$ \\
$Q_{uB}$        & $(\bar q_p \sigma^{\mu\nu} u_r) \widetilde H \, B_{\mu\nu}$ \\
$Q_{dG}$        & $(\bar q_p \sigma^{\mu\nu} T^A d_r) H\, G_{\mu\nu}^A$ \\
$Q_{dW}$         & $(\bar q_p \sigma^{\mu\nu} d_r) \tau^I H\, W_{\mu\nu}^I$ \\
$Q_{dB}$        & $(\bar q_p \sigma^{\mu\nu} d_r) H\, B_{\mu\nu}$
\end{tabular}
\end{minipage}
\begin{minipage}[t]{5.4cm}
\renewcommand{\arraystretch}{1.5}
\begin{tabular}[t]{c|c}
\multicolumn{2}{c}{$7:\psi^2H^2 D$} \\
\hline
$Q_{H l}^{(1)}$      & $(H^\dag i\overleftrightarrow{D}_\mu H)(\bar l_p \gamma^\mu l_r)$\\
$Q_{H l}^{(3)}$      & $(H^\dag i\overleftrightarrow{D}^I_\mu H)(\bar l_p \tau^I \gamma^\mu l_r)$\\
$Q_{H e}$            & $(H^\dag i\overleftrightarrow{D}_\mu H)(\bar e_p \gamma^\mu e_r)$\\
$Q_{H q}^{(1)}$      & $(H^\dag i\overleftrightarrow{D}_\mu H)(\bar q_p \gamma^\mu q_r)$\\
$Q_{H q}^{(3)}$      & $(H^\dag i\overleftrightarrow{D}^I_\mu H)(\bar q_p \tau^I \gamma^\mu q_r)$\\
$Q_{H u}$            & $(H^\dag i\overleftrightarrow{D}_\mu H)(\bar u_p \gamma^\mu u_r)$\\
$Q_{H d}$            & $(H^\dag i\overleftrightarrow{D}_\mu H)(\bar d_p \gamma^\mu d_r)$\\
$Q_{H u d}$ + h.c.   & $i(\widetilde H ^\dag D_\mu H)(\bar u_p \gamma^\mu d_r)$\\
\end{tabular}
\end{minipage}

\vspace{0.25cm}

\begin{minipage}[t]{4.75cm}
\renewcommand{\arraystretch}{1.5}
\begin{tabular}[t]{c|c}
\multicolumn{2}{c}{$8:(\bar LL)(\bar LL)$} \\
\hline
$Q_{ll}$        & $(\bar l_p \gamma_\mu l_r)(\bar l_s \gamma^\mu l_t)$ \\
$Q_{qq}^{(1)}$  & $(\bar q_p \gamma_\mu q_r)(\bar q_s \gamma^\mu q_t)$ \\
$Q_{qq}^{(3)}$  & $(\bar q_p \gamma_\mu \tau^I q_r)(\bar q_s \gamma^\mu \tau^I q_t)$ \\
$Q_{lq}^{(1)}$                & $(\bar l_p \gamma_\mu l_r)(\bar q_s \gamma^\mu q_t)$ \\
$Q_{lq}^{(3)}$                & $(\bar l_p \gamma_\mu \tau^I l_r)(\bar q_s \gamma^\mu \tau^I q_t)$
\end{tabular}
\end{minipage}
\begin{minipage}[t]{5.25cm}
\renewcommand{\arraystretch}{1.5}
\begin{tabular}[t]{c|c}
\multicolumn{2}{c}{$8:(\bar RR)(\bar RR)$} \\
\hline
$Q_{ee}$               & $(\bar e_p \gamma_\mu e_r)(\bar e_s \gamma^\mu e_t)$ \\
$Q_{uu}$        & $(\bar u_p \gamma_\mu u_r)(\bar u_s \gamma^\mu u_t)$ \\
$Q_{dd}$        & $(\bar d_p \gamma_\mu d_r)(\bar d_s \gamma^\mu d_t)$ \\
$Q_{eu}$                      & $(\bar e_p \gamma_\mu e_r)(\bar u_s \gamma^\mu u_t)$ \\
$Q_{ed}$                      & $(\bar e_p \gamma_\mu e_r)(\bar d_s\gamma^\mu d_t)$ \\
$Q_{ud}^{(1)}$                & $(\bar u_p \gamma_\mu u_r)(\bar d_s \gamma^\mu d_t)$ \\
$Q_{ud}^{(8)}$                & $(\bar u_p \gamma_\mu T^A u_r)(\bar d_s \gamma^\mu T^A d_t)$ \\
\end{tabular}
\end{minipage}
\begin{minipage}[t]{4.75cm}
\renewcommand{\arraystretch}{1.5}
\begin{tabular}[t]{c|c}
\multicolumn{2}{c}{$8:(\bar LL)(\bar RR)$} \\
\hline
$Q_{le}$               & $(\bar l_p \gamma_\mu l_r)(\bar e_s \gamma^\mu e_t)$ \\
$Q_{lu}$               & $(\bar l_p \gamma_\mu l_r)(\bar u_s \gamma^\mu u_t)$ \\
$Q_{ld}$               & $(\bar l_p \gamma_\mu l_r)(\bar d_s \gamma^\mu d_t)$ \\
$Q_{qe}$               & $(\bar q_p \gamma_\mu q_r)(\bar e_s \gamma^\mu e_t)$ \\
$Q_{qu}^{(1)}$         & $(\bar q_p \gamma_\mu q_r)(\bar u_s \gamma^\mu u_t)$ \\
$Q_{qu}^{(8)}$         & $(\bar q_p \gamma_\mu T^A q_r)(\bar u_s \gamma^\mu T^A u_t)$ \\
$Q_{qd}^{(1)}$ & $(\bar q_p \gamma_\mu q_r)(\bar d_s \gamma^\mu d_t)$ \\
$Q_{qd}^{(8)}$ & $(\bar q_p \gamma_\mu T^A q_r)(\bar d_s \gamma^\mu T^A d_t)$\\
\end{tabular}
\end{minipage}

\vspace{0.25cm}

\begin{minipage}[t]{3.75cm}
\renewcommand{\arraystretch}{1.5}
\begin{tabular}[t]{c|c}
\multicolumn{2}{c}{$8:(\bar LR)(\bar RL)+\hbox{h.c.}$} \\
\hline
$Q_{ledq}$ & $(\bar l_p^j e_r)(\bar d_s q_{tj})$
\end{tabular}
\end{minipage}
\begin{minipage}[t]{5.5cm}
\renewcommand{\arraystretch}{1.5}
\begin{tabular}[t]{c|c}
\multicolumn{2}{c}{$8:(\bar LR)(\bar L R)+\hbox{h.c.}$} \\
\hline
$Q_{quqd}^{(1)}$ & $(\bar q_p^j u_r) \epsilon_{jk} (\bar q_s^k d_t)$ \\
$Q_{quqd}^{(8)}$ & $(\bar q_p^j T^A u_r) \epsilon_{jk} (\bar q_s^k T^A d_t)$ \\
$Q_{lequ}^{(1)}$ & $(\bar l_p^j e_r) \epsilon_{jk} (\bar q_s^k u_t)$ \\
$Q_{lequ}^{(3)}$ & $(\bar l_p^j \sigma_{\mu\nu} e_r) \epsilon_{jk} (\bar q_s^k \sigma^{\mu\nu} u_t)$
\end{tabular}
\end{minipage}
\begin{minipage}[t]{5.5cm}
\vskip 18mm
$\begin{aligned}
H^\dag i\overleftrightarrow{D}_\mu H  &\equiv H^\dag iD_\mu H - (i D_\mu H^\dag)H\\
H^\dag i\overleftrightarrow{D}^I_\mu H  &\equiv H^\dag i\tau^I D_\mu H - (i D_\mu \tau^I H^\dag)H
  \end{aligned}
$

\end{minipage}

\end{center}
\caption{\label{op59}
The $\mathcal{L}^{(6)}$ operators built from Standard Model fields which conserve baryon number, as given in
Ref.~\cite{Grzadkowski:2010es,Jenkins:2013zja,Jenkins:2013wua,Alonso:2013hga}. The operators are divided into eight Classes: $X^3$, $H^6$, etc. Operators with $+\hbox{h.c.}$
in the table heading also have Hermitian conjugates, as does the $\psi^2H^2D$ operator $Q_{Hud}$. The subscripts $p,r,s,t$ are flavour indices which are suppressed on the left hand sides
of the sub-tables.}
\end{table}
%
%
%

\clearpage
\providecommand{\href}[2]{#2}\begingroup\raggedright\endgroup

\end{document}